\documentclass[prx,aps,showpacs,nofootinbib]{revtex4-2}

\usepackage{amsmath, amssymb, amsthm, amsfonts, latexsym, graphicx}
\usepackage{float}
\usepackage{dsfont}
\usepackage{verbatim} 
\usepackage[colorlinks]{hyperref}
\usepackage{url}
\usepackage{xcolor}
\usepackage{subcaption}
\usepackage{booktabs}
\usepackage{multirow}
\usepackage{units}
\usepackage[version=3]{mhchem}
\usepackage[stable]{footmisc}
\usepackage{appendix}

\usepackage{mathbbol} 
\usepackage{soul}
\usepackage{bm}

\newcommand{\bs}{\boldsymbol}	

\let\phi=\varphi

\let\epsilon=\varepsilon

\newcommand{\eq}[1]{Eq.~\hyperref[eq:#1]{(\ref*{eq:#1})}}
\renewcommand{\sec}[1]{\hyperref[sec:#1]{Section~\ref*{sec:#1}}}
\newcommand{\app}[1]{\hyperref[app:#1]{Appendix~\ref*{app:#1}}}
\newcommand{\tab}[1]{\hyperref[tab:#1]{Table~\ref*{tab:#1}}}
\newcommand{\fig}[1]{\hyperref[fig:#1]{Figure~\ref*{fig:#1}}}
\newcommand{\figa}[2]{\hyperref[fig:#1]{Figure~\ref*{fig:#1}#2}}
\newcommand{\figx}[2]{\hyperref[fig:#1]{Figure~\ref*{fig:#1}(#2)}}
\newcommand{\thm}[1]{\hyperref[thm:#1]{Theorem~\ref*{thm:#1}}}
\newcommand{\lem}[1]{\hyperref[lem:#1]{Lemma~\ref*{lem:#1}}}
\newcommand{\cor}[1]{\hyperref[cor:#1]{Corollary~\ref*{cor:#1}}}
\newcommand{\defn}[1]{\hyperref[def:#1]{Definition~\ref*{def:#1}}}
\newcommand{\alg}[1]{\hyperref[alg:#1]{Algorithm~\ref*{alg:#1}}}
\newcommand{\prob}[1]{\hyperref[prob:#1]{Problem~\ref*{prob:#1}}}

\begin{document}

\title{Relativistic Linear Response in Quantum-Electrodynamical Density Functional Theory}
\author{Lukas Konecny}
\email{lukas.konecny@uit.no} 
\affiliation{Hylleraas Centre for Quantum Molecular Sciences, Department of Chemistry, UiT The Arctic University of Norway, N-9037 Troms{\o}, Norway}
\affiliation{Max Planck Institute for the Structure and Dynamics of Matter, Center for Free Electron Laser Science, Luruper Chaussee 149, 22761 Hamburg, Germany}
\author{Valeriia P. Kosheleva}
\email{valeriia.kosheleva@mpsd.mpg.de}
\affiliation{Max Planck Institute for the Structure and Dynamics of Matter, Center for Free Electron Laser Science, Luruper Chaussee 149, 22761 Hamburg, Germany}
\author{Heiko Appel}
\email{heiko.appel@mpsd.mpg.de}
\affiliation{Max Planck Institute for the Structure and Dynamics of Matter, Center for Free Electron Laser Science, Luruper Chaussee 149, 22761 Hamburg, Germany}
\author{Michael Ruggenthaler}
\email{michael.ruggenthaler@mpsd.mpg.de}
\affiliation{Max Planck Institute for the Structure and Dynamics of Matter, Center for Free Electron Laser Science, Luruper Chaussee 149, 22761 Hamburg, Germany}
\author{Angel Rubio}
\email{angel.rubio@mpsd.mpg.de}
\affiliation{Max Planck Institute for the Structure and Dynamics of Matter, Center for Free Electron Laser Science, Luruper Chaussee 149, 22761 Hamburg, Germany}
\affiliation{Center for Computational Quantum Physics (CCQ), The Flatiron Institute, 162 Fifth Avenue, New York, New York 10010, USA}


\begin{abstract}

We present the theoretical derivation and numerical implementation of the linear response equations
for relativistic quantum electrodynamical density functional theory (QEDFT).
In contrast to previous works based on the Pauli--Fierz Hamiltonian,
our approach describes electrons interacting with photonic cavity modes at the four-component Dirac-Kohn-Sham level, derived from fully relativistic QED through a series of established approximations.
Moreover, we show that a new type of spin--orbit-like (SO) cavity-mediated interaction appears
under the relativistic description of the coupling of matter with quantized cavity modes.
Benchmark calculations performed for atoms of group 12 elements (Zn, Cd, Hg) demonstrate how
a relativistic treatment enables the description of exciton polaritons which arise from
the hybridization of formally forbidden singlet--triplet transitions with cavity modes.
For atoms in cavities tuned on resonance with a singlet--triplet transition we discover a significant interplay between SO effects and coupling to an off-resonant intense singlet-singlet transition. This dynamic relationship highlights the crucial role of \textit{ab initio} approaches in understanding cavity quantum electrodynamics.
Finally, using the mercury porphyrin complex as an example, we show that relativistic linear response QEDFT provides computationally feasible first-principles calculations of polaritonic states in large heavy element-containing molecules of chemical interest.

%
\end{abstract}
\date{\today}
\maketitle

\section{Introduction}
The strength of light--matter coupling, normally weak when a material interacts with vacuum modes of free space, can be greatly enhanced by placing the material into an optical cavity.
If a cavity mode is in resonance with an excitation of the material, new hybrid light--matter states called polaritons emerge.~\cite{Ebbesen2016,Ruggenthaler2018, frisk2019ultrastrong,basov2020polariton}
Depending on the nature of the excitation, different types of polaritonic states exist such exciton--polaritons and vibrational polaritons resulting from the coupling of light to electronic
and vibrational transitions, respectively.
As a result a new field of cavity quantum electrodynamical (QED) materials~\cite{Flick2015,Ebbesen2016,Ruggenthaler2018,Schlawin2022,Ruggenthaler2023} has emerged bringing together various platforms for manipulating and engineering quantum materials with electromagnetic fields.
It encompasses a wide spectrum of different fields including quantum optics~\cite{Tannoudji1997}, polaritonic chemistry~\cite{Ebbesen2016,Feist2017,Ruggenthaler2018,Li2020,Sidler2022,mandal2023theoretical,simpkins2023control,Ruggenthaler2023}, the generation of light-induced states of matter through classical fields~\cite{Basov2017,Buzzi2018} or quantum fields emanating from a cavity~\cite{Wang2019,Kiffner2019,Li:205140:2020,Ashida2021}. This rapidly growing field has attracted significant attention from both theorists and experimentalists, leading to the exploration of various effects under
strong light--matter coupling.
%
Strong light--matter interaction in a cavity enables the modification of chemical reaction landscapes~\cite{Hutchison2012,Anoop2016,li2021cavity,Garcia-Vidal2021,Schaefer2022,ahn2023modification} and ground states of matter~\cite{Flick2018,Latini2021,Rokaj2023,Bostrom2022}, enhances charge and energy transfer~\cite{Coles2014,Orgiu2015,schachenmayer2015cavity,Zhong2016,Feist2015,Stranius2018}, and allows for selective manipulation of electronic states~\cite{Stranius2018}, the modification of electron--phonon coupling and superconductivity~\cite{Cotlet2016,Sentef2018,Curtis2019,Schlawin2019}, controlling of excitons~\cite{Latini2019,Levinsen2019,Forg2019} while achieving exciton-polariton condensation~\cite{Kasprzak2006,Keeling2020}. In recent studies, Quantum Hall systems, whether operating in the integer regime~\cite{Hagenmuller2010,Scalari2012,Li2018,Keller2020,rokaj2022polaritonic} or the fractional regime~\cite{Ravets2018,Smolka2014}, have exhibited remarkable phenomena when confined within optical cavities. These systems have shown what is known as ultra-strong coupling with the light field, resulting in significant changes in their transport properties~\cite{Paravicini-Bagliani2019}. Furthermore, there is a growing interest in exploring the implications of coupling with chiral electromagnetic fields, and this topic is currently being actively investigated~\cite{Petersen2014,Lodahl2017,Zhang2019,Hubener2021}.
This regime of the strong coupling between matter and light can be achieved due to collective effects, when a large number of molecules is placed in a cavity and coherently interact with its modes, or for micro- and nano-cavities that confine light to very small length scales~\cite{Ebbesen2016,chikkaraddy2016single,kavokin2017microcavities,frisk2019ultrastrong,Ruggenthaler2023}.

To describe the interaction of the material with vacuum fluctuations i.e. cavity modes, the quantum description of light is important making QED the theory of choice.
The standard practical approach is provided by models of cavity QED,
which describe a system of photons strongly
coupled to a material approximated as a few-level system,
e.g. the Jaynes--Cummings~\cite{Jaynes1963} and the Dicke~\cite{Dicke1954} model.
 However, if we want to capture subtle 
 and complex changes of the material inside a cavity, a first principles description of the matter subsystem becomes necessary.~\cite{Ruggenthaler2018}. 

To bridge the worlds of quantum chemistry and quantum optics, new computational methods that simultaneously allow \textit{ab initio} description of matter while treating transverse photons as dynamical variables, were developed~\cite{Ruggenthaler2018, mandal2023theoretical, Foley2023, Ruggenthaler2023}. 
The first such method was quantum electrodynamical density functional theory (QEDFT) proposed for time-dependent~\cite{Ruggenthaler2011,Ruggenthaler2014,tokatly2013} and ground-state/coupled-vacuum properties~\cite{Ruggenthaler2015,Penz2023}. Besides non-interacting auxiliary systems in the Kohn-Sham formulation of QEDFT also alternative, explicitly correlated approaches were introduced~\cite{nielsen2018dressed}.
The first computational implementations of QEDFT were reported by Flick et al. for time-dependent situations~\cite{Flick2015, Flick2017}, for ground-state properties~\cite{Flick2018} and finally as a linear response theory of coupled matter--photon system and formulated on a real-space grid.~\cite{Flick2019} 
It motivated subsequent implementations based on Gaussian orbitals~\cite{Yang2021, Liebenthal2023}
as well as a development of a real-space Sternheimer formalism~\cite{Welakuh2022}.
Moreover, the work of Yang et al.~\citep{Yang2021} also considers various simplified cases
based on the neglect of the dipole self-energy terms,
on the Tamm--Dancoff approximation (TDA) for electrons~\cite{Hirata1999} as well as
on an analogous approximation for photons,
together with the different combinations of these approximations.
In addition to QEDFT, \textit{ab initio} theories combining QED with wave function based quantum
chemical methods were developed
starting with quantum electrodynamical Hartree--Fock (QED-HF) theory~\cite{buchholz2019reduced,Haugland2020, Buchholz2020},
and ranging to quantum electrodynamical coupled clusters (QED-CC)~\cite{Haugland2020, Mordovina2020} and
quantum electrodynamical configuration interaction singles (QED-CIS)~\cite{McTague2022},
as well as Green function-based approaches~\cite{Melo2016} and reduced density-matrix theories~\cite{buchholz2019reduced,Buchholz2020}.

Most of these realizations of QEDFT were based on the non-relativistic Pauli--Fierz Hamiltonian.
However, there exists a plethora of phenomena called relativistic effects
that are not covered by the non-relativistic, Schr\"{o}dinger or Pauli equation-based treatment and are
described correctly only within relativistic theories based on the Dirac equation.~\cite{Pyykko2012}
The most prominent examples include the color of gold~\cite{Pyykko1979, Romaniello2007}, the liquid state of mercury~\cite{Calvo2013},
and the voltage of the lead--acid battery~\cite{Ahuja2011}.
Further manifestations of relativistic effects in atomic, molecular, and solid state systems involve 
bound-electron $g$ factor~\cite{Kosheleva2022,Glazov2019},
electron absorption and X-ray spectra~\cite{Laskowski2010, List2016, Vidal2020, Kasper2020, Konecny2022},
NMR~\cite{Hrobarik2011-NMR, Hrobarik2012, Vicha2016-2, Vicha2020},
pNMR~\cite{Martin2015, Novotny2016, Mondal2017, Cherry2017, Komorovsky2023}
and EPR spectroscopies~\cite{Malkin2005, Hrobarik2011-EPR, Gohr2015, Bolvin2016, Misenkova2022, Komorovsky2023},
bond lengths,~\cite{Cuyacot2022}
reaction mechanisms~\cite{Demissie2016},
phosphorescence and decay pathways~\cite{Moitra2021},
properties of superheavy elements~\cite{Pershina2015, Giuliani2019},
band structures~\cite{Kadek2023},
band gap opening,~\cite{Rashba1988, Gmitra2009, Zhu2011}
and the stability of structural phases~\cite{Hermann2010}.
Relativistic quantum chemistry~\cite{Dyall2007, Reiher2014} addresses the relativistic effects by describing electrons at the level
of the Dirac equation. The gold standard is represented by the four-component (4c) Dirac--Coulomb Hamiltonian that combines
the one-electron Dirac equation with the instantaneous Coulomb interaction among electrons and nuclei and electrons with each other.~\cite{Saue2011}
This Hamiltonian contains relativistic kinetic energy as well as one-electron spin--orbit coupling and is sometimes
extended to include Breit terms.~\cite{Saue2011, Kelley2013,  Sun2023, Hoyer2023}
Moreover, there is an ongoing research effort focused on including the effects of relativistic QED 
in quantum chemistry to account for radiative corrections~\cite{Mackenzie1980, Shabaev2013, Aucar2014, Schwerdtfeger2015, Pasteka2017, Malyshev2022, Sunaga2022, Inoue2023}.
At the same time, researchers are developing approximate two-component Hamiltonians that
would capture the most important relativistic effects at a lower computational cost.~\cite{Kutzelnigg2005, Liu2007, Ilias2007, Knecht2022}

QEDFT with relativistic Dirac equation-based treatment of electronic structure thus enables the correct description
of molecules containing any element across the periodic table when strongly coupled to cavity modes (under certain stability conditions detailed in Sec.~\ref{sec:Approximations}).
Moreover, the 4c description includes SOC variationally thus allowing first-principles
access to singlet--triplet transitions that are forbidden according to non-relativistic theories.
While spin--orbit coupling can be added perturbatively, usually up to the
first order in $\frac{Z}{c^2}$ ($Z$ being the atomic number and $c$ the
speed of light), the fully relativistic treatment accounts for 
the higher order effect that can become prominent for high-$Z$
systems.~\cite{Liao2023}
In addition, the linear response properties such as the coupling strengths between the singlet and triplet states or the phosphorescence parameters
are correctly obtained from the first-order perturbation theory~\cite{Fransson2016} instead of
requiring the more demanding quadratic response as they would with perturbative SOC.~\cite{Aagren1996}
Another case for relativistic QEDFT is provided by cavity engineering of
the energy ratio between singlet--singlet and singlet--triplet excited states.
This was suggested to improve the yields of singlet fission~\cite{Climent2022} and 
intersystem crossing~\cite{Stranius2018, Eizner2019}, particularly
for heavy element-containing systems where the rate of these processes is enhanced
by the increased spin–orbit coupling.~\cite{Kena2007, Hertzog2019}
Additionally, the inclusion of SOC in QEDFT theory will enable to compute spin--orbit related phenomena in complex nanostructures and solids.~\cite{Xu2014, YangSH2021}

At the same time, relativistic QEDFT calculations close the gap between the stable and non-perturbative low-energy sector of QED described by the Pauli--Fierz quantum field theory~\cite{spohn2004dynamics,hiroshima2019ground,Ruggenthaler2023} and a fully second-quantized QED formulation in various flavours~\cite{ryder1996quantum,mandl2010quantum,takaesu2009spectral}. Even without cavities and their changes in the modes of the electromagentic field, there are many fundamental questions still to be answered within QED as a general framework to describe on the most fundamental level the interaction between charged particles and light.~\cite{baez2014introduction}

In this work, we introduce a \textit{relativistic} quantum-electrodynamical density functional theory (QEDFT) for molecules that combines four-component Dirac--Kohn--Sham treatment of electrons with quantized
description of photonic modes.
We first present the theoretical foundation of the developed method where we derive the relativistic Hamiltonian
for the coupled light--matter system in the long-wavelength approximation.
We demonstrate the emergence of a novel spin-orbit-like cavity-mediated interaction within the framework of relativistic coupling between matter and quantized cavity modes.
Then in Sec.~\ref{sec:LRtheory} we formulate a linear response equation in the framework of QEDFT for its excitation energies that allows the first-principles calculation of
polaritonically modified spectra and present its numerical implementation. As an example, we calculate  the absorption spectra of Zn, Cd, and Hg atoms
focusing on the singlet--triplet transitions in a cavity. Via careful analysis of the absorption spectra we show that there is a significant interplay between spin-orbit effects and off-resonant coupling to the intense singlet-singlet transition. This interplay highlights the crucial role of the \textit{ab initio} approach in understanding cavity quantum electrodynamics.
Finally, we considered a large heavy element complex
to demonstrate the applicability of our method to systems of chemical interest.

\section{Theory}
The SI system of units is used throughout the paper unless specified otherwise. For additional notations and conventions see Appendix A.
\subsection{Light-matter coupling in fully-relativistic limit: total Hamiltonian of a system}
We consider a system of electrons minimally coupled to photons in the presence of external classical fields. 
The full QED Hamiltonian of such a system in Coulomb gauge is formally given as~\cite{greiner,Ruggenthaler2014}
\begin{align}
\begin{split}
\hat{H}_\mathrm{QED}(t)
& =
\int d {\bf r} :\hat{\Bar{\psi}}({\bf r}) (-i\hbar c \boldsymbol{\gamma}\cdot\boldsymbol{\nabla} + \alpha^{0}mc^2 ) \hat{\psi}({\bf r}) : 
+ \frac{1}{c} \int d {\bf r} :\hat{j}^0({\bf r}): a_0^\mathrm{ext} ({\bf r},t) \\
& + \frac{1}{c^2} \int d {\bf r} d {\bf r}' \frac{j_0^\mathrm{ext}({\bf r},t) :\hat{j}^0({\bf r}'):}{4\pi\epsilon_0|{\bf r}-{\bf r'}|}  + \frac{1}{2c^2} \int d {\bf r}  d {\bf r}' \frac{:\hat{j}_0({\bf r}) \hat{j}^0({\bf r}'):}{4\pi\epsilon_0|{\bf r}-{\bf r'}|}  \\
& - \frac{1}{c} \int d {\bf r} :\hat{\mathbf{j}}({\bf r}) (\mathbf{a}^\mathrm{ext} ({\bf r},t)  + \hat{\mathbf{A}}({\bf r})): 
-\frac{1}{c} \int d \mathbf{r} \mathbf{j}^{ \text { ext }}({\bf r},t) \hat{\mathbf{A}}({\bf r})\\
& +  \frac{\epsilon_0}{2} \int d {\bf r} : (\hat{\mathbf{E}}_{\perp}^2({\bf r}) + c^2\hat{\mathbf{B}}^2({\bf r})):.
\label{H_PF}
\end{split}
\end{align}
Here $c$ is the speed of the light in free-space vacuum, $\hbar$ is the reduced Planck constant, $\epsilon_0$ is the vacuum permittivity, $m$ is the bare mass of the electron, $\alpha^{\mu}=\gamma^{0} \gamma^{\mu}$ are the  Dirac matrices in the standard representation (see Appendix A).
Here all the operators are in the Schr\"odinger picture and $::$ denotes their normal ordering.

The electron-positron field operator $\hat{\psi}({\bf r})$ is given by~\cite{peskin,greiner}
\begin{align}
\begin{split}
\hat{\psi}({\bf r})&=
\sum_{\mu= \pm \frac{1}{2}} \int d{\bf p}\left(\hat{a}_{{\bf p} \mu}\psi^{(+)}_{{\bf p}\mu}({\bf r})
+\hat{b}^{\dagger}_{{\bf p} \mu} \psi^{(-)}_{{\bf p}\mu}({\bf r})\right),
\end{split}
\end{align}
where $\hat{a}_{{\bf p} \mu}$ is an annihilation operator of the electron with momentum ${\bf p} = (p_x, p_y, p_z)$ and helicity $\mu$, operator $\hat{b}^{\dagger}_{{\bf p} \mu}$ is a creation operator for the positron with the same momenta and helicity.
Here $\psi^{(+)}_{{\bf p}\mu}({\bf r})$ and $\psi^{(-)}_{{\bf p}\mu}({\bf r})$ are the solutions of the free Dirac equation for the electron and positron, respectively.
The electron-positron field operator $\hat{\psi}({\bf r})$ formally obeys the usual Heisenberg equation of motion (here we indicate with label $\mathrm{H}$ the Heisenberg picture)
\begin{equation}
\frac{d}{d t} \hat{\psi}_{\mathrm{H}}({\bf r},t) =\frac{i}{\hbar}\left[ \hat{H}_\mathrm{QED, H}(t),\hat{\psi}_{\mathrm{H}}({\bf r},t)\right].
\end{equation}
\\
\indent
The Dirac four-current operator $\hat{J}({\bf r})$ is constructed as $\hat{J}({\bf r}) = \left(\hat{j}_0({\bf r}), \hat{\mathbf{j}}({\bf r}) \right)$ with the components having the following form
\begin{align}
\hat{j}_0({\bf r}) &= e c :\hat{\Bar{\psi}}({\bf r}) \gamma_{0} \hat{\psi}({\bf r}):, \\
\hat{\mathbf{j}}({\bf r}) &= e c :\hat{\Bar{\psi}}({\bf r}) \boldsymbol{\gamma} \hat{\psi}({\bf r}):,
\end{align}
where $\hat{\Bar{\psi}}({\bf r}) \equiv \hat{\psi}^{\dagger}({\bf r})\gamma^{0}$.
\\
\indent
The quantized vector potential $\hat{\mathbf{A}}({\bf r})$ in Coulomb gauge is given by~\cite{greiner} \footnote{We note that in comparison with the usual definition of a vector potential, we here have an extra multiplication by $c$ in order to have the units of $A_{\mu}$ in Volts.}
\begin{equation}
\hat{\mathbf{A}}({\bf r})=\sqrt{\frac{c^2 \hbar }{\epsilon_0(2 \pi)^3}} \int \frac{\mathrm{d} \mathbf{k}}{\sqrt{2 \omega_{\mathbf{k}}}} \sum_{\lambda=1}^2 \left[\hat{a}(\mathbf{k}, \lambda) \mathrm{e}^{\mathrm{i} \mathbf{k}\cdot \mathbf{r}} \boldsymbol{\epsilon}(\mathbf{k}, \lambda)+\hat{a}^{\dagger}(\mathbf{k}, \lambda) \mathrm{e}^{-\mathrm{i} \mathbf{k} \cdot \mathbf{r}}\boldsymbol{\epsilon}^{*}(\mathbf{k}, \lambda)\right],
\label{A_quant}
\end{equation}
where $\hat{a}(\mathbf{k}, \lambda)$ and $\hat{a}^{\dagger}(\mathbf{k}, \lambda)$ are the creation and annihilation operators of a photon with momentum $\mathbf{k}$ and helicity $\lambda$, respectively; $\omega_{\mathbf{k}}=c|\mathbf{k}|$ is the energy of the photon and $\boldsymbol{\epsilon}(\mathbf{k}, \lambda)$ is the transverse dimensionless polarization vector that obeys~\cite{greiner}
\begin{equation}
\mathbf{k} \cdot \boldsymbol{\epsilon}(\mathbf{k}, \lambda)= \boldsymbol{\epsilon}^*(\mathbf{k}, 1) \cdot \boldsymbol{\epsilon}(\mathbf{k}, 2)=0.
\label{polarization}
\end{equation}
 We note that in a Coulomb gauge the field $\hat{\mathbf{A}}({\bf r})$ is fully transverse, i. e. $\boldsymbol{\nabla}\cdot \hat{\mathbf{A}}({\bf r}) =0 $.
\\
\indent
The vector potential $\hat{\mathbf{A}}({\bf r})$ formally obeys the following equation of motion
\begin{equation}
\frac{d}{d t} \hat{\mathbf{A}}_{\mathrm{H}}({\bf r},t) =\frac{i}{\hbar}\left[ \hat{H}_\mathrm{QED, H}(t), \hat{\mathbf{A}}_{\mathrm{H}}({\bf r},t)\right].
\end{equation}
The last term in Eq.~\eqref{H_PF} is the free photon Hamiltonian $\hat{H}_{\text {Ph,free }}$ which creates and annihilates only transverse photons.
Using Eq.~\eqref{A_quant} one can rewrite it as
\begin{equation}
    \hat{H}_{\text {Ph,free }} =\frac{\epsilon_0}{2} \int d {\bf r} : (\hat{\mathbf{E}}_{\perp}^2({\bf r}) + c^2\hat{\mathbf{B}}^2({\bf r})): = \sum_{\lambda=1}^2 \int \mathrm{d} \mathbf{k} \hbar \omega_{\mathbf{k}}\hat{a}^{\dagger}(\mathbf{k}, \lambda) \hat{a}(\mathbf{k}, \lambda).
\end{equation}
Here $\hat{\mathbf{E}}_{\perp}({\bf r})=\left[-\frac{\partial \hat{\mathbf{A}}_{\rm H}({\bf r},t)
}{c\partial t} \right]_{\rm S}$ with the subindex S denoting the Schr\"odinger picture, and $\hat{\mathbf{B}}({\bf r})= \frac{1}{c} \boldsymbol{\nabla} \times \hat{\mathbf{A}}({\bf r})$, are quantized transverse electric and magnetic fields, respectively~\cite{greiner},
that expressed via photonic creation and annihilation operators assume the form
\begin{align}
\hat{\mathbf{E}}_{\perp}({\bf r})&=\sqrt{\frac{\hbar }{\epsilon_0(2 \pi)^3}} \int \frac{\mathrm{d} \mathbf{k}i  \omega_{\mathbf{k}}}{\sqrt{2 \omega_{\mathbf{k}}}} \sum_{\lambda=1}^2 \left[\hat{a}(\mathbf{k}, \lambda) \mathrm{e}^{\mathrm{i} \mathbf{k}\cdot \mathbf{r}} \boldsymbol{\epsilon}(\mathbf{k}, \lambda)-\hat{a}^{\dagger}(\mathbf{k}, \lambda) \mathrm{e}^{-\mathrm{i} \mathbf{k} \cdot \mathbf{r}}\boldsymbol{\epsilon}^{*}(\mathbf{k}, \lambda)\right],\\
\hat{\mathbf{B}}({\bf r})&=\sqrt{\frac{\hbar }{\epsilon_0(2 \pi)^3}} \int \frac{\mathrm{d} \mathbf{k} }{\sqrt{2 \omega_{\mathbf{k}}}} \sum_{\lambda=1}^2 \left[\hat{a}(\mathbf{k}, \lambda) \mathrm{e}^{\mathrm{i} \mathbf{k}\cdot \mathbf{r}} [\mathbf{k} \times \boldsymbol{\epsilon}(\mathbf{k}, \lambda)] +\hat{a}^{\dagger}(\mathbf{k}, \lambda) \mathrm{e}^{-\mathrm{i} \mathbf{k} \cdot \mathbf{r}}[\mathbf{k} \times \boldsymbol{\epsilon}^{*}(\mathbf{k}, \lambda)]\right].
\end{align}
\\
\indent
In our system, we also introduce external (with respect to the system's electrons plus photons) classical magnetic vector potential $\mathbf{a}^\mathrm{ext} ({\bf r},t)$ and scalar potential $a^\mathrm{ext}_0 ({\bf r},t)$ as well as classical external four-currents $J^\mathrm{ext}({\bf r},t) = \left(j_0^\mathrm{ext}({\bf r},t)=c\rho^\mathrm{ext}({\bf r},t), \mathbf{j}^\mathrm{ext}({\bf r},t) \right)$ with $\rho^\mathrm{ext}({\bf r},t)$ being a charge density. 
Using Eq.~\eqref{A_quant} one can rewrite the energy due to coupling to a classical external charge current $\mathbf{j}^\mathrm{ext}({\bf r},t)$
as~\cite{Ruggenthaler2015}
\begin{equation}
\frac{1}{c} \int d \mathbf{r} \mathbf{j}^{ \text { ext }}({\bf r},t) \hat{\mathbf{A}}({\bf r})=\int d \mathbf{k}\hbar \omega_{\mathbf{k}}\left(\hat{a}(\mathbf{k}, \lambda) [j^{\text { ext }}]^{*}(\mathbf{k}, \lambda,t)+j^{\text { ext }}(\mathbf{k}, \lambda,t) \hat{a}^{\dagger}(\mathbf{k}, \lambda)\right),
\label{coupling_external_current}
\end{equation}
with the expansion coefficients
$j^{\text { ext }}(\mathbf{k}, \lambda, t)=[j^{\text { ext }}(-\mathbf{k}, \lambda, t)]^*= \left(\frac{1}{2} \omega_{\mathbf{k}}^3 \epsilon_0 \hbar (2\pi)^3\right)^{-1 / 2} \int d \mathbf{r} \boldsymbol{\epsilon}^*(\mathbf{k}, \lambda) \cdot \mathbf{j}^{ \text { ext }}(\mathbf{r},t) \mathrm{e}^{ -\mathrm{i} \mathbf{k} \cdot \mathbf{r}}$. We note, that with these expansion coefficients, one can recover the transverse part of the external current since $\boldsymbol{\epsilon}(\mathbf{k}, \lambda)$ is a transverse vector
\begin{equation}
\mathbf{j}^{ \text { ext }}_{\perp}({\bf r},t)=\sqrt{\frac{ \hbar \epsilon_0}{(2\pi)^3}} \int \frac{\mathrm{d} \mathbf{k} \omega_{\mathbf{k}}^2}{\sqrt{2 \omega_{\mathbf{k}}}} \sum_{\lambda=1}^2 \left[j^{\text { ext }}(\mathbf{k}, \lambda,t)\mathrm{e}^{\mathrm{i} \mathbf{k}\cdot \mathbf{r}} \boldsymbol{\epsilon}(\mathbf{k}, \lambda)+[j^{\text { ext }}(\mathbf{k}, \lambda,t)]^* \mathrm{e}^{-\mathrm{i} \mathbf{k} \cdot \mathbf{r}}\boldsymbol{\epsilon}^{*}(\mathbf{k}, \lambda)\right].
\label{j_exp}
\end{equation}
\subsection{Approximations}
\label{sec:Approximations}
We note that so far, we have chosen implicitly circularly polarised vectors $\boldsymbol{\epsilon}(\mathbf{k}, \lambda)$.
However, for the quantization of the electromagnetic field, one can choose any pair of polarization vectors obeying \eqref{polarization}, for example, linearly polarised ones. We also note that circular and linear representations are connected by a canonical transformation that preserves the form of the equations. %
Therefore,  without loss of generality, we restrict ourselves to the case of linear polarisation, i.e., real transverse polarization vectors.
Then the coupling term from Eq.~\eqref{coupling_external_current} reads as
\begin{equation}
\frac{1}{c} \int d \mathbf{r} \mathbf{j}^{ \text { ext }}({\bf r},t) \hat{\mathbf{A}}({\bf r})=\int d \mathbf{k}\hbar \omega_{\mathbf{k}}j^{\text { ext }}(\mathbf{k}, \lambda, t)\left(\hat{a}(\mathbf{k}, \lambda) + \hat{a}^{\dagger}(\mathbf{k}, \lambda)\right),
\label{coupling_external_current_real_pol}
\end{equation}
with the real expansion coefficients
\begin{equation}
j^{\text { ext }}(\mathbf{k}, \lambda, t)= [j^{\text { ext }}(\mathbf{k}, \lambda, t)]^* = \left(\frac{1}{2} \omega_{\mathbf{k}}^3 \epsilon_0 \hbar (2\pi)^3\right)^{-1 / 2} \int d \mathbf{r} \boldsymbol{\epsilon}(\mathbf{k}, \lambda) \cdot \mathbf{j}^{ \text { ext }}(\mathbf{r}, t) \exp^{-\mathrm{i} \mathbf{k} \cdot \mathbf{r}}.
\end{equation}
Next, we would like to work with a stable vacuum, i.e. the energy gap between the positive and negative energy solutions of the Dirac equation is nonzero. If the intensity of the external fields is small compared to the Schwinger limit, and if one considers the long-wavelength approximation then no pair production occurs~\cite{reinhardt1977quantum,Selsto2009,Deckert2010}. The energy gap then substitutes the boundedness-from-below (minimal-energy state) of non-relativistic theories. 
Since no pair creation is present, in addition to charge conservation, particle-number conservation is satisfied, allowing us to proceed to a first-quantized formulation for electrons.
%
Specifically due to the long-wavelength approximation, Eq.~\eqref{A_quant} can be written as
\begin{equation}
\hat{\mathbf{A}}=\sqrt{\frac{\hbar c^2}{\epsilon_0(2 \pi)^3}} \int \frac{\mathrm{d} \mathbf{k}}{\sqrt{2 \omega_{\mathbf{k}}}} \sum_{\lambda=1}^2 \boldsymbol{\epsilon}(\mathbf{k}, \lambda)\left[\hat{a}(\mathbf{k}, \lambda) + \hat{a}^{\dagger}(\mathbf{k}, \lambda)\right],
\label{A_quant_lin_dip}
\end{equation}
and then the Hamiltonian of Eq.~\eqref{H_PF}
takes a form
\begin{align}
\begin{split}
\hat{H}_\mathrm{QED}^\mathrm{Dip}(t)
&\approxeq\sum_{l=1}^N[-i\hbar c \boldsymbol{\alpha}_l\cdot\boldsymbol{\nabla}_l + \beta_l mc^2
+  e a_0^\mathrm{ext} ({\bf r}_l,t)]+  \frac{e}{c} \sum_{l=1}^N\int d {\bf r}  \frac{j_0^\mathrm{ext}({\bf r},t) }{4\pi\epsilon_0|{\bf r}-{\bf r}_l|} \\
& +  \frac{1}{2} \sum_{l \neq m}^N \frac{e^2}{4 \pi \varepsilon_0\left|{\bf r}_l-{\bf r}_m\right|} -e \sum_{l=1}^N  (\boldsymbol{\alpha}_l\cdot\mathbf{a}^\mathrm{ext} (t)  + \boldsymbol{\alpha}_l\cdot\hat{\mathbf{A}}) \\
& +   \sum_{\lambda=1}^2 \int \mathrm{d} \mathbf{k} \hbar \omega_{\mathbf{k}}
[
\hat{a}^{\dagger}(\mathbf{k}, \lambda) \hat{a}(\mathbf{k}, \lambda)
-
j^{\text { ext }}(\mathbf{k}, \lambda, t)\left(\hat{a}(\mathbf{k}, \lambda) + \hat{a}^{\dagger}(\mathbf{k}, \lambda)\right)
],
\label{H_PF_non_q}
\end{split}
\end{align}
where $N$ is the number of electrons in the system. 
%
%
Assume we are dealing with a cavity, in which case the free-space mode expansion \eqref{A_quant_lin_dip} can be converted into the corresponding expansion for the photonic structure e. g. Fabry--Pérot cavity~\cite{Rokaj2018}. Since we are working in a dipole approximation, i.e. no momentum is transferred in a cavity, the actual spatial mode structure and the momentum matching are no longer important. As a result, one can modify free-space frequencies, coupling strength, and polarizations to fit a given cavity structure without breaking fundamental symmetries~\cite{Ruggenthaler2023}
%
%
%
%
%
\begin{equation}
\hat{\mathbf{A}}= \sqrt{\hbar c^2}\sum_{\alpha}^{M_p} g_{\alpha}\boldsymbol{\epsilon}_{\alpha}\hat{Q}_{\alpha}.
\label{A_cavity}
\end{equation}
Here we have assumed a discretized continuum of $M_p$ modes labeled by $\alpha$, with each $\alpha$ corresponding to a different frequency $\omega_{\alpha}$, coupling strength $g_{\alpha}$, and polarization $\boldsymbol{\epsilon}_{\alpha}$. The modes are given in terms of generalized coordinates $\hat{Q}_{\alpha}$ such as
\begin{align}
\label{eq:AnnihilationOperator}
& \hat{a}_{\alpha}=\sqrt{\frac{\omega_{\alpha}}{2}}\left(\hat{Q}_{\alpha}+
\frac{1}{\omega_{\alpha}}
\frac{\partial}{\partial \hat{Q}_{\alpha}}\right), \\
\label{eq:CreationOperator}
& \hat{a}_{\alpha}^{\dagger}=\sqrt{\frac{\omega_{\alpha}}{2}}\left(\hat{Q}_{\alpha}-\frac{1}{\omega_{\alpha}}\frac{\partial}{\partial \hat{Q}_{\alpha}}\right),
\\
& \left[\hat{Q}_{\alpha},-i \frac{\partial}{\partial \hat{Q}_{\alpha'}}\right]=i\delta_{\alpha \alpha'},
\\
& \left[\hat{a}_{\alpha}, \hat{a}_{\alpha'}^{\dagger}\right]=\delta_{\alpha \alpha'},
\label{qandp}
\end{align}
where  $-i \frac{\partial}{\partial \hat{Q}_{\alpha}}$ is conjugate momenta of a generalized coordinate, $\hat{a}_{\alpha}$ and $\hat{a}_{\alpha}^{\dagger}$ are annihilation and creation of the photon in a mode $\alpha$, respectively.
We note that in the case of free space with a quantization volume $l^3$, the quantities from Eq.~\eqref{A_cavity} correspond to $\mathbf{k}_{\mathbf{m}}=2 \pi \mathbf{m} / l$, $\alpha \equiv\left(\mathbf{k}_{\mathbf{m}}, \lambda\right), \omega_\alpha=c\left|\mathbf{k}_{\mathbf{m}}\right|$ and $g_\alpha=\sqrt{1 / \epsilon_0 l^3}$, where $\mathbf{m} \in \mathbb{Z}^3$. In the case of general photonic structures the respective quantities can be determined from e.g. macroscopic QED~\cite{svendsen2023molecules}.
\begin{figure}
    \centering
    \includegraphics[width=0.4\textwidth]{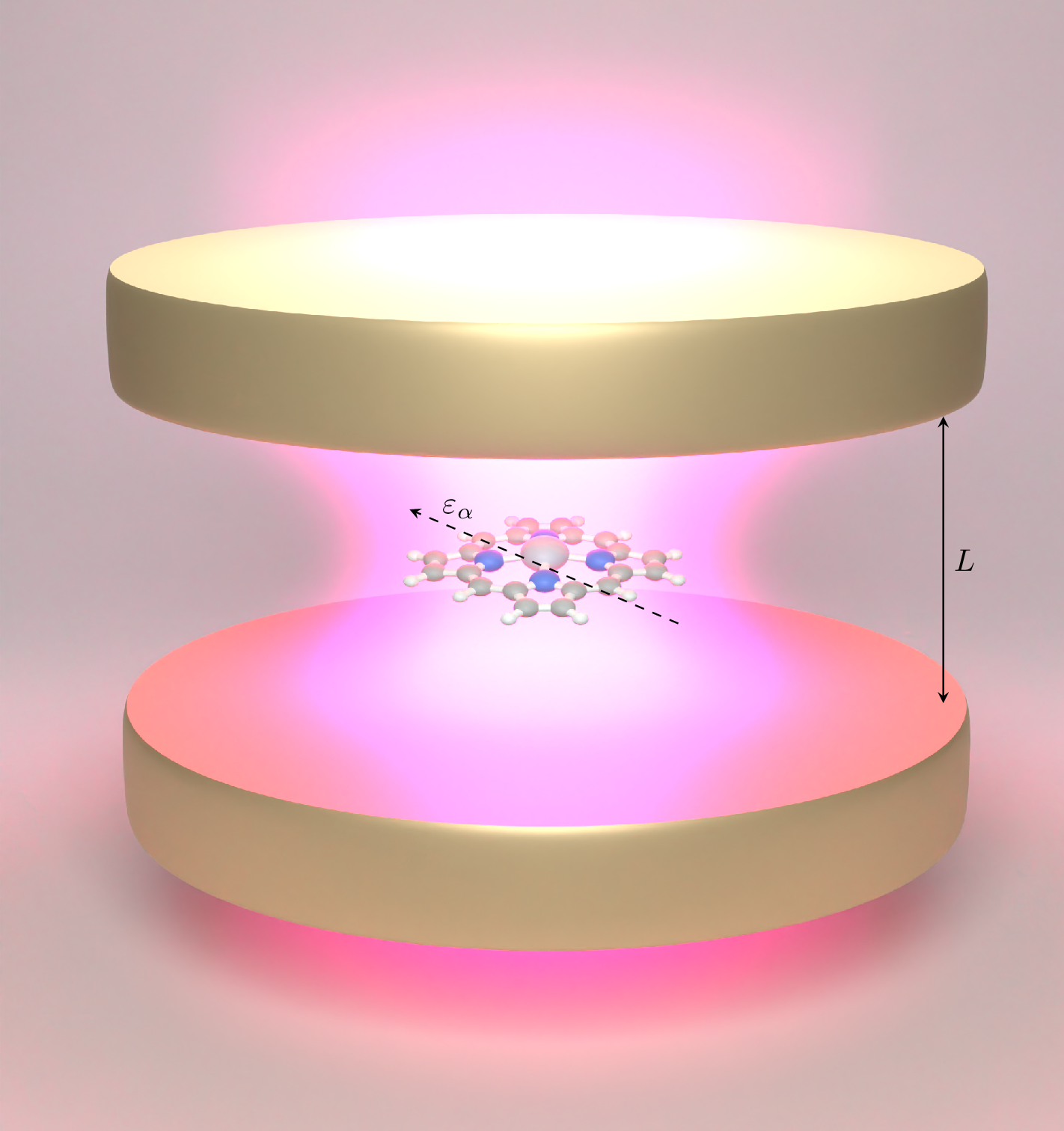}
    \caption{Schematic representation of a Fabry--P\'{e}rot cavity. We assume that molecules of interest are localized around a cavity's center, far away from the cavity mirrors. Then one can approximate the main frequency due to the distance between mirrors $L$ by $\omega_n=c \frac{\pi}{L}|\bf{n}|$ with $\bf{n}$ parallel to L. The influence of the continuum free-space modes (perpendicular to L) is incorporated in the observed mass of the particles. The coupling strength $g_{\alpha \equiv (n,\lambda) }$ for the two independent polarization directions $\lambda$ of the polarization vector $\bf{\epsilon}_{\alpha \equiv (n,\lambda) }$ increases proportionally to the $\sqrt{\frac{1}{V}}$, where $V$ is an effective volume of a cavity.}
    \label{fig:cavity}
\end{figure}
Using Eqs.~\eqref{A_cavity}-\eqref{qandp} one can rewrite Eq.~\eqref{H_PF_non_q} as
\begin{align}
\begin{split}
\hat{H}_\mathrm{QED,cavity}^\mathrm{Dip}(t)
&\approxeq\sum_{l=1}^N[-i\hbar c \boldsymbol{\alpha}_l\cdot\boldsymbol{\nabla}_l + \beta_l mc^2
+  e a_0^\mathrm{ext} ({\bf r}_l,t)] \\
& +  \frac{1}{2} \sum_{l \neq m}^N \frac{e^2}{4 \pi \varepsilon_0\left|{\bf r}_l-{\bf r}_m\right|} 
-e \sum_{l=1}^N  (\boldsymbol{\alpha}_l\cdot\mathbf{a}^\mathrm{ext} (t)  + \boldsymbol{\alpha}_l\cdot\hat{\mathbf{A}}) \\
& +   \sum_{\alpha=1}^{M_{p}}
\left(
-\frac{\hbar}{2} \frac{\partial^2}{\partial \hat{Q}_{\alpha}^2}
+\frac{\hbar \omega_\alpha^2}{2} \hat{Q}_{\alpha}^2 
-\sqrt{2\hbar^2 \omega_\alpha^3}j_{\alpha}^{\mathrm{ext}}(t)\hat{Q}_{\alpha}
\right),
\label{H_PF_pandq}
\end{split}
\end{align}
where 
%
%
%
\begin{equation}
j_{\alpha}^{\mathrm{ext}}(t)= [j_{\alpha}^{\mathrm{ext}}(t)]^* = \left(\frac{1}{2} \omega_{\mathbf{k}}^3  \hbar\right)^{-1 / 2}\frac{1}{\epsilon_0g_{\alpha}}\boldsymbol{\epsilon}_{\alpha} \cdot \mathbf{j}^{ \text { ext }}(t).
\end{equation}
Note, that we probe our system with an external perturbation $\mathbf{a}^\mathrm{ext} ({\bf r}_l,t)$ that does not generate longitudinal currents, i.e. $j_0^\mathrm{ext}({\bf r},t) =0$. 
\\
\indent
In the dipole approximation, it is possible to eliminate the term $e \sum_{l=1}^N  (\boldsymbol{\alpha}_l\cdot\mathbf{a}^\mathrm{ext} (t)  + \boldsymbol{\alpha}_l\cdot\hat{\mathbf{A}})$ from Eq.~\eqref{H_PF_pandq}
by performing the unitary length-gauge transformation which is defined as~\cite{Rokaj2018,tokatly2013,Schafer2020}
\begin{equation}
\hat{H}_{\mathrm{L}}=\hat{U}^{\dagger} \hat{H} \hat{U} -i \hbar \hat{U}^{\dagger}(\partial_t \hat{U}) , \quad \hat{U}=e^{
\frac{\mathrm{i}}{\hbar c} \left[  \hat{\mathbf{A}} 
+ \mathbf{a}^\mathrm{ext} (t) 
\right]\cdot \boldsymbol{\mu}},
\label{Lgauge}
\end{equation}
where $\boldsymbol{\mu}=\sum_{i=1}^N e\mathbf{r}_i$ is the total dipole operator of the electrons in a system, and index $\mathrm{L}$ indicates the length gauge. 
Using Eq.~\eqref{Lgauge} one can rewrite the Hamiltonian in length gauge as
\begin{align}
\begin{split}
\hat{U}^{\dagger} \hat{H}_\mathrm{QED,cavity}^\mathrm{Dip}(t)  \hat{U}= \hat{H}_\mathrm{L}(t)
&=\sum_{l=1}^N[-i\hbar c \boldsymbol{\alpha}_l\cdot\boldsymbol{\nabla}_l + \beta_l mc^2
+  e a_0^\mathrm{ext} ({\bf r}_l,t)] \\
& +  \frac{1}{2} \sum_{l \neq m}^N \frac{e^2}{4 \pi \varepsilon_0\left|{\bf r}_l-{\bf r}_m\right|} 
-  \sum_{l=1}^N v^{\text{ext}}({\bf r}_l,t)  \\
& +   \sum_{\alpha=1}^{M_{p}}
\left(\frac{\hbar}{2}\left(
\omega_\alpha \hat{q}_\alpha
- \frac{g_{\alpha}\boldsymbol{\epsilon}_{\alpha} \cdot \boldsymbol{\mu}}{\sqrt{\hbar}}
\right)^2
-\frac{\hbar }{2} 
\frac{\partial^2}{\partial \hat{q}_{\alpha}^2}
+i\sqrt{2\hbar^2 \omega_\alpha}j_{\alpha}^{\mathrm{ext}}(t)
\frac{\partial}{\partial \hat{q}_{\alpha}}
\right),
\label{H_PF_pandqL}
\end{split}
\end{align}
where $v^{\text{ext}}({\bf r}_l,t) = e\mathbf{E}^\mathrm{ext}(t) \cdot {\bf r}_l$ is the external potential generated by the external electric field $\mathbf{E}^\mathrm{ext}(t) = -\tfrac{\partial}{c \partial t} \mathbf{a}^\mathrm{ext} (t)$.
Here we have performed the additional canonical transformations $-i 
\frac{\partial}{\partial_{\hat{Q}_{\alpha}}} \mapsto -\omega_\alpha \hat{q}_\alpha, \hat{Q}_{\alpha} \mapsto
-i \frac{\partial}{\omega_\alpha\partial_{\hat{q}_{\alpha}}}$, which leave the commutation relations unchanged, i. e. $\left[\hat{Q}_{\alpha},-i \frac{\partial}{\partial \hat{Q}_{\alpha'}}\right]
=
\left[-i \frac{\partial}{\omega_\alpha\partial_{\hat{q}_{\alpha}}},-\omega_\alpha \hat{q}_\alpha\right]
=
i\delta_{\alpha \alpha'}$. We want to note that $\hat{q}_\alpha$ does not correspond to a pure photonic quantity when we couple the modes to the electronic system. It is connected to the auxiliary displacement field of the macroscopic Maxwell equations and is usually referred to as a displacement coordinate~\cite{Rokaj2018,Schafer2020}. 
\\
\indent
We are interested in studying the electronic structure properties of molecules/atoms coupled to photon modes. Then $e a_0^\mathrm{ext} ({\bf r}_l,t) \rightarrow v_{\mathrm{nucl}} ({\bf r}_l) = -\frac{1}{4 \pi \epsilon_0} \sum_{j=1}^{N_\mathrm{n}} \frac{Z_j e^2}{\left|\mathbf{r}_l-\mathbf{R}_j\right|}$ is a Coulomb potential describing the interaction of $N_\mathrm{n}$ nuclei with the bound electrons.
Yet we perform another time-dependent unitary transformation $\hat{U}(t) =e^{\left[
\frac{\mathrm{i}}{\hbar }  
 \sum_{\alpha=1}^{M_{p}}
 \hat{q}_{\alpha}\sqrt{2\hbar^2 \omega_\alpha}j_{\alpha}^{\mathrm{ext}}(t)
\right]}$ 
aiming to eliminate the last term in Eq.~\eqref{H_PF_pandqL}. Then the Hamiltonian \eqref{H_PF_pandqL} takes the final form
\begin{align}
\begin{split}
 \hat{H}_\mathrm{L}(t)
&=\sum_{l=1}^N[-i\hbar c \boldsymbol{\alpha}_l\cdot\boldsymbol{\nabla}_l + \beta_l mc^2
+ v_{\mathrm{nucl}} ({\bf r}_l)] \\
& +  \frac{1}{2} \sum_{m \neq l}^N \sum_{l=1}^N \frac{e^2}{4 \pi \varepsilon_0\left|{\bf r}_l-{\bf r}_m\right|} 
-  \sum_{l=1}^N v^{\text{ext}}({\bf r}_l,t)  \\
& +  \sum_{\alpha=1}^{M_{p}}
\left(\frac{\hbar}{2}\left(
\omega_\alpha \hat{q}_\alpha
- \frac{g_{\alpha}\boldsymbol{\epsilon}_{\alpha} \cdot \boldsymbol{\mu}}{\sqrt{\hbar}}
\right)^2
-\frac{\hbar }{2} 
\frac{\partial^2}{\partial \hat{q}_{\alpha}^2}
+\sqrt{2\hbar^2 \omega_\alpha}\left[\partial_tj_{\alpha}^{\mathrm{ext}}(t)\right]
\hat{q}_{\alpha}
\right),
\label{H_PF_pandqL_final_final}
\end{split}
\end{align}
where we disregarded the physically irrelevant time-dependent energy/phase shift $-\hbar \left[j_{\alpha}^{\mathrm{ext}}(t)\right]^2$.

%

A particular consequence of the Dirac Hamiltonian is the emergence of the spin--orbit coupling (SOC).
While SOC can be added ad hoc into a non-relativistic theory based on the Schr\"{o}dinger equation,
it arises naturally only in the Dirac equation due to its multi-component structure.
The SOC can be made explicit when expressing the Dirac equation in a two-component form,
or when performing expansion in different orders of the speed of light~\cite{Dyall2007}. There are formally different ways of obtaining such an expansion~\cite{gesztesy1983efficient,frohlich1993gauge}. In our case, the most convenient form is from the relativistic energy correction, such that the potential from the nuclei gives rise
to one-electron SOC contributions
\begin{equation}
\label{eq:1eSOC}
\Delta E_\mathrm{1e\, SOC}
=
\frac{1}{4 m^2 c^2}
\left\langle
\left( \sum_{l=1}^{N} \boldsymbol{\sigma}_l \cdot \hat{\mathbf{p}}_{l} \right) \Psi^{L}
\right| \sum_{m=1}^N v_{\mathrm{nucl}} ({\bf r}_m) \left.
\left( \sum_{l'=1}^{N} \boldsymbol{\sigma}_{l'} \cdot \hat{\mathbf{p}}_{l'} \right) \Psi^{L} \right\rangle
,
\end{equation}
while the electron--electron Coulomb 
interactions leads to two-electron SOC contributions
\begin{equation}
\label{eq:2eSOC}
\Delta E_\mathrm{2e\, SOC}
=
\frac{1}{4 m^2 c^2}
\left\langle
\left( \sum_{l=1}^{N} \boldsymbol{\sigma}_l \cdot \hat{\mathbf{p}}_{l} \right) \Psi^{L}
\right| \frac{1}{2} \sum_{m' \neq m}^N \sum_{m=1}^N \frac{e^2}{4 \pi \varepsilon_0\left|{\bf r}_m-{\bf r}_{m'}\right|} \left.
\left( \sum_{l'=1}^{N} \boldsymbol{\sigma}_{l'} \cdot \hat{\mathbf{p}}_{l'} \right) \Psi^{L} \right\rangle
,
\end{equation}
where $\hat{\mathbf{p}} = -i \hbar \boldsymbol{\nabla}$ and $\Psi^{L}$ is the respective ``large'' component of the Dirac wavefunction.
In the presence of SOC the spin operator no longer commutes with the Hamiltonian meaning
that the eigenstates of the Hamiltonian are not eigenstates of the spin operator,
referred to as spin no longer being a good quantum number.
This results for example in non-zero transition amplitudes
between states of different spin symmetry.
Following the same strategy as done for the standard terms of Eqs.~\eqref{eq:1eSOC} and \eqref{eq:2eSOC} also for the new cavity-coupling term, we obtain cavity-mediated SOC contributions from the relativistic energy correction as
\begin{align}
\begin{split}
\Delta E &= \frac{1}{4 m^2 c^2}\left\langle \left( \sum_{l=1}^{N} \boldsymbol{\sigma}_l \cdot \hat{\mathbf{p}}_{l} \right) \Psi^{L} \left| \sum_{\alpha =1}^{M_p} \frac{\hbar}{2}\left( - \frac{2\omega_{\alpha} \hat{q}_{\alpha}g_{\alpha} \boldsymbol{\epsilon}_{\alpha} \cdot \boldsymbol{\mu}}{\sqrt{\hbar}}
+ \frac{g_{\alpha}^2 (\boldsymbol{\epsilon}_{\alpha} \cdot \boldsymbol{\mu})^2}{\hbar}
\right)  \left( \sum_{l'=1}^{N} \boldsymbol{\sigma}_{l'} \cdot \hat{\mathbf{p}}_{l'} \right) \Psi^{L} \right. \right\rangle \\
& = 
-\left\langle \Psi^{L} \left| \sum_{l=1}^{N}\sum_{\alpha =1}^{M_p}
\frac{\hat{q}_{\alpha}\sqrt{\hbar^3}\omega_\alpha g_\alpha e }{8 m^2 c^2 } \left(\boldsymbol{\epsilon}_{\alpha} \cdot\left(\hat{\mathbf{p}}_{l} \times \boldsymbol{\sigma}_l\right)+\left(\hat{\mathbf{p}}_{l} \times \boldsymbol{\sigma}_l\right) \cdot \boldsymbol{\epsilon}_{\alpha}\right)\right| \Psi^{L} \right\rangle +\\
&+
\left\langle \Psi^{L} \left|\sum_{l=1}^{N} \sum_{k=1}^{N} \sum_{\alpha =1}^{M_p} \frac{g_\alpha^2 \hbar e^2}{16m^2 c^2}\left(\boldsymbol{\epsilon}_{\alpha} \cdot \mathbf{r}_k\right) \left[\boldsymbol{\epsilon}_{\alpha} \cdot\left(\hat{\mathbf{p}}_{l} \times \boldsymbol{\sigma}_l\right)
+
\left(\hat{\mathbf{p}}_{l} \times \boldsymbol{\sigma}_l\right)\cdot \boldsymbol{\epsilon}_{\alpha}
\right] \right| \Psi^{L} \right\rangle + \ldots
\end{split}
\end{align}
%
%
where
the ellipsis stands for other terms that contain a correction to kinetic energy
and a Darwin-like term.
Among others, we find a cavity-induced SOC term of the form
\begin{align}
\label{eq:CavitySOC}
-\frac{e \hbar}{8 m^2c^2}\sum_{l=1}^{N} \left(\hat{\mathbf{p}}_l \cdot (\boldsymbol{\sigma}_l \times \hat{\mathbf{E}}) +  (\boldsymbol{\sigma}_l \times \hat{\mathbf{E}})\cdot \hat{\mathbf{p}}_l    \right),
\end{align}
and the electric field operator of the cavity in the length gauge is given by 
\begin{align}
\hat{\mathbf{E}}
=
\sum_{\alpha =1}^{M_p} \left( \sqrt{\hbar} g_{\alpha} \omega_{\alpha}\hat{q}_{\alpha}
- \frac{g_\alpha^2}{2} \boldsymbol{\mu}\cdot\boldsymbol{\varepsilon}_\alpha
\right) \boldsymbol{\epsilon}_{\alpha}
.
\end{align}
Note that the electric dipole moment operator $\boldsymbol{\mu}$ contains a sum over all electrons making the
second term a collective effect.
Here the the first part of the electric field is proportional to the coupling strength $g_{\alpha}$ and usually dominates for single-molecule situations in the weak to strong coupling regime~\cite{Ruggenthaler2018,frisk2019ultrastrong}. However, for collective-coupling situation, where it is the response of the total ensemble of molecules, the second term proportional to $g_{\alpha}^2$ can become dominant~\cite{sidler2023unraveling,schnappinger2023cavity}. One can therefore imagine that even a collectively-coupled ensemble of molecules could show enhanced SOC effects.
As a final remark, in a similar way to Eq.~\eqref{eq:CavitySOC} the term $v^{\text{ext}}({\bf r}_l,t)$ in Eq.~\eqref{H_PF_pandqL_final_final} can transiently lead to a light-induced SOC due to external classical fields.

\subsection{Kohn--Sham QEDFT equations of motion}

After specifying the Hamiltonian one can find the time evolution of a system from a given initial state
\linebreak
$\Psi_0\left(x_1 \ldots, x_{N}; q_1,\ldots, q_{M_p}\right)$ by solving the coupled electron--photon
equation of motion
\begin{equation}
i \hbar \partial_t \Psi\left(x_1  \ldots, x_{N}; q_1,\ldots, q_{M_p};t\right)= \hat{H}_{\text{L}}(t)  \Psi\left(x_1 \ldots,x_{N} ; q_1,\ldots, q_{M_p};t\right),
\label{eq_sch}
\end{equation}
where $x_i \equiv\left(\mathbf{r}_i, \tau_i\right)$ denotes spatial coordinate $\mathbf{r}_i$ and the four spin-components $ \tau_i$ of $i$-th Dirac electron.
While solutions of Eq.~\eqref{eq_sch} contain all the information about the given system,
it becomes numerically intractable for realistic many-electron systems and therefore
approximate methods have to be employed.
In this work, our method of choice is density functional theory in the form of QEDFT and
in notation we follow the formalism outlined in Ref.~\citenum{Flick2019}.
The electron--photon system is described by the electronic density $n\left( \mathbf{r}, t\right) = N\sum_{\tau, \tau_2,...,\tau_N} \int |\Psi(\mathbf{r}, \tau,x_2,...,x_N)|^2 \mathrm{d}\mathbf{r}_2 ... \mathrm{d}\mathbf{r}_N$ and the expectation value of photonic coordinates $q_\alpha\left( t\right)$ as basic variables~\cite{tokatly2013,Welakuh2022,Flick2019,Ruggenthaler2014} which uniquely determine its state.
We work in the Kohn-Sham (KS) formalism, where we consider an auxiliary system of independent particles
that reproduces the density of the real system and is described by molecular orbitals $\varphi_j(x,t)$, with index $j$ enumerating the $N_\mathrm{occ}$ occupied orbitals.
The electron density is then calculated as~\cite{Ullrich}
\begin{equation}
n(\mathbf{r}, t)= \sum_{j=1}^{N_\mathrm{occ}}\sum_{\tau=1}^{4}\left|\varphi_{j}(\mathbf{r}, \tau,t)\right|^2
.
\label{density}
\end{equation}
Then the equations of motion for the basic variables in terms of the auxiliary system read as~\cite{Flick2019, Ruggenthaler2014}
\begin{align}
\label{eq:el_EOM}
i\hbar \partial_t \varphi_j(x,t)
& =
\hat{H}^{\text{KS}}(t) \varphi_j(x,t)
, \\
\label{eq:ph_EOM}
\left(\frac{\partial^2}{\partial t^2} + \omega_\alpha^2 \right) q_\alpha (t) 
& = 
-\sqrt{2}\partial_t j_{\alpha, \mathrm{KS}}\left(\left[n, q_\alpha, t \right]\right),
\end{align}
with the single-particle KS Hamiltonian $\hat{H}^{\text{KS}}(t)$
\begin{equation}
\hat{H}^{\text{KS}}(t)=-i\hbar c \boldsymbol{\alpha}\cdot\boldsymbol{\nabla} + \beta mc^2 
+  v_{\mathrm{KS}}\left(\left[v, n, q\right] ; {\bf r},t \right),
\label{H_KS}
\end{equation}
and the Kohn--Sham current $j_{\alpha, \mathrm{KS}}\left(\left[n, q_\alpha, t \right]\right)$ defined below in Eq.~\eqref{eq:KScurrent}.
The Kohn--Sham potential $v_{\mathrm{KS}}\left(\left[v, n, q\right] ; {\bf r},t \right)$
in Eq.~\eqref{H_KS} is defined as
\begin{equation}
v_{\mathrm{KS}}\left(\left[v, n, q \right] ; {\bf r},t \right)
= v({\bf r},t) +  v_{\mathrm{Mxc}}\left(\left[ n, q\right] ; {\bf r},t\right).
\end{equation}
Here the external potential $v({\bf r},t)$ according to Eq.~\eqref{H_PF_pandqL_final_final} is given by
\begin{equation}
v({\bf r},t) = v_{\mathrm{nucl}} ({\bf r}) + v^{\text{ext}}({\bf r},t), 
\label{v_tot_ext}
\end{equation}
and $v_{\mathrm{Mxc}}\left(\left[ n, q\right] ; {\bf r},t\right)$ is the mean field exchange-correlation potential
\begin{equation}
v_{\mathrm{Mxc}}\left(\left[ n, q\right] ; {\bf r},t\right)
= v_{\mathrm{M}}\left(\left[ n, q\right] ; {\bf r},t\right)  + v_{\mathrm{xc}}\left(\left[ n, q\right] ; {\bf r},t\right).
\end{equation}
Since we consider a system of electrons coupled to photonic modes, one can split the mean-field potential $v_{\mathrm{M}}\left(\left[ n, q\right] ; {\bf r},t\right)$ into a contribution which comes solely from the field generated by electrons, the Hartree potential $v_{\mathrm{H}}([n] ; {\bf r},t)$,
and mean field generated by photons $v_{\mathrm{P}}\left(\left[n, q\right] ;  {\bf r},t\right)$
\begin{equation}
v_{\mathrm{M}}\left(\left[ n, q\right] ; {\bf r},t\right)
= v_{\mathrm{H}}([n] ; {\bf r},t)  + v_{\mathrm{P}}\left(\left[n, q\right] ;  {\bf r},t\right),
\end{equation}
where
\begin{equation}
\label{vP}
v_{\mathrm{P}}\left(\left[n, q\right] ;  {\bf r},t\right)
=\sum_{\alpha=1}^{M_p}\left( 
eg_{\alpha}
\int d {\bf r}^{\prime} n\left({\bf r}^{\prime},t\right)
\boldsymbol{\varepsilon}_\alpha \cdot {\bf r}^{\prime} 
-
\sqrt{\hbar}\omega_{\alpha}q_{\alpha}(t)
\right)
eg_{\alpha}\boldsymbol{\varepsilon}_\alpha \cdot {\bf r}.
\end{equation}
Similarly, one can split the exchange-correlation potential $v_{\mathrm{xc}}\left(\left[ n, q\right] ; {\bf r},t\right)$ into a sum of electron--electron $v_{\mathrm{xc}}^{e-e}([n] ; {\bf r},t)$ and photon--electron $v_{\mathrm{xc}}^{\mathrm{e-P}}\left(\left[n, q\right] ;  {\bf r},t\right)$ exchange-correlation potentials,
\begin{equation}
\label{eq:vxc_terms}
v_{\mathrm{xc}}\left(\left[ n, q\right] ; {\bf r},t\right)
= v_{\mathrm{xc}}^{e-e}([n] ; {\bf r},t)  + v_{\mathrm{xc}}^{\mathrm{e-P}}\left(\left[n, q\right] ;  {\bf r},t\right),
\end{equation}
respectively.
Note that in density functional approximations employed in practical calculations, the potentials $v_{\mathrm{xc}}^{e-e}([n] ; {\bf r},t)$ and $v_{\mathrm{xc}}^{\mathrm{e-P}}([n, q] ;  {\bf r},t)$ can further dependent on the gradient of the density or on spin densities, as well as include an admixture of the exact exchange.  
Finally, we introduce KS current $j_{\alpha, \mathrm{KS}}\left(\left[n, q_\alpha, t \right]\right)$ in Eq.~\eqref{eq:ph_EOM}
\begin{align}
\label{eq:KScurrent}
\begin{split}
\partial_t j_{\alpha, \mathrm{KS}}\left(\left[n, q_{\alpha}, \partial_t j_{\alpha}^{\text{ext}}\right] ;t\right) 
&=\partial_tj_{\alpha,Mxc}\left(\left[n, q_{\alpha}\right] ;t\right)
+ \partial_t j^{\text{ext}}_{\alpha}(t)\\
&= \partial_tj_{\alpha,M}\left(\left[n\right] ;t\right)
+\partial_tj_{\alpha,xc}\left(\left[n, q_{\alpha}\right] ;t\right)+ \partial_t j^{\text{ext}}_{\alpha}(t)\\
&= -\frac{\omega_\alpha e}{\sqrt{2\hbar}}\int d \mathbf{r} g_{\alpha}\boldsymbol{\epsilon}_{\alpha} \cdot \mathbf{r} n(\mathbf{r})+ \partial_t j^{\text{ext}}_{\alpha}(t),
\end{split}
\end{align}
where the exchange-correlation current $\partial_tj_{\alpha,xc}(t)$ is zero in the case of dipole coupling and only the mean-field Kohn-Sham current $\partial_tj_{\alpha,M}\left(\left[n\right] ;t\right)$ contributes~\cite{Flick2015,Flick2019}.

\subsection{Linear response equations for relativistic QEDFT}
\label{sec:LRtheory}
%

%
%
In the present work, we assume that the external probe potential $v^{\text{ext}}({\bf r},t)$ and current $j_{\alpha}^{\text{ext}}(t)$ are sufficiently weak
to allow the use of time-dependent perturbation theory for the description of their effect on a molecule coupled to photon modes.
Our goal is to develop a solution of Eqs.~\eqref{eq:el_EOM} and \eqref{eq:ph_EOM} to the first order of perturbation theory with the external potential and current acting as the perturbation.
Note that we do not assume that the coupling to the cavity modes is weak, only that the external
perturbation to the coupled cavity-matter system is small.
We will proceed in the spirit of standard time-dependent (Rayleigh--Schr\"{o}dinger) perturbation theory following our earlier work~\cite{Konecny2019}. 
We express the time-dependent spin-orbitals as a perturbation series
\begin{equation}
\label{phi_orders}
    \varphi_i(x, t) = \varphi_i^{(0)}(x, t) + \lambda \varphi_i^{(1)}(x, t) + \lambda^2 \varphi_i^{(2)}(x, t) + \ldots 
    ,
\end{equation}
with $\lambda$ being the perturbation parameter and $\varphi_i^{(k)}(x, t)$ the $k$-th order correction to the $i$-th molecular spinor.
Furthermore, the $k$-th order corrections are expanded in the basis of ground-state (canonical) molecular spinorbitals
\begin{equation}
    \varphi_i^{(k)}(x, t)
    =
    d^{(k)}_{pi}(t) \varphi_p(x) e^{-i\varepsilon_i t}
    ,
\end{equation}
where $\varphi_p(x)$ are the canonical spin-orbitals obtained as the solution of time-independent SCF procedure for $\hat{H}^{\text{KS}}(t)$ from Eq.~\eqref{H_KS},
and $d^{(k)}_{pi}(t)$ are the time-dependent expansion coefficients. The exponential $e^{-i\varepsilon_i t}$ results from the solution for free evolution 
$\varphi_i (x) e^{-i\varepsilon_i t}$ where the weak perturbation then modifies
$\varphi_i (x)$ to $\sum_k d^{(k)}_{pi}(t) \varphi_p(x)$.

Since the Kohn--Sham Hamiltonian $H^\mathrm{KS}(t)$ depends on the electron density and in turn on the molecular orbitals $\varphi_i({\bf r}, t)$, as well as on the photonic coordinate $q_\alpha (t)$,
it is also modified by the perturbation and can be expanded in the orders of $\lambda$
\begin{equation}
    H^\mathrm{KS} = H^\mathrm{KS, (0)} + \lambda H^\mathrm{KS, (1)} + \lambda^2 H^\mathrm{KS, (2)} + \ldots
\end{equation}
By inserting the expansion for the Hamiltonian and the molecular orbitals into the time-dependent Kohn--Sham equation, we obtain differential
equations for the expansion coefficients of all orders.
Similarly for photons, we consider the perturbation expansion of the displacement coordinate
\begin{equation}
\label{q_orders}
    q_\alpha (t) = q^{(0)}_\alpha (t) + \lambda q^{(1)}_\alpha (t) + \lambda^2 q^{(2)}_\alpha (t) + \ldots
\end{equation}
and formulate and solve EOMs for the different orders in the expansion.
\\
\indent
In Eqs. \eqref{phi_orders}, \eqref{q_orders} the zeroth order refers to the ground state solution, i.e. the external probe potential $v^{\text{ext}}({\bf r},t)$ and current $j^{\text{ext}}_{\alpha}(t)$ are absent. 
The stationary solution of Eq.~\eqref{eq:ph_EOM} has the form 
\begin{equation}
\label{q_stat}
q_\alpha^{(0)} = \frac{eg_{\alpha}}{\omega_{\alpha}\sqrt{\hbar}}\int d\mathbf{r}\boldsymbol{\epsilon}_{\alpha} \cdot \mathbf{r} n(\mathbf{r}),    
\end{equation}
which causes the photonic potential $v_\mathrm{p}$ to vanish
(as can be verified by direct substitution of Eq.~\eqref{q_stat} into Eq.~\eqref{vP}).
The coupling to photon modes is still present via the photon--electron XC potential $v_{\mathrm{xc}}^{\mathrm{e-P}}\left(\left[n, q\right] ;  {\bf r}_l,t\right)$,
see Eq.~\eqref{eq:vxc_terms}, which is, however, part of the unperturbed Hamiltonian
whose eigenfunctions are the canonical orbitals.
Therefore, the zeroth order electronic EOM has the usual form
\begin{equation}
i \partial_t d^{(0)}_{pi}(t) = H^\mathrm{KS, (0)}_{pq} d^{(0)}_{qi}(t),    
\end{equation}
with corresponding solutions
\begin{equation}
d^{(0)}_{pi}(t) = \delta_{pi}.    
\end{equation}
Here, $\delta_{pi}$ is the Kronecker delta resulting from the fact that the zeroth order KS Hamiltonian is diagonal in the basis of ground state molecular orbitals with molecular orbital energies on the diagonal.
Note that in practical calculations we employ the pRPA which neglects $v_{\mathrm{xc}}^{\mathrm{e-P}}\left(\left[n, q\right] ;  {\bf r}_l,t\right)$ after Eq.~\eqref{eq:vxc_terms}
and, in addition, start from the uncoupled ground state.
However, the derivation presented in the following text is general and holds also with
the full photon--electron exchange--correlation potential included.

In first order, the electronic EOM takes the form
\begin{align}
\begin{split}
    \label{eq:d1CoefEOM}
    i \partial_t d^{(1)}_{ai}(t)
    = &
    A_{ai,bj} (\omega_\mathrm{ext})\, d^{(1)}_{bj}(t) + B_{ai,bj} (\omega_\mathrm{ext}) \, d^{(1)*}_{bj}(t) \\
    & + \frac{e^2 g_\alpha^2}{2} (\mathbf{r}_{ai} \cdot \boldsymbol{\epsilon}_\alpha) (\mathbf{r}_{jb} \cdot \boldsymbol{\epsilon}_\alpha)\, d^{(1)}_{bj}(t)
    + \frac{e^2 g_\alpha^2}{2} (\mathbf{r}_{ai} \cdot \boldsymbol{\epsilon}_\alpha) (\mathbf{r}_{bj} \cdot \boldsymbol{\epsilon}_\alpha)\, d^{(1)*}_{bj}(t) \\
    & - e g_\alpha \sqrt{\hbar} \omega_\alpha (\mathbf{r}_{ai} \cdot \boldsymbol{\epsilon}_\alpha) q^{(1)}_\alpha (t) \\
    & + P_\mathrm{ai} e^{-i \omega_\mathrm{ext} t} + P_\mathrm{ai}^* e^{i \omega_\mathrm{ext} t}
    ,
\end{split}
\end{align}
where the right-hand side contains the electronic kernel, the interaction with the cavity modes, and the interaction
with the external field.
Specifically, in Eq.~\eqref{eq:d1CoefEOM}, the first line on the right-hand side contains the usual Coulomb,
exchange, and exchange--correlation coupling matrices
\begin{subequations}
\label{eq:CouplingMatrices}
\begin{align}
A_{ai,bj} (\omega_\mathrm{ext})
& =
(\epsilon_a - \epsilon_i) \delta_{ab}\delta_{ij}
+
K_{\mu\nu\lambda\tau} (\omega_\mathrm{ext}) C_{\mu a}^* C_{\nu i} C_{\lambda j}^* C_{\tau b}
, \\
B_{ai,bj} (\omega_\mathrm{ext})
& =
K_{\mu\nu\lambda\tau} (\omega_\mathrm{ext}) C_{\mu a}^* C_{\nu i} C_{\lambda b}^* C_{\tau j}
, \\
\label{eq:XCkernel}
K_{\mu\nu\lambda\tau} (\omega_\mathrm{ext})
& =
K_{\mu\nu\lambda\tau}^\mathrm{HF}(\xi) + K_{\mu\nu\lambda\tau}^\mathrm{XC}(\xi) (\omega_\mathrm{ext})
+
K_{\mu\nu\lambda\tau}^\mathrm{e-P} (\omega_\mathrm{ext})
,
\end{align}
\end{subequations}
with molecular orbital energies $\epsilon_p\ (p = a,i)$
and coefficients $C_{\mu p}$ obtained from the solution of the
ground-state SCF procedure for the time-independent Dirac--Coulomb equation,
and $K_{\mu\nu\lambda\tau}$ being matrix elements of the response kernel
with HF, XC, e--P denoting its Hartree--Fock, electron--electron exchange--correlation,
and photon--electron parts,
respectively, in hybrid functionals mixed with weight $\xi$. We note that, strictly speaking, in the case of a hybrid auxiliary systems the respective exchange-correlation contributions of the electron-electron and electron-photon parts get modified, but we will not indicate this for notational simplicity~\cite{baer2018time}.
The XC kernels can be in general non-adiabatic, denoted by their dependence
on $\omega_\mathrm{ext}$.
The second and the third line contain terms arising from the mean-field coupling to the photon field,
with the dipole self-energy terms in the second line and in the third one the interaction
with the displacement field of the cavity modes $q^{(1)}_\alpha (t)$.
Finally, the last line contains the coupling to the external classical electromagnetic field
representing the probe field used to perform absorption spectroscopy on the molecule,
i.e. the perturbation.

For photons, the EOM in the first-order reads
\begin{equation}
    \label{eq:1stOrderMaxwell}
    \left(\frac{\partial^2}{\partial t^2} + \omega_\alpha^2 \right) q^{(1)}_\alpha (t)
    =
    -\sqrt{2} \left[ (\partial_t j^{(1)}_{\alpha, \mathrm{KS}}\left(\left[n, q_\alpha, t \right]\right)
    + (\partial_t j^\mathrm{ext}(t)) \right]
    ,
\end{equation}
where the first-order contribution to the current comes only from its dependence on the (first-order) electron density in the form
\begin{equation}
    \partial_t j^{(1)}_{\alpha, \mathrm{KS}}\left(\left[n, t \right]\right)
    =
    - \frac{g_\alpha  \omega_\alpha e \int d^3 r n^{(1)}(r,t) (\mathbf{r} \cdot \boldsymbol{\epsilon}_\alpha) }{\sqrt{2 \hbar}}
    .
\end{equation}
By expressing the first-order electron density via the canonical molecular orbitals
\begin{equation}
    n^{(1)}(r,t)
    =
    \varphi_j^\dagger(x) d^{(1)}_{bj}(t) \varphi_b(x)
    + d^{(1)*}_{bj}(t) \varphi_b^\dagger(x) \varphi_j(x)
    ,
\end{equation}
we obtain the EOM for $q^{(1)}_\alpha (t)$ coupled to electronic coefficients $d^{(1)}_{ai}(t)$ and $d^{(1)*}_{ai}(t)$ which reads
\begin{equation}
    \label{eq:q2ndOrderDE}
    \left(\frac{\partial^2}{\partial t^2} + \omega_\alpha^2 \right) q^{(1)}_\alpha (t)
    =
    \frac{g_\alpha  \omega_\alpha e}{\sqrt{\hbar}} \left[ (\mathbf{r}_{jb} \cdot \boldsymbol{\epsilon}_\alpha)\, d^{(1)}_{bj}(t)
    + (\mathbf{r}_{bj} \cdot \mathbf{\epsilon}_\alpha)\, d^{(1)*}_{bj}(t)
    \right]
    - \sqrt{2} \partial_t j^\mathrm{ext}(t)
    .
\end{equation}
Eq.~\eqref{eq:q2ndOrderDE} together with Eq.~\eqref{eq:d1CoefEOM} and its complex conjugate constitute a system
of coupled differential equations for variables $q^{(1)}_\alpha (t)$, $d^{(1)}_{ai}(t)$, and $d^{(1)*}_{ai}(t)$.
Note that since $q_\alpha(t)$ is real, we do not need a separate equation for $q^{(1)*}_\alpha (t)$.
However, while Eq.~\eqref{eq:d1CoefEOM} is a first-order differential equation, Eq.~\eqref{eq:q2ndOrderDE} is
a second-order equation. We can transform it to a first-order equation by defining a new variable $p^{(1)}_\alpha (t)$ as
\begin{equation}
    \label{eq:p1stOrderDef}
    p^{(1)}_\alpha (t)
    =
    -i \frac{\partial}{\partial t} q^{(1)}_\alpha (t)
    ,
\end{equation}
which obeys a first-order differential equation
\begin{equation}
    \label{eq:p1stOrderEOM}
    i\frac{\partial}{\partial t} p^{(1)}_\alpha (t)
    =
    - \omega_\alpha^2 q^{(1)}_\alpha (t)
    + \frac{g_\alpha e \omega_\alpha}{\sqrt{\hbar}} \left[ (\mathbf{r}_{jb} \cdot \boldsymbol{\epsilon}_\alpha)\, d^{(1)}_{bj}(t)
    + (\mathbf{r}_{bj} \cdot \boldsymbol{\epsilon}_\alpha)\, d^{(1)*}_{bj}(t)
    \right]
    - \sqrt{2} \partial_t j^\mathrm{ext}(t)
    .
\end{equation}
Thus we have replaced a second-order differential equation~\eqref{eq:q2ndOrderDE} with two first-order differential
equations \eqref{eq:p1stOrderDef} and \eqref{eq:p1stOrderEOM}
The imaginary unit in Eq.~\eqref{eq:p1stOrderDef} was added for convenience in order to obtain a first-order EOM
containing $i \partial_t$ in accord with Eq.~\eqref{eq:d1CoefEOM}.

The final complete system of coupled differential equations for first-order corrections to the electronic and photonic
degrees of freedom is
\begin{subequations}
\label{eq:4coupledDiffEq}
\begin{align}
    i \partial_t d^{(1)}_{ai}(t)
    & =
    ( A_{ai,bj} + \Delta_{ai,bj} ) d^{(1)}_{bj}(t) + ( B_{ai,bj} + \Delta'_{ai,bj} ) d^{(1)*}_{bj}(t)
    - L_{ai,\alpha} q^{(1)}_\alpha (t)
    + P_\mathrm{ai} e^{-i \omega_\mathrm{ext} t} + P_\mathrm{ai}^* e^{i \omega_\mathrm{ext} t} , \\
    - i \partial_t d^{(1)*}_{ai}(t)
    & =
    ( A^*_{ai,bj} + \Delta^*_{ai,bj} ) d^{(1)*}_{bj}(t) + ( B^*_{ai,bj} + \Delta^{\prime *}_{ai,bj} ) d^{(1)}_{bj}(t)
    - L^*_{ai,\alpha} q^{(1)}_\alpha (t)
    + P^*_\mathrm{ai} e^{i \omega_\mathrm{ext} t} + P_\mathrm{ai} e^{- i \omega_\mathrm{ext} t} , \\
    i\frac{\partial}{\partial t} p^{(1)}_\alpha (t)
    & =
    + Q_{\alpha, bj}^{\prime *} d^{(1)}_{bj}(t)
    + Q'_{\alpha, bj} d^{(1)*}_{bj}(t)
    - \omega_\alpha^2 q^{(1)}_\alpha (t)
    - J e^{-i \omega_\mathrm{ext} t} - J^* e^{i \omega_\mathrm{ext} t} , \\
    -i \frac{\partial}{\partial t} q^{(1)}_\alpha (t)
    & =
   p^{(1)}_\alpha (t)
   ,
\end{align}
\end{subequations}
where we introduced the notation
\begin{subequations}
\label{eq:ElPhCouplingTerms}
\begin{align}
    \Delta_{ai,bj}  & = \frac{e^2 g_\alpha^2}{2} (\mathbf{r}_{ai} \cdot \boldsymbol{\epsilon}_\alpha) (\mathbf{r}_{jb} \cdot \boldsymbol{\epsilon}_\alpha) , \\
    \Delta'_{ai,bj} & = \frac{e^2 g_\alpha^2}{2} (\mathbf{r}_{ai} \cdot \boldsymbol{\epsilon}_\alpha) (\mathbf{r}_{bj} \cdot \boldsymbol{\epsilon}_\alpha) , \\
    L_{ai,\alpha}   & = - e g_\alpha \sqrt{\hbar} \omega_\alpha (\mathbf{r}_{ai} \cdot \boldsymbol{\epsilon}_\alpha) , \\
    Q'_{ai,\alpha}   & = \frac{g_\alpha e \omega_\alpha}{\sqrt{\hbar}} (\mathbf{r}_{jb} \cdot \boldsymbol{\epsilon}_\alpha) ,
\end{align}
\end{subequations}
and where we considered the external current to have a time dependence given by
$\sqrt{2} \partial_t j^\mathrm{ext}(t) = J e^{-i \omega_\mathrm{ext} t} + J^* e^{i \omega_\mathrm{ext} t}$.
The explicit dependence of coupling matrices on $\omega_\mathrm{ext}$ resulting from
such dependence in the exchange--correlation kernel was omitted for brevity.
The first-order equations of motion for the coupled electron--photon system,
Eqs.~\eqref{eq:4coupledDiffEq},
can be solved by different approaches such as the method of undetermined coefficients
shown in Appendix~\ref{sec:AppendixLRderivation},
or by the direct solution of the system of equations as a matrix differential equation,
followed by a similarity transformation shown in Appendix~\ref{sec:AppendixMatrixDiffEq}.
Both of these approaches lead to an algebraic equation corresponding to an eigenvalue
problem whose eigenvalues are excitation energies and eigenvectors transition vectors
of the coupled light--matter system.
Alternatively, one can arrive at the final equation for excitation energies
by considering density--density linear response function as presented at the non-relativistic
level of theory in Ref.~\citenum{Flick2019}.

The linear response equation obtained from 
Eqs.~\eqref{eq:4coupledDiffEq} has the form of the Casida equation~\cite{Casida1995, Casida2009, Ullrich},
of linear-response time dependent density functional theory (TDDFT) extended
by blocks describing photons
\begin{equation}
\label{eq:CavityCasida}
\begin{pmatrix}
\mathbf{A} + \mathbf{\Delta}      & \mathbf{B} + \mathbf{\Delta'}    & -\mathbf{L}     & -\mathbf{L}     \\
\mathbf{B}^* + \mathbf{\Delta'}^* & \mathbf{A}^* + \mathbf{\Delta}^* & -\mathbf{L}^{*} & -\mathbf{L}^{*} \\
-\mathbf{Q}                       & -\mathbf{Q}^*                    & \bs{\omega}     & \mathbf{0}      \\
-\mathbf{Q}                       & -\mathbf{Q}^*                    & \mathbf{0}      & \bs{\omega}     \\
\end{pmatrix}
\begin{pmatrix}
\mathbf{X}_n \\
\mathbf{Y}_n \\
\mathbf{M}_n \\
\mathbf{N}_n \\
\end{pmatrix}
=
\Omega_n
\begin{pmatrix}
\mathbf{1} &  \mathbf{0} & \mathbf{0} &  \mathbf{0} \\
\mathbf{0} & -\mathbf{1} & \mathbf{0} &  \mathbf{0} \\
\mathbf{0} &  \mathbf{0} & \mathbf{1} &  \mathbf{0} \\
\mathbf{0} &  \mathbf{0} & \mathbf{0} & -\mathbf{1} \\
\end{pmatrix}
\begin{pmatrix}
\mathbf{X}_n \\
\mathbf{Y}_n \\
\mathbf{M}_n \\
\mathbf{N}_n \\
\end{pmatrix}
,
\end{equation}
where the terms $\mathbf{A}$ and $\mathbf{B}$ are the electronic coupling matrices introduced
in Eqs.~\eqref{eq:CouplingMatrices},
the electron--photon coupling and self-energy terms $\mathbf{L}$,
$Q_{jb,\alpha}
=
\frac{1}{2\omega_\alpha}Q'_{jb,\alpha}
=
\frac{1}{2} \frac{g_\alpha e}{\sqrt{\hbar}}(\mathbf{r}_{jb} \cdot \boldsymbol{\epsilon}_\alpha)$,
$\mathbf{\Delta}$, and $\mathbf{\Delta'}$ are defined in Eqs.~\eqref{eq:CouplingMatrices}.
The solutions of Eq.~\eqref{eq:CavityCasida} are the eigenvalues $\Omega_n$ corresponding
to excitation energies of the coupled light-matter system
and the eigenvectors that contain the electronic excitation and deexcitation amplitudes
$\mathbf{X}_n$ and $\mathbf{Y}_n$ parametrizing $d^{(1)}_{ai}$ and $d^{(1)*}_{ai}$,
together with the photonic creation and annihilation amplitudes $\mathbf{M}_n$ and $\mathbf{N}_n$,
parametrizing $q^{(1)}_\alpha$ and $p^{(1)}_\alpha$
(for this interpretation see Appendix~\ref{sec:AppendixMatrixDiffEq}).
The linear response equation~\eqref{eq:CavityCasida} is formulated in the basis of the ground state
molecular orbitals where the dimensions of the terms in the equation expressed via
the number of occupied orbitals $N_\mathrm{occ}$, the number of virtual orbitals $N_\mathrm{vir}$,
and the number of photon modes $N_\mathrm{ph}$ are
\begin{itemize}
\item $\mathbf{A}$, $\mathbf{B}$, $\mathbf{\Delta}$, $\mathbf{\Delta'}$:
      matrices of dimension $N_\mathrm{vir}N_\mathrm{occ} \times N_\mathrm{vir}N_\mathrm{occ}$,
\item $\mathbf{L}$, $\mathbf{Q}$: matrices of dimension $N_\mathrm{vir}N_\mathrm{occ} \times N_\mathrm{ph}$,
\item $\bs{\omega}$: diagonal matrix of dimension $N_\mathrm{ph} \times N_\mathrm{ph}$,
\item $\mathbf{X}_n$, $\mathbf{Y}_n$: vectors of dimension $N_\mathrm{vir}N_\mathrm{occ}$,
\item $\mathbf{M}_n$, $\mathbf{N}_n$: vectors of dimension $N_\mathrm{ph}$.
\end{itemize}
Note that while the transition vectors $\mathbf{X}_n$, $\mathbf{Y}_n$ are originally defined as matrices
of dimension $N_\mathrm{vir} \times N_\mathrm{occ}$ ($X_{ai}$, $Y_{ai}$ where $a \in \mathrm{vir}$, $i \in \mathrm{occ}$), 
for the purposes of solution and computer implementation of Eq.~\eqref{eq:CavityCasida}, they are unrolled into one-dimensional vectors
($X_{\kappa}$, $Y_{\kappa}$, where $\kappa$ runs over all products $ai$).
Similarly, the terms $\mathbf{A}$, $\mathbf{B}$, $\mathbf{\Delta}$, $\mathbf{\Delta'}$ originally defined as rank 4 tensors,
and the terms $\mathbf{L}$, $\mathbf{Q}$ defined as rank 3 tensors,
became matrices by considering virtual--occupied pair $ai$ to label a single dimension.
The matrix on the left-hand side of Eq.~\eqref{eq:CavityCasida} is not symmetric which has
consequences for the calculation of spectra as detailed in Section~\ref{sec:Spectrum}
and Appendix~\ref{sec:SpectrumDetails}.
Previous works on linear response QEDFT either transformed Eq.~\eqref{eq:CavityCasida} into
an equation of half the original dimension with eigenvalue $\Omega_n^2$ made possible
due to real-valued orbitals used in the non-relativistic theory,~\cite{Flick2019}
or worked with a symmetric equation obtained due to approximations made in the electron--photon
Hamiltonian~\cite{Yang2021, Liebenthal2023}.
In Appendix~\ref{sec:AppendixMatrixDiffEq} we show that Eq.~\eqref{eq:CavityCasida} can be symmetrized
by a similarity transformation to a complex version of this symmetric equation, due to
equivalent approximations made in the KS potential and current in the practical implementation.


\subsection{Calculation of spectra}
\label{sec:Spectrum}

The frequency-dependent polarizability~\cite{Norman2018} that defines the absorption
spectrum is given by (compare with Eq.~\eqref{eq:CavitySternheimer})
\begin{equation}
\label{eq:alphaFromSternheimer}
\bm{\alpha}(\omega_\mathrm{ext})
    =
    \begin{pmatrix}
    \mathbf{P}^* & \mathbf{P} & \mathbf{0} & \mathbf{0}
    \end{pmatrix}
    \begin{bmatrix}
    \begin{pmatrix}
    \mathbf{A} + \mathbf{\Delta}     & \mathbf{B} + \mathbf{\Delta}'             & -\mathbf{L}     & -\mathbf{L}   \\
    \mathbf{A}^* + \mathbf{\Delta}^* & \mathbf{B}^* + \mathbf{\Delta}^{\prime *} & -\mathbf{L}^*   & -\mathbf{L}^* \\
    -\mathbf{Q}                      & -\mathbf{Q}^*                             & \bs{\omega}     & \mathbf{0}    \\
    -\mathbf{Q}                      & -\mathbf{Q}^*                             & \mathbf{0}      & \bs{\omega}   \\
    \end{pmatrix}
    -
    \omega_\mathrm{ext}
    \begin{pmatrix}
    \mathbf{1} &  \mathbf{0} & \mathbf{0} &  \mathbf{0} \\
    \mathbf{0} & -\mathbf{1} & \mathbf{0} &  \mathbf{0} \\
    \mathbf{0} &  \mathbf{0} & \mathbf{1} &  \mathbf{0} \\
    \mathbf{0} &  \mathbf{0} & \mathbf{0} & -\mathbf{1} \\
    \end{pmatrix}
    \end{bmatrix}^{-1}
    \begin{pmatrix}
    \mathbf{P}   \\
    \mathbf{P}^* \\
    \mathbf{0}   \\
    \mathbf{0}
    \end{pmatrix}
    ,
\end{equation}
where $\mathbf{P}$ is the matrix representation of the electric dipole moment operator.
We can replace the matrix $[\ldots]^{-1}$ in Eq.~\eqref{eq:alphaFromSternheimer}
by its spectral decomposition from eigenvectors obtained by solving Eq.~\eqref{eq:CavityCasida}.
However, the electron--photon Casida equation, Eq.~\eqref{eq:CavityCasida}, corresponds to a generalized eigenvalue problem
of a non-Hermitian non-symmetric matrix (unlike the electronic Casida equation containing only the Hermitian electron--electron
block
$\begin{pmatrix}
    \mathbf{A} & \mathbf{B} \\
    \mathbf{B}^* & \mathbf{A}^*
\end{pmatrix}$
).
Therefore, the eigenproblem has different left and right eigenvectors, a fact that has to be taken
into account when performing the spectral decomposition.
In this case, the expression for the frequency-dependent linear polarizability tensor derived in
Appendix~\ref{sec:SpectrumDetails} has the from
\begin{equation}
\label{eq:PolarizabilityFinal}
  \bm{\alpha}(\omega)
  =
  \sum_n
  \left[
  \frac{t^\mathrm{R*}_{n} t^\mathrm{L}_{n}}{\Omega_n+\omega}
  -
  \frac{t^\mathrm{R}_n t^\mathrm{L*}_n}{\omega-\Omega_n}
  \right]
  ,
\end{equation}
where index $n$ runs over positive energy eigenvalues and the left and right transition dipole moments are
\begin{subequations}
\label{eq:RLtransDipMoments}
\begin{align}
    t^\mathrm{R}_n
    & =
    \mathbf{P}^\dagger \mathbf{X}^\mathrm{R}_n + \mathbf{P} \mathbf{Y}^\mathrm{R}_n , \\
    t^\mathrm{L}_n
    & =
    \mathbf{P}^\dagger \mathbf{X}^\mathrm{L}_n - \mathbf{P} \mathbf{Y}^\mathrm{L}_n ,
\end{align}
\end{subequations}
with $\mathbf{X}^\mathrm{R/L}_n$ and $\mathbf{Y}^\mathrm{R/N}_n$ corresponding to electronic
parts of the right/left $n$-th eigenvector, respectively.
An empirical broadening parameter $\gamma$ can be added to obtain spectra with finite peak widths by replacing
$\omega$ with $\omega+i\gamma$ in Eq.~\eqref{eq:PolarizabilityFinal}.
Note, however, that the empirical broadening parameter is only a convenience used when
representing a cavity by a single or few modes. Physical radiative broadening can be described
in linear response QEDFT by considering coupling to a dense spectrum of photon modes~\cite{Flick2019}.
The absorption spectrum of a molecule in a cavity is then calculated using the usual formula for the dipole strength function
\begin{equation}
\label{eq:rspFunction}
S(\omega)
=
\frac{4\pi\omega}{3 c} \Im\, \textrm{Tr} \left[ \boldsymbol{\alpha}(\omega) \right]
,
\end{equation}
where $c$ is the speed of light, $\Im$ denotes the imaginary part, and Tr is the trace over the Cartesian components.
Note that while this is the formula used in QEDFT literature, the issue of calculating a proper rotationally averaged
spectral function for the molecule-in-cavity set-up comparable to experiments is more subtle and requires further work~\cite{Sidler2020,schnappinger2023cavity}.


\subsection{Numerical solver}
\label{sec:Solver}

Eq.~\eqref{eq:CavityCasida} is thus an eigenproblem for a large matrix
whose construction and solution comprises the bulk of the computational
effort of relativistic linear response QEDFT.
The equation was implemented into the relativistic quantum
chemistry DFT package ReSpect~\cite{ReSpect} and is built upon an earlier
four-component molecular linear response TDDFT code~\cite{Komorovsky2019}
that solves the Casida equations for electrons.
The linear response QEDFT equation, Eq.~\eqref{eq:CavityCasida}, is solved using an iterative
Davidson--Olsen solver~\cite{Davidson1975,Olsen1990} as is common in TDDFT implementations,
here extended to consider the photon part as well.
The solver expresses the solution vector $\mathbf{Z} = (\mathbf{X}\ \mathbf{Y}\ \mathbf{M}\ \mathbf{N})^\mathrm{T}$ as a linear combination of
so-called trial vectors, i.e. $\mathbf{Z} = \sum_{\tau} c_\tau \mathbf{t}_\tau$. Moreover, the solvers developed for
electronic relativistic linear response TDDFT employ a paired structure of the trial vectors~\cite{Saue2003linear, Bast2009relativistic, Komorovsky2019}, where each trial vector
$\mathbf{t}_\tau = (\mathbf{t}^X_\tau \ \mathbf{t}^Y_\tau)^\mathrm{T}$,
a pair vector $\tilde{\mathbf{t}}_\tau = (\mathbf{t}^{Y*}_\tau \ \mathbf{t}^{X*}_\tau)^\mathrm{T}$ is considered.
However, since the calculation of polarizability requires the left eigenvectors, these also
have to be correctly described by the trial basis. To this end, trial vectors with a special
internal structure informed by the analytic relationship between the left and right eigenvectors
have to be added as well. Such an extension was first suggested by Komorovsky, Cherry, and
Repisky~\cite{Komorovsky2019} for open-shell TDDFT, and includes adding a trial vector
$\mathbf{t}_\tau^\mathrm{L} = (\mathbf{t}^X_\tau \ -\mathbf{t}^Y_\tau)^\mathrm{T}$ for
each vector $\mathbf{t}_\tau \equiv \mathbf{t}_\tau^\mathrm{R}$ as well as a vector
$\tilde{\mathbf{t}}_\tau^\mathrm{L} = (\mathbf{t}^{Y*}_\tau \ -\mathbf{t}^{X*}_\tau)^\mathrm{T}$
for its pair $\tilde{\mathbf{t}}_\tau \equiv \tilde{\mathbf{t}}_\tau^\mathrm{R}$.
In the case of linear response QEDFT describing a coupled electron--photon system, the special structure of paired and left trial vectors is translated to the photon part as well giving
\begin{equation}
\begin{pmatrix}
\mathbf{X} \\
\mathbf{Y} \\
\mathbf{M} \\
\mathbf{N} \\
\end{pmatrix}
=
\sum_\tau
c_\tau^\mathrm{R}
\begin{pmatrix}
\mathbf{t}_\tau^X \\
\mathbf{t}_\tau^Y \\
\mathbf{t}_\tau^M \\
\mathbf{t}_\tau^N \\
\end{pmatrix}
+
\sum_\tau
\tilde{c}_\tau^\mathrm{R}
\begin{pmatrix}
\mathbf{t}_\tau^{Y*} \\
\mathbf{t}_\tau^{X*} \\
\mathbf{t}_\tau^{N*} \\
\mathbf{t}_\tau^{M*} \\
\end{pmatrix}
+
\sum_\tau
c_\tau^\mathrm{L}
\begin{pmatrix}
\mathbf{t}_\tau^X \\
-\mathbf{t}_\tau^Y \\
\mathbf{t}_\tau^M \\
-\mathbf{t}_\tau^N \\
\end{pmatrix}
+
\sum_\tau
\tilde{c}_\tau^\mathrm{L}
\begin{pmatrix}
\mathbf{t}_\tau^{Y*} \\
-\mathbf{t}_\tau^{X*} \\
\mathbf{t}_\tau^{N*} \\
-\mathbf{t}_\tau^{M*} \\
\end{pmatrix}
,
\end{equation}
as the ansatz for the solution.
The expansion coefficients $c_\tau$ and $\tilde{c}_\tau$ are determined by solving Eq.~\eqref{eq:CavityCasida} projected
onto the subspace defined by the trial vectors. New trial vectors are generated in each iteration by preconditioning
the residue vectors where each desired eigenvalue can contribute up to two trial vectors ($\mathbf{t}$ and its pair $\tilde{\mathbf{t}}$)
in each iteration. The subspace is expanded until convergence determined by the norm of the residue vector for the particular
frequency is reached. The solutions for the desired number of eigenvalues are sought in a common basis composed of trial vectors
contributed by all the eigenvalues.
New trial vectors are orthogonalized i) with respect to the previously added trial vectors using Gram--Schmidt orthogonalization
and Kahan--Partlett ``twice is enough'' algorithm; and ii) within the $(\mathbf{t}, \tilde{\mathbf{t}})$ pair using an analytic formula.
The vectors that are reduced to a length below a set threshold during the orthogonalization are discarded from the trial basis.
This protocol achieves robust convergence of the solver across different molecular systems and the number of desired eigenvalues.
To cross-validate our results, we implemented the symmetrized version of the equation as well
and confirmed that these two equations lead to the same spectra.

\section{Computational details}

In the benchmark calculations presented below we consider the series of
group 12 atoms (Zn, Cd, Hg),
and a large mercury porphyrin (Hg@porphyrin).
The geometry of the Hg porphyrin was taken from Ref.~\citenum{Fransson2016}
and corresponds to a non-planar configuration of \ce{C_{4\textit{v}}} symmetry.

All spectra were calculated using the relativistic spectroscopy DFT program ReSpect.~\cite{ReSpect}
The reference orbitals for the linear response calculations were obtained from
the ground state uncoupled to photons.
The relativistic four-component molecular spinorbitals for electrons
were described Gaussian-type orbital (GTO) basis sets where
the scalar GTOs were used for the large component and the basis for
the small-component basis was generated by imposing the restricted kinetically
balanced (RKB) relation.~\cite{Stanton1984}
The selected scalar basis sets were the uncontracted Dyall's VDZ basis sets for metals (Zn, Cd, Hg)~\cite{dyall2007-4d, dyall2010-5d, Dyall2023-database1, Dyall2023-database2}
and the uncontracted Dunning's cc-pVDZ basis sets~\cite{Dunning1989} for light elements.
and the B3LYP XC approximation of the density functional.~\cite{slater1951, vosko1980, Becke1988, Lee1988, Stephens1994}
The numerical integration of the noncollinear exchange--correlation potential and kernel was done with an
adaptive molecular grid of medium size (program default).
In the 4c calculations, atomic nuclei of finite size were approximated by a Gaussian charge distribution model.

The electron--photon correlation is treated under the photon random phase approximation (pRPA)~\cite{Flick2019,Welakuh2022}
which amounts to disregarding the photon--electron exchange--correlation potential $v_{\mathrm{xc}}^{\mathrm{e-P}}\left(\left[n, q\right] ;  {\bf r}_l,t\right)$
and kernel $K_{\mu\nu\lambda\tau}^\mathrm{e-P}$
in Eqs.~\eqref{eq:vxc_terms} and \eqref{eq:XCkernel}, respectively.
Moreover, we consider the adiabatic approximation in the construction of the XC kernels, i.e. all 
dependencies on $\omega_\mathrm{ext}$ in Eqs.~\eqref{eq:CouplingMatrices} are dropped.
At present, we restrict ourselves to closed-shell systems for which the adiabatic noncollinear
XC kernel has the form given in Eqs.~24-26c in Ref.~\citenum{Konecny2019}.
The eigenvalue equation was solved iteratively for the first 50 or 100 excitation energies
for systems coupled to a single cavity mode.
The spectra were subsequently evaluated from the excitation energies and transition moments
obtained from the left and right eigenvectors, in the case of coupling to a single cavity
mode, empirical Lorentzian broadening was used to obtain finite-width peaks.

\section{Results and discussion}
\label{sec:Results}


\subsection{Atoms with singlet--triplet transitions}

First, we consider the case of atoms of group 12 metals, Zn, Cd, and Hg
whose spectra were previously studied at the level of four-component Dirac--Coulomb Hamiltonian
using linear response~\cite{Gao2005time, Bast2009relativistic} and real-time TDDFT~\cite{Repisky2015}.
The first two excited states correspond to the triplet state \ce{^3P} and the
singlet state \ce{^1P}, where
the transition from the \ce{^1S0} ground state to the excited triplet state
is forbidden in theories neglecting spin--orbit coupling (SOC).
With SOC considered, the transition to \ce{^3P1} becomes allowed
while the transitions to the other two triplet components (\ce{^3P0}, \ce{^3P2})
remain forbidden due to the $\Delta J$ selection rule within the dipole approximation.
The excitation energies for the allowed \ce{^1S0} $\rightarrow$ \ce{^3P1} and \ce{^1S0} $\rightarrow$ \ce{^1P1}
transitions are summarized in Table S1 in the Supplemental Material
while the calculated values reproduce reference experimental data~\cite{Sansonetti2005}
reasonably well given the functional and basis set size.
Both the \ce{^1S0} $\rightarrow$ \ce{^3P1} and \ce{^1S0} $\rightarrow$ \ce{^1P1} transitions
are composed of three degenerate spectral lines.
In the orbital picture, the highest occupied shell corresponds to $n$s orbitals
and the lowest unoccupied to $n$p while after SOC is taken into account the p
shell splits into $n$\ce{p_{1/2}} and $n$\ce{p_{3/2}} levels.
There are in total 12 possible single-electron excitations between $n$s and
$n$\ce{p_{1/2}} and $n$\ce{p_{3/2}} orbitals and these constitute the dominant
contributions to TDDFT transition vectors.
However, these 12 simple configurations cannot be straightforwardly identified
with the components of the singlet and triplet excited states as their transition
vectors contain more than one significant orbital pair contribution.

To investigate the effect of the strong coupling to photons on these atomic spectra,
we solve the linear response QEDFT equation for a single atom coupled to a single
effective cavity mode with the cavity frequency scanning the region around
the singlet--triplet (S--T) and singlet--singlet (S--S) transitions.
By focusing on a single efficient cavity mode, it is implied that the remaining spectrum of modes is accounted for by working with the renormalized (observable)
masses of the electrons~\cite{Ruggenthaler2023, Svendsen2023}.

\begin{figure}
\centering
\caption{Absorption spectra of Hg atom in a cavity strongly coupled ($g_\alpha = \unit[0.01]{au}$) to photonic modes, non-relativistic one-component (1c) vs relativistic four-component (4c) calculations.}
  \begin{subfigure}{0.32\textwidth}
    \includegraphics[width=\textwidth]{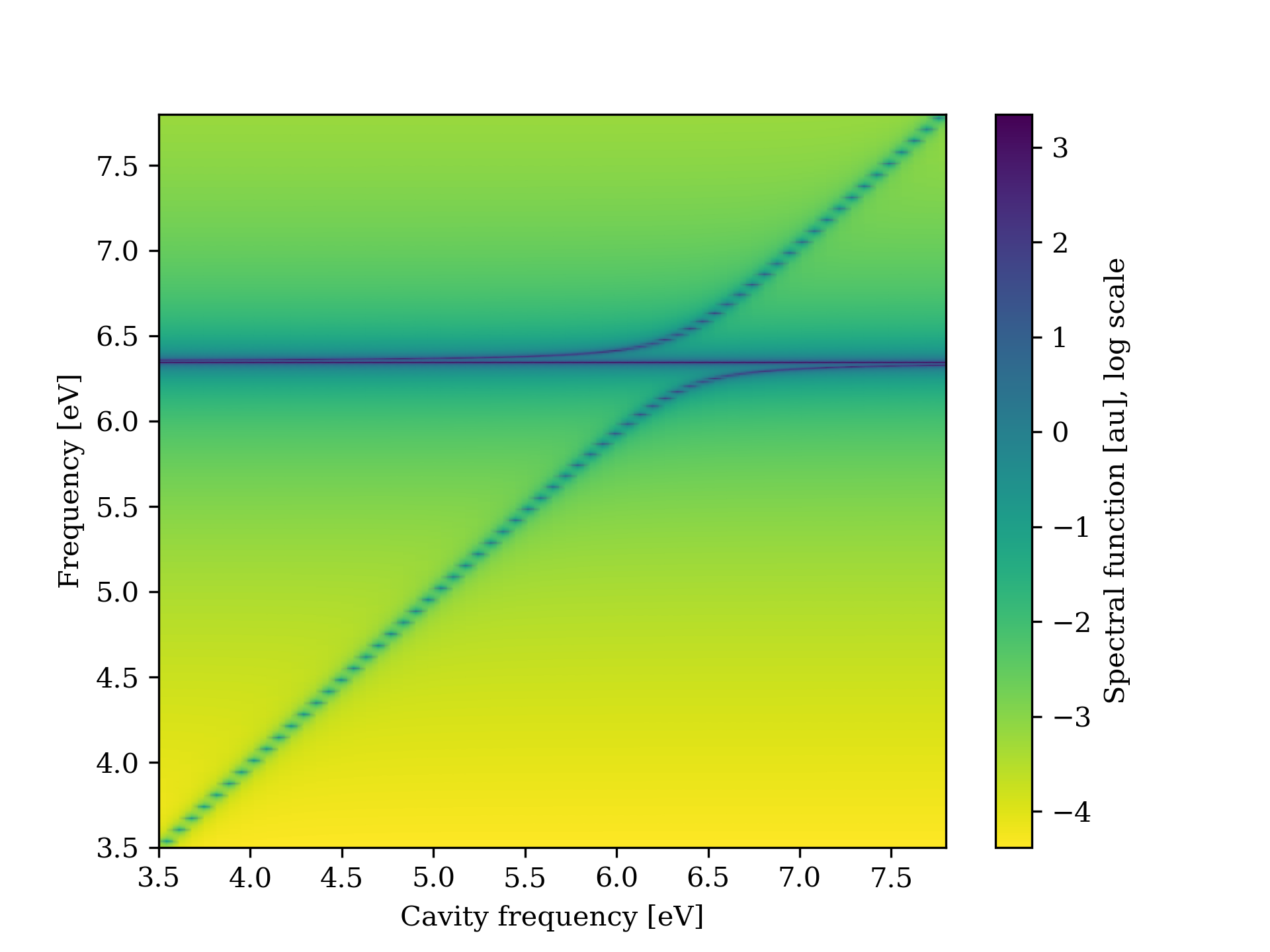}
    \caption{Hg 1c}
    \label{fig:Group12_1cVs4c:Hg1c}
  \end{subfigure}
  \begin{subfigure}{0.32\textwidth}
    \includegraphics[width=\textwidth]{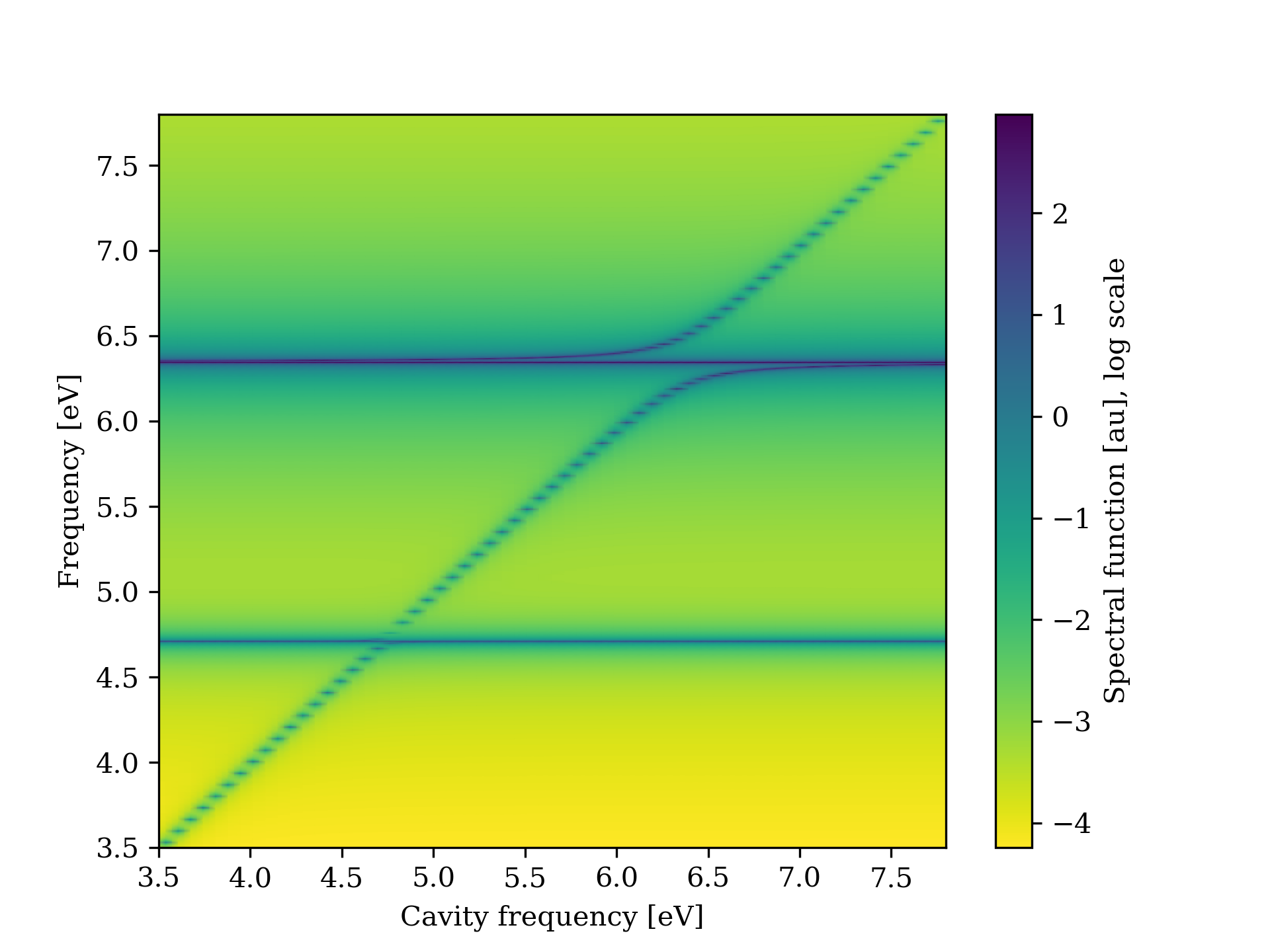}
    \caption{Hg 4c}
    \label{fig:Group12_1cVs4c:Hg4c}
  \end{subfigure}
  \begin{subfigure}{0.32\textwidth}
    \includegraphics[width=\textwidth]{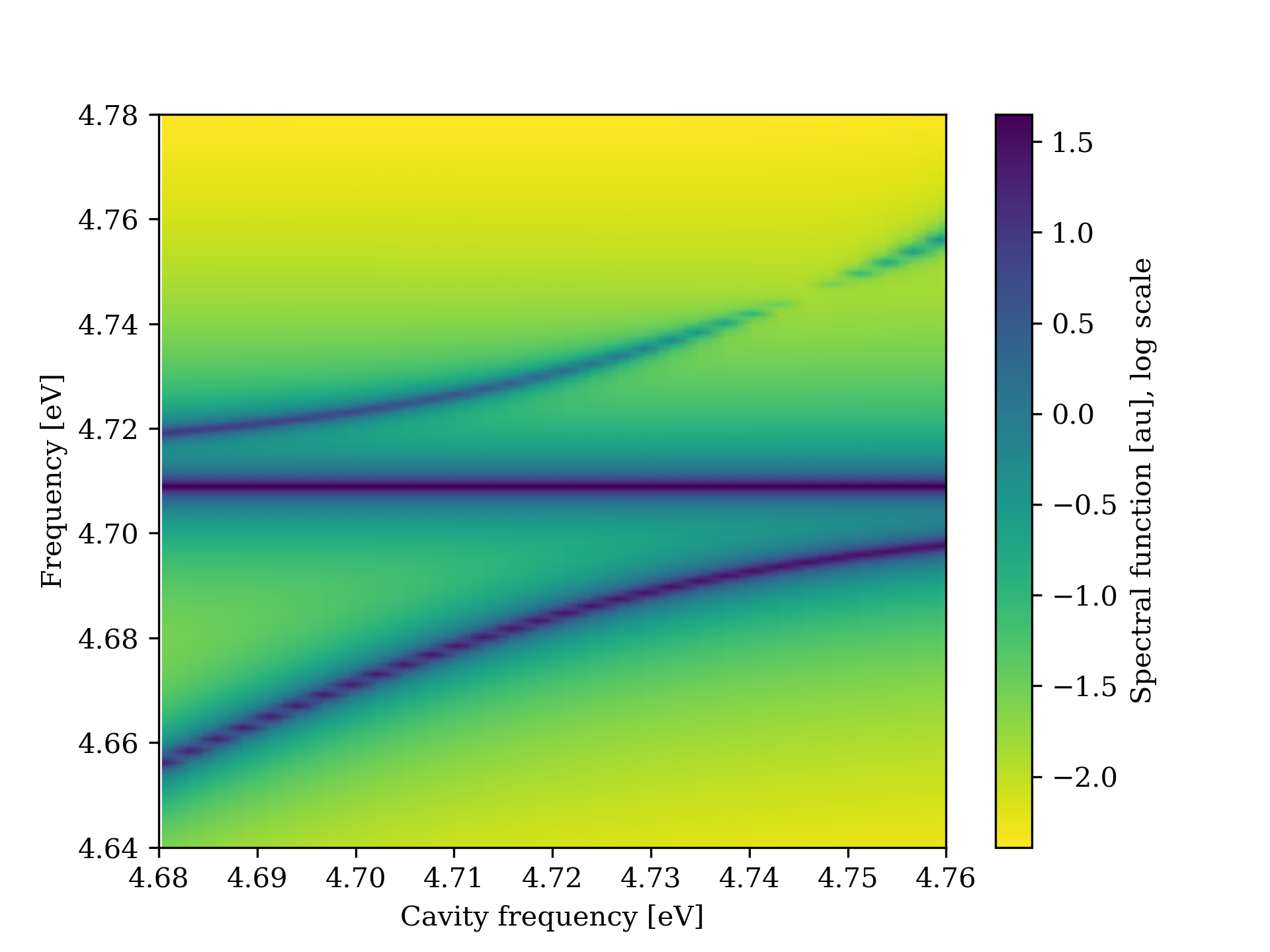}
    \caption{Hg 4c zoom}
    \label{fig:Group12_1cVs4c:Hgzoom}
  \end{subfigure}
  \label{fig:Group12_1cVs4c}
\end{figure}

The results of series of calculations with different cavity frequencies are excitation energies
and corresponding transition moments from which it is possible to calculate absorption spectra.
Such spectra are depicted for Hg atom in Figure~\ref{fig:Group12_1cVs4c} showing both non-relativistic
one-component (1c) and fully relativistic four-component (4c) calculations (results for Zn and Cd
atoms are in the Supplemental Material, Figure S1).
The $x$-axis shows the cavity frequency, the $y$-axis corresponds to excitation energy while
the magnitude of the spectral function is depicted as the color gradient in logarithmic
($\log_{10}$) scale to allow comparison between the S--T and S--S transitions of vastly
different magnitudes. The spectra were artificially broadened using Lorentzian
broadening with broadening parameter $\gamma = \unit[0.00002]{au}$ (see text after
Eqs.~\eqref{eq:RLtransDipMoments}). Only excited states with nonzero transition dipole moments are
visible as signals in the spectra, i.e. \ce{^1P1} and \ce{^3P1}, while the other components of the
triplet state with total angular momentum $J=0$ and $J=2$ are dark.

The plots depict a common scenario wherein light-matter mixing increases as the cavity frequency
approaches a resonance with an excitation energy. Both the \ce{^3P1} and \ce{^1P1} excited states
are originally triply degenerate with the degenerate levels having perpendicular transition
dipole moments. Therefore, the photon state mixes with only one of the three states -- the one
with the transition dipole moment parallel with the photon polarization -- and the uncoupled excitation
energies are still visible due to the two remaining levels of both \ce{^3P1} and \ce{^1P1}.
When compared with the non-relativistic picture, the relativistic description causes a shift
in the energy of the excited singlet state -- the non-relativistic spectra were artificially shifted
so that the S--S excitation energies would be aligned. The specific shifts were $\unit[0.16]{eV}$,
$\unit[0.43]{eV}$, and $\unit[1.57]{eV}$, for Zn, Cd, and Hg, respectively.
Furthermore, the inclusion of relativity introduces a transition to the excited triplet
state. The intensity of the S--T transition as well as the spectral shifts are more prominent
for heavier atoms.
As the transition to the excited triplet state becomes allowed, a mixing with the
cavity mode is observed near the resonant frequency. Since the coupling to the cavity mode is
proportional to the transition dipole moment, which for the S--T transition
increases with atomic number, the Rabi splitting of the polaritons originating from the
triplet state is the largest for mercury.
From the zoomed-in insets in the dispersion plots in Figure~\ref{fig:Group12_1cVs4c} we see
that the maximum (50:50) light--matter mixing does not occur
at exactly the S--T resonant frequency, but rather at a higher frequency.
The maximum light-matter mixing around the S--S resonance
also occurs at a frequency shifted from the resonance by approximately the same
amount as the around the S--T transition.

We calculated absorption spectra with cavities tuned
to these maximum mixing frequencies, i.e. $\unit[4.0204]{eV}$, $\unit[3.744]{eV}$, and
$\unit[4.724]{eV}$, for Zn, Cd, and Hg, respectively. The spectra for Zn and Hg are
depicted in Figure~\ref{fig:Group12_AllIn1} while the spectrum for Cd is available
in the Supplemental Material, Figure S2.
As discussed above, two S--T transitions with perpendicular transition dipole are still
present as excitation energies and visible in the spectra of Cd and Hg as a line with 2/3
of the intensity seen in the uncoupled system (for Zn, the extremely weak line corresponding
to the uncoupled S--T transitions is obscured by peak broadening).
In the spectral region around the S--T transition, the spectra contain very
intense polaritonic peaks with the increase of the transition moment with respect to the
original S--T transition ranging from a striking 50-fold increase for Zn to about 20\% for Hg
(compared to the intensity of one of the three degenerate states meaning that the peak
composed of two degenerate uncoupled S--T states is more intense than the polaritonic peak
in Hg).

While it is tempting to suggest an interpretation based on cavity enhancement of the S--T transition,
it is important to note that the peak appears also in the one-component non-relativistic (1c)
spectra. Since the 1c spectra are evaluated from the non-relativistic linear response calculations
that do not contain the S--T transition, the peak must correspond to the lower polariton
originating from the mixing of cavity mode with S--S exciton. Even if the cavity frequency is
far from the resonance with the S--S transition and the polaritonic state is light-like ($>90\%$),
its intensity is sufficient to be comparable with the S--T transition in the relativistic (4c)
calculations.
This is in agreement with the 2D spectra in Figure~\ref{fig:Group12_1cVs4c} where the
intensity of the lower polaritonic branch in the 1c calculations is more intense (Zn)
or comparable (Hg, also Cd) with the S--T transition in the 4c spectra.

In the 4c calculations, the effects of the coupling to the low-intensity S--T transition and
the intense but off-resonant S--S transition combine. Besides the spectral signature in the
S--T region discussed above, in the region around the S--S transition, one
of the originally three degenerate spectral lines is blueshifted and its response
vector contains a small but non-negligible ($10^{-3}$) photonic contribution.
Most of the intensity increase of the polaritonic peaks arising from the S--T transition
is accounted for by the corresponding decrease in intensity of the shifted S--S transition.
The decrease in intensity of the shifted S--S line is in fact larger
than the gain of the S--T transition, with the gain being about 80\% of the decrease.
The Thomas--Reiche--Kuhn sum rule is then preserved by states higher in the spectrum
when one state of three degenerate states is also slightly shifted in energy and its
intensity is now higher. This repeats for all triply degenerate states at higher energies.
The sum of photonic contributions from the lower and upper S--T polariton
and the shifted S--S line is 1 with the accuracy of $10^{-4}$--$10^{-5}$ and
lower for the higher lying states.
Note that in pRPA QEDFT with single cavity mode, the total sum of photon contributions
should be 1 and only one extra peak can appear in the spectrum of an atom or molecule
coupled to the cavity.

\begin{figure}
\centering
\caption{Absorption spectra of Hg atom in a cavity with different coupling strength $g_\alpha$
compared to a two-level Jaynes--Cummings model (black lines) considering only the S--T transition.}
  \begin{subfigure}{0.32\textwidth}
    \includegraphics[width=\textwidth]{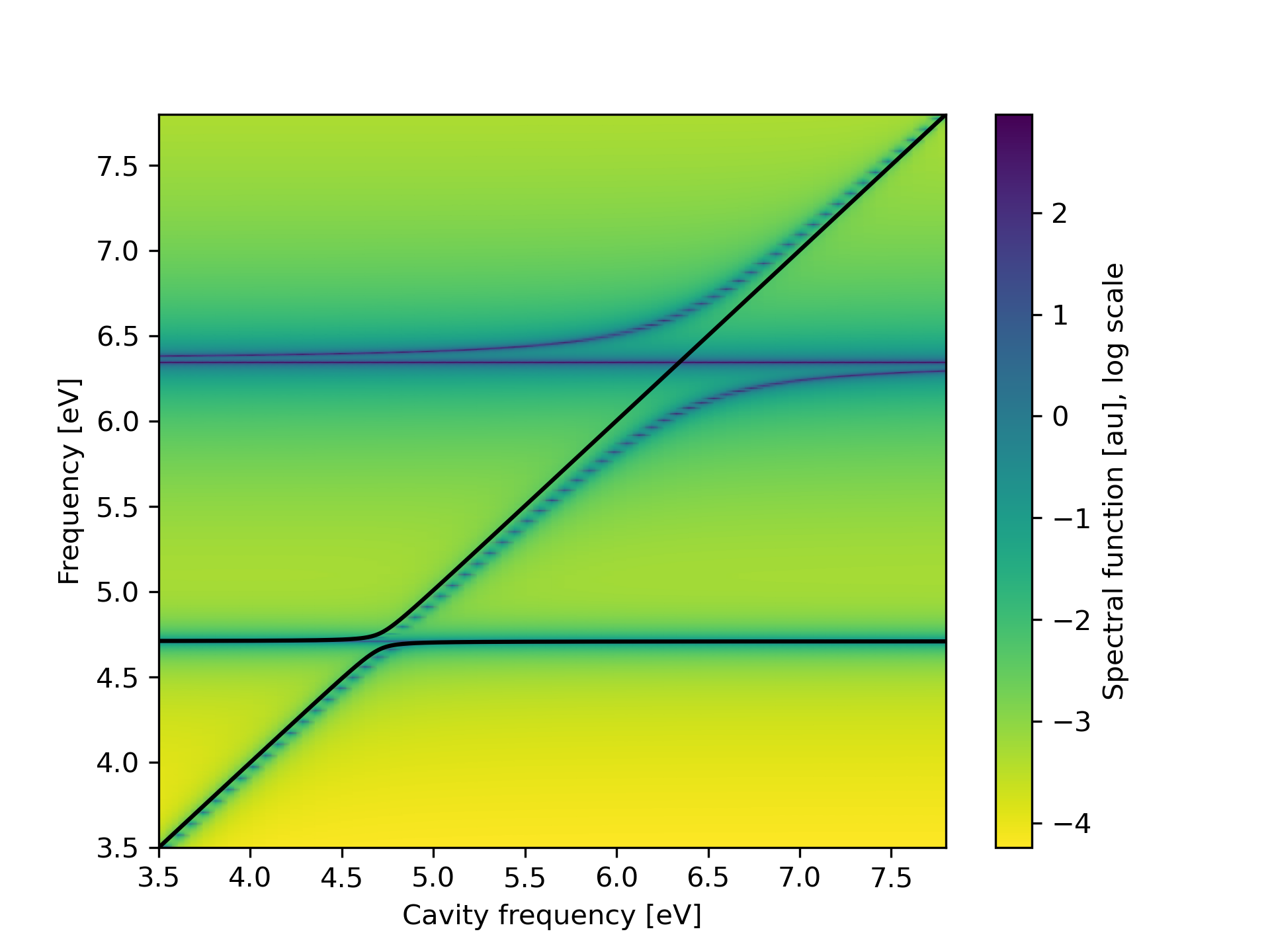}
    \caption{$g_\alpha = 0.02$ au}
  \end{subfigure}
  \begin{subfigure}{0.32\textwidth}
    \includegraphics[width=\textwidth]{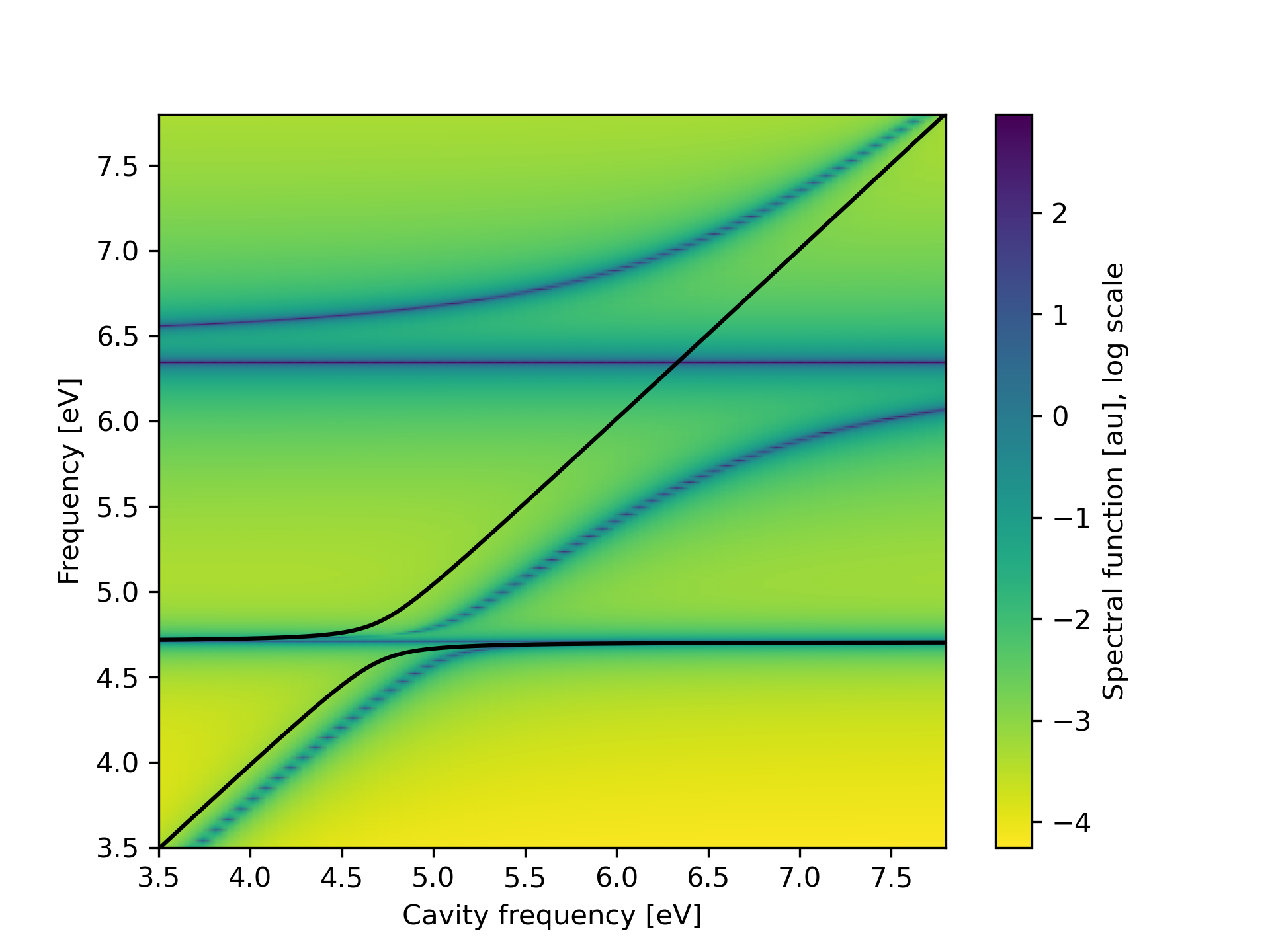}
    \caption{$g_\alpha = 0.05$ au}
  \end{subfigure}
  \begin{subfigure}{0.32\textwidth}
    \includegraphics[width=\textwidth]{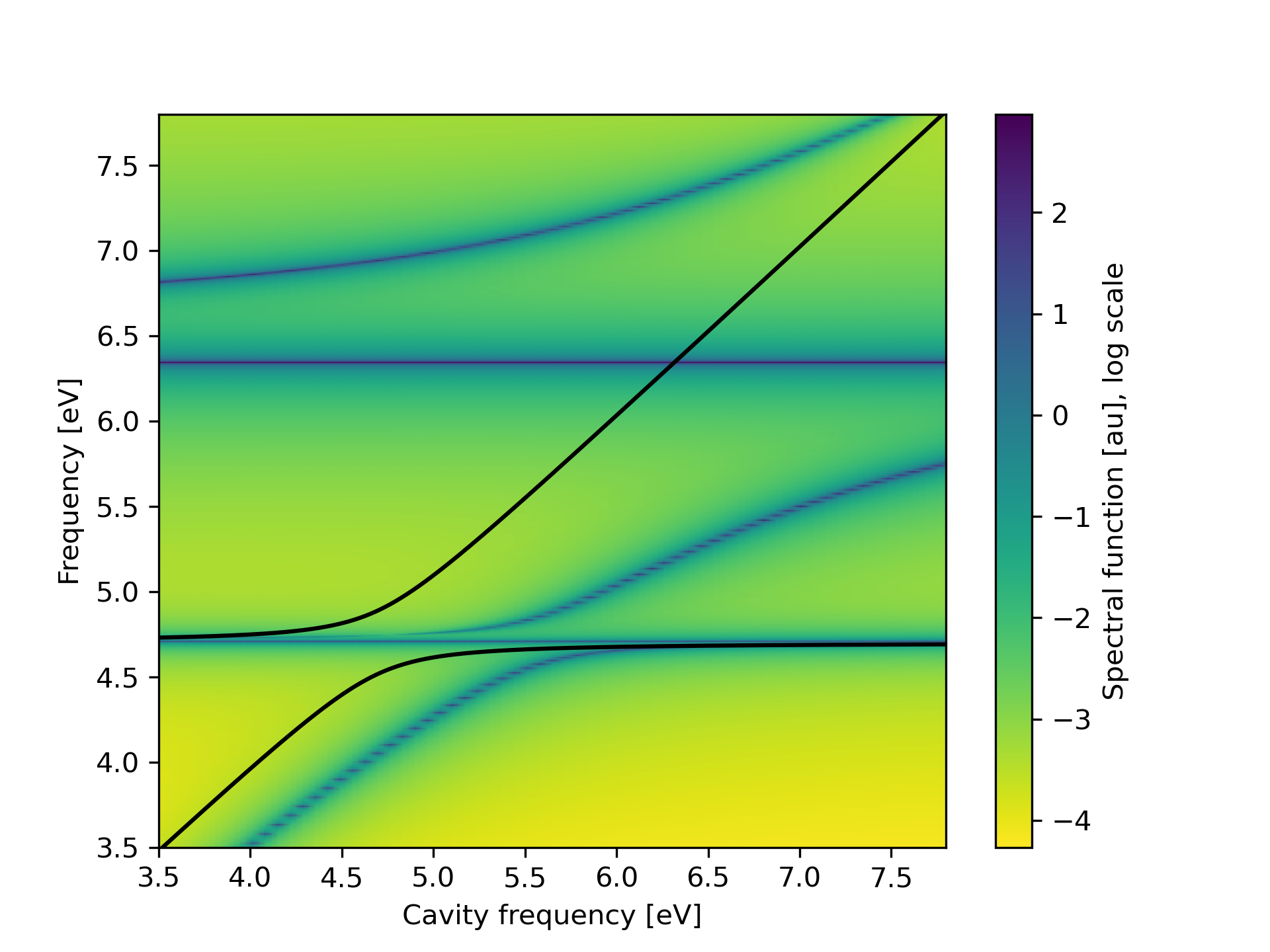}
    \caption{$g_\alpha = 0.08$ au}
  \end{subfigure}
  \label{fig:Hg_QEDFTvsJC}
\end{figure}

These results show the importance of the off-resonant coupling where the photon mode
is coupled not only to the resonant S--T exciton but also to the off-resonant yet
high-transition-moment S--S exciton as well as \emph{all higher-lying states}.
In turn, this demonstrates the importance of \textit{ab initio} QEDFT treatment: a simple
two-state model built from the S--T excitation energy and corresponding transition
dipole moment would produce two polaritonic states whose intensity would sum to the
original S--T intensity.
Another effect beyond the two-state model is the shift of S--T resonance (50:50 light-matter
mixing) toward higher frequencies. This becomes more pronounced as the coupling strength
increases as shown in Figure~\ref{fig:Hg_QEDFTvsJC}. Here, the results of QEDFT calculations
are compared with Jaynes--Cummings (JC) energies
$E_{\pm} = (\hbar \omega_\alpha - \frac{1}{2}\hbar\Delta) \pm \frac{1}{2} \hbar \sqrt{\Delta^2 + \Omega^2}$, where $\Delta = \omega_\alpha - \omega_\mathrm{S-T}$ and
$\Omega = 2 g_\alpha \sqrt{\frac{\omega_\mathrm{\alpha}}{2\hbar}}  \left\langle ^3\mathrm{P}_1 | \boldsymbol{\mu} \cdot \boldsymbol{\epsilon}_{\alpha} | ^1\mathrm{S}_0 \right\rangle$ with
S--T excitation energy $\omega_\mathrm{S-T}$ and transition dipole moment $\left\langle ^3\mathrm{P}_1 | \boldsymbol{\mu} \cdot \boldsymbol{\epsilon}_{\alpha} | ^1\mathrm{S}_0 \right\rangle$ taken
from a TDDFT calculation without coupling to cavity.

While models can be extended to account for more than two states, it is impossible
to \textit{a priori} say how many excited states should be included, particularly
for systems with dense spectra such as molecules. Therefore, \textit{ab initio} calculations
are an indispensable tool for qualitative and quantitative predictions of properties
of strongly bound light--matter systems, as well as for guiding the development of models.
The quest to extend their capabilities to include further effects, such as relativity
and spin--orbit coupling, is thus an important direction in polaritonic chemistry.

\begin{figure}
 \centering
 \caption{Absorption spectra of Zn and Hg atoms in cavities strongly coupled ($g_\alpha = \unit[0.01]{au}$) to a cavity set to effective resonance with the singlet--triplet (S--T) transition defined by 50:50 light--matter mixing of the lower polariton rather than by
 the numerical value compared to reference spectra of free atoms without cavities. The region around the low-intensity S--T transition is magnified to ease reading by a factor specified in each figure.}
   \begin{subfigure}{0.49\textwidth}
     \includegraphics[width=\textwidth]{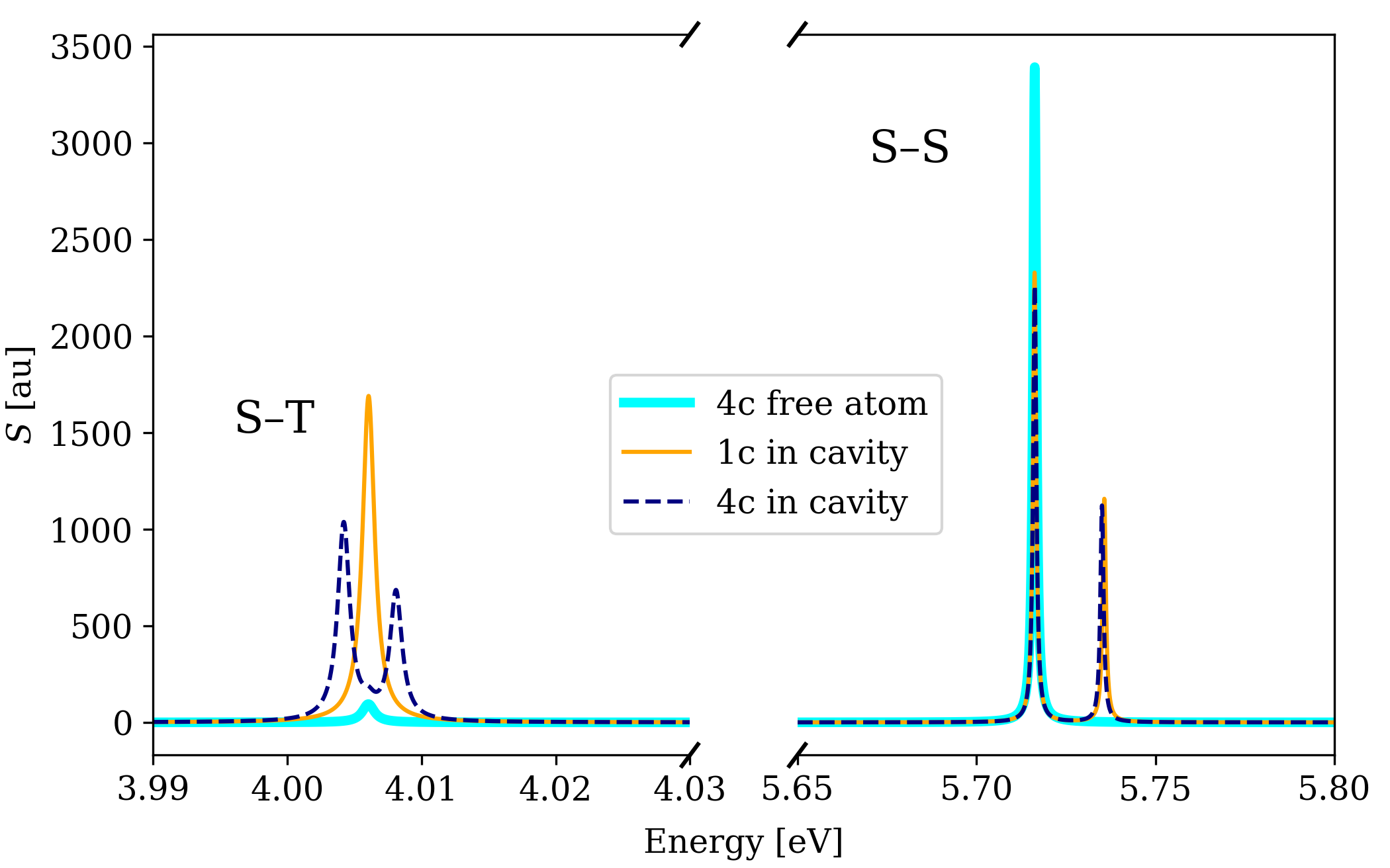}
     \caption{Zn}
   \end{subfigure}
  \begin{subfigure}{0.49\textwidth}
    \includegraphics[width=\textwidth]{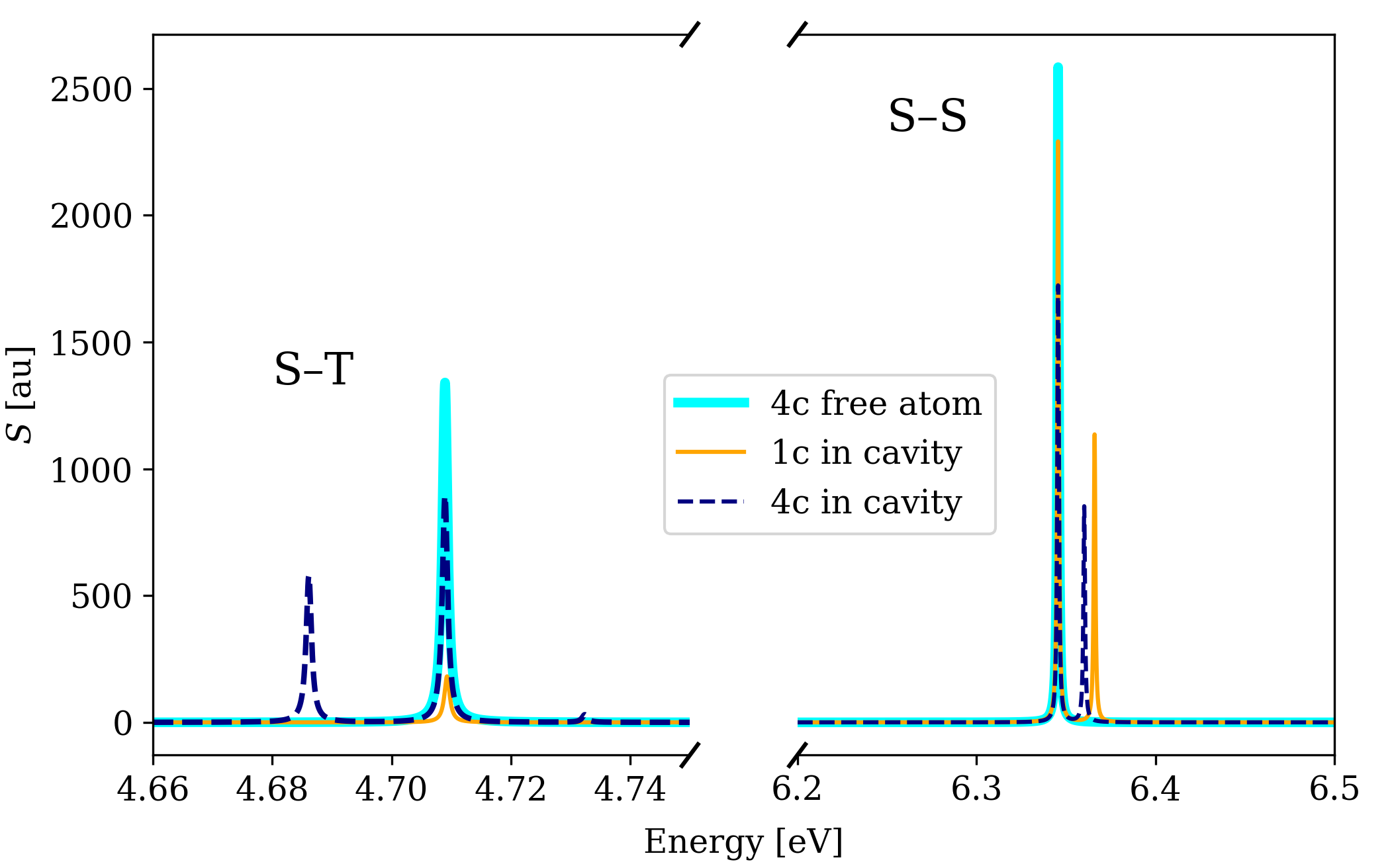}
    \caption{Hg}
    \label{fig:Group12_AllIn1:Hg}
  \end{subfigure}
  \label{fig:Group12_AllIn1}
\end{figure}

\subsection{Large molecular metal complex}

To demonstrate that our implementation is not restricted to atoms,
but applicable also to large molecules containing heavy elements,
we calculate the electron absorption spectrum of a mercury porphyrin inside a Fabry--P{\'e}rot cavity.
Porphyrins are a class of molecules with important biological roles as well as technological
applications and there is a large number of both experimental~\cite{Edwards1971} and theoretical~\cite{Baerends2002, Day2008} works devoted to
their study including the modification of their properties in optical cavities.~\cite{Kena2007, Avramenko2022, Sun2022}
Moreover, the absorption spectra of porphyrin complexes of group 12 atoms including the Hg porphyrin
considered here, were studied before at the four-component relativistic level of theory
by Fransson et al.~\cite{Fransson2016} 
Note that unlike the reference work~\cite{Fransson2016} we do not explicitly consider
point group symmetry in our calculations thus performing an all-atom all-electron
calculation in the ground state.

\begin{figure}
  \centering
  \caption{Comparison of relativistic (4c) electron absorption spectra
           (both band spectra, i.e. spectral function $S$ on the left
           vertical axis, and line spectra as oscillator strength $f$
           on the right vertical axis, only lines with $f>10^{-5}$ are plotted.)
           of Hg porphyrin complex in free space and in cavity set to
           resonance with the porphyrin B line with different coupling
           strengths $g_\alpha$ [au]. The Lorentzian spectra were obtained
           using broadening parameter $\gamma = 0.002$.
           Inset: the structure of the molecule.
           }
  \includegraphics[width=0.7\textwidth]{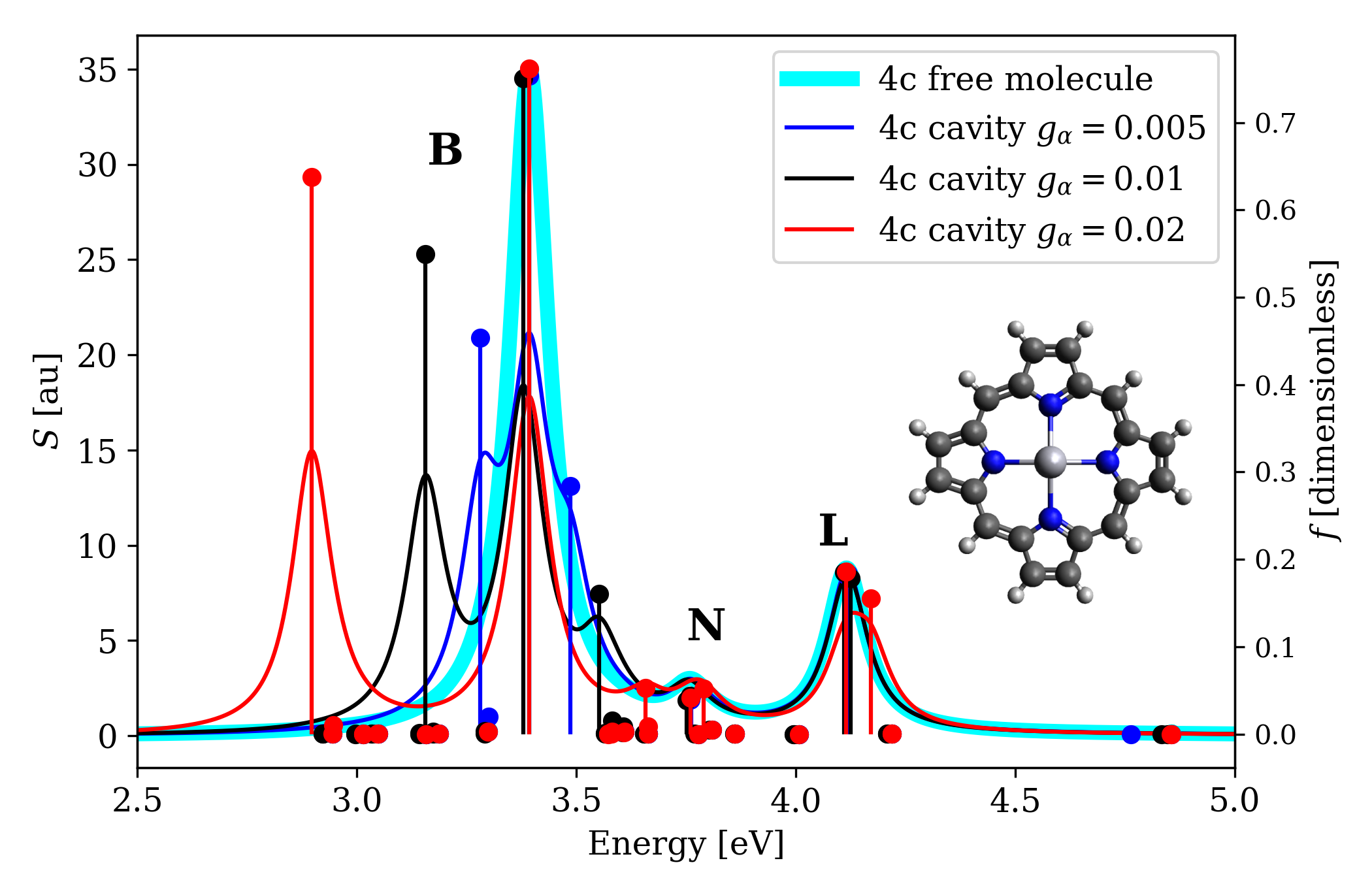}
  \label{fig:HgP-B3LYP_g005}
\end{figure}

The spectrum of the free Hg-substituted porphyrin is dominated by the strong B-band
at $\unit[3.39]{eV}$, while other notable features are the N- and L-bands at 3.76
and $\unit[4.12]{eV}$, respectively with the Q-band being very weak and not visible
in this spectrum.
All of these spectral lines arise from doubly degenerate excited states with perpendicular
transition dipole moments, i.e. $y$ and $z$ for the molecule oriented such that
its \ce{C4} symmetry axis coincides with the $x$ axis.
In the calculations of polaritonically modified spectra we consider the Hg porphyrin complex
embedded in a Fabry--P\'{e}rot cavity with an effective single mode of frequency that is in resonance with the B band excitation energy with progressively stronger coupling constant
$g_\alpha$.
The resulting spectra together with the spectrum of the uncoupled system are depicted
in Figure~\ref{fig:HgP-B3LYP_g005}. While for the weakest coupling $g_\alpha = 0.005$
(dark blue line), the dominating modification of the spectrum due to the cavity is the
splitting of one of the two degenerate B lines into an upper and lower polariton,
for the stringer values of the coupling, other lines in the spectrum start to be
affected as well. Particularly for $g_\alpha = 0.02$ we see breaking of the degeneracy
in the form of frequency and intensity shifts of other spectral lines including the N
and L lines. Therefore, besides demonstrating the readiness of our method for the
treatment of chemically relevant molecules, these results also stress the importance
of ab initio QED methods.

\section{Conclusions}

In this article we presented the first implementation of a relativistic \textit{ab initio} quantum electrodynamical method combining
relativistic description of electronic structure with the treatment of transverse photons as dynamical variables
in the form of linear response QEDFT based on the four-component Dirac--Kohn--Sham Hamiltonian.
We presented a detailed theory derivation of the final electron--photon Hamiltonian starting from relativistic QED
while keeping track of all approximations made on the way.
For the linear response equations we showed how a spectral function can be calculated from a non-hermitian
Casida-like equation by considering both the left and right eigenvectors, as well as showed how this equation
can be transformed to the symmetric equation often considered in \textit{ab initio} QEDFT literature.

We applied the developed method to the calculation of polaritonically modified spectra of heavy-element atoms
with singlet--triplet transitions and demonstrated that the intensity of the transition can be enhanced by the
coupling to the cavity as well showed the coupling between the singlet--singlet and singlet--triplet transitions
via the cavity modes that results in non-resonant splitting of spectral lines as well the transfer of intensity
between them. Of specific interest is that the cavity even in the long-wavelength limit can influence the SOC due to a fully relativistic (four-component) treatment of the electrons.
Our results showing the interplay of off-resonant coupling and spin--orbit driven
transition demonstrate the need for accurate \textit{ab initio} methods such as QEDFT to describe
cavity modification of molecular properties as well as help to guide the formulation of
approximate models -- in the cases presented a minimal two-state model would be insufficient
in describing the cavity modified of spectra around the singlet--triplet resonance.
Moreover, the linear response QEDFT methodology developed with the primary motivation of describing polaritonic
chemistry in cavities allows to address other situations when electrons are coupled to photons as well.
Such an example is the coupling to vacuum modes that presents a way of calculating radiative lifetimes from first
principles thereby resolving one of the shortcomings of electronic-structure linear-response theories
that add this variable as an empirical
parameter.
Finally by addressing the spectrum of mercury porphyrin in the cavity we demonstrated that our program is capable
of handling even large molecules of chemical interest.

\textit{Ab initio} QED methods have emerged recently as a first principles approach for the prediction and interpretation of
phenomena occurring under strong coupling of electronic and vibrational degrees of freedom of a material to the
photonic modes of an optical cavity. In this work, we expanded the toolbox of \textit{ab initio} QED methods further by
bringing them into the relativistic domain. Considering the advances in understanding also collective coupling effects~\cite{Sidler2020,sidler2023unraveling,schnappinger2023cavity} and the here presented changes in the SOC, we think that this toolbox can help to design cavity setups such that SOC and other relativistic effects can be enhanced. This also conceptually is very interesting since it highlights that the transverse photons, usually discarded when deriving relativistic quantum chemistry equations from full QED, can be decisive for real atomic and molecular systems. For instance, taking into account (approximately) the continuum of modes can be used to study non-perturbatively very fundamental effects of quantum field theories, such as the mass renormalization of the electrons~\cite{Ruggenthaler2023}.
We also showed how relativistic description of electronic structure of a material
inside a photonic structure reveals the presence of a new type of spin--orbit interaction
mediated by cavity photons.
This opens a new avenue in the fields of cavity QED and polaritonic chemistry with a potential
for follow-up works focused on singlet--triplet transitions in singlet fission systems,
the exploration of the new SOC term in solid state systems,
or the development of effective Hamiltonians.

\begin{appendices}

\section{Notations and conventions}
%
%
%
We use the following definitions: $x^{\mu}$ for contravariant four-vector, $x_{\mu}=g_{\mu \nu} x^{\nu}$ for covariant four-vector, $g_{\mu \nu}$ is the metric tensor
$$
g_{\mu \nu}=g^{\mu \nu}=\left(\begin{array}{rrrr}
1 & 0 & 0 & 0 \\
0 & -1 & 0 & 0 \\
0 & 0 & -1 & 0 \\
0 & 0 & 0 & -1
\end{array}\right),
$$
and $\alpha^{\mu}=\gamma^{0} \gamma^{\mu}$ are the Dirac matrices in their standard 
representation
$$
\alpha^{0}=\left(\begin{array}{cc}
\mathds{1}_{2 \times 2} & 0 \\
0 & \mathds{1}_{2 \times 2}
\end{array}\right), \quad
\boldsymbol{\alpha}=\left(\begin{array}{cc}
0 & \boldsymbol{\sigma} \\
\boldsymbol{\sigma} & 0
\end{array}\right), 
$$
$$
\gamma^{0} \equiv \beta=\left(\begin{array}{cc}
\mathds{1}_{2 \times 2} & 0 \\
0 & -\mathds{1}_{2 \times 2}
\end{array}\right), \quad
\boldsymbol{\gamma}=\left(\begin{array}{cc}
0 & \boldsymbol{\sigma} \\
-\boldsymbol{\sigma} & 0
\end{array}\right),
$$
and the $2 \times 2$ Pauli matrices

$$
 \quad \sigma_{x}=\left(\begin{array}{cc}
0 & 1 \\
1 & 0
\end{array}\right), \quad \sigma_{y}=\left(\begin{array}{cc}
0 & -i \\
i & 0
\end{array}\right), \quad \sigma_{z}=\left(\begin{array}{rr}
1 & 0 \\
0 & -1
\end{array}\right).
$$

Here $\mathds{1}_{2 \times 2}$ is a unit $2 \times 2$ matrix
$$\mathds{1}_{2 \times 2}=\left(\begin{array}{ll}
1 & 0 \\
0 & 1
\end{array}\right).
$$
We use the italic style $(r)$ for scalars, boldface $(\bf{r})$ for three vectors, and Roman style $(\mathrm{r})$ for four-vectors and their components. 
Four vectors have the form $\mathrm{p}=\left(p_{0}, \bf{p}\right)$ and coordinates in four space are also denoted as $\mathrm{r}=\left(t, \bf{r}\right)$. 
The scalar product of four-vectors is defined as $\mathrm{k} \mathrm{p}=k^{\mu} p_{\mu}=k^{0} p_{0}-\bf{k \cdot p}$. The summing over repeated indices is assumed by default.
%
%
%
Fock space vectors and matrices are also denoted by Roman boldface with $\mathbf{1}$ and $\mathbf{0}$ denoting the unit and zero matrix, respectively.

\section{Solution of first order EOMs by undetermined coefficients}
\label{sec:AppendixLRderivation}

In the method of undetermined coefficients, we use the following ansatz for the electronic equation
\begin{subequations}
\label{eq:AnsatzEl}
\begin{align}
    d^{(1)}_{ai}(t)   & = X_{ai} e^{-i \omega_\mathrm{ext} t} + Y^*_{ai} e^{i \omega_\mathrm{ext} t} , \\
    d^{(1)*}_{ai}(t) & = X^*_{ai} e^{i \omega_\mathrm{ext} t} + Y_{ai} e^{-i \omega_\mathrm{ext} t} ,
\end{align}
\end{subequations}
motivated by the inhomogeneity in Eq.~\eqref{eq:d1CoefEOM}.
Similarly, we formulate an ansatz for the photonic displacement coordinate in the form
\begin{equation}
\label{eq:AnsatzPh_q}
    q^{(1)}_\alpha (t) = \tilde{M}_{\alpha} e^{-i \omega_\mathrm{ext} t} + \tilde{M}^*_{\alpha} e^{i \omega_\mathrm{ext} t},
\end{equation}
from which follows
\begin{equation}
\label{eq:AnsatzPh_p}
    p^{(1)}_\alpha (t) = - \omega_\mathrm{ext} \tilde{M}_{\alpha} e^{-i \omega_\mathrm{ext} t} + \omega_\mathrm{ext} \tilde{M}^*_{\alpha} e^{i \omega_\mathrm{ext} t}
    ,
\end{equation}
and where the real nature of $q^{(1)}_\alpha (t)$ was taken into account.

We consider only a single inhomogeneity in the whole system of equations at a time, i.e., the external field
or the external current. The ans{\"a}tze will then contain its respective frequency (i.e. $\omega_\mathrm{ext}$ or $\omega'_\mathrm{ext}$).
By inserting expressions from Eqs.~\eqref{eq:AnsatzEl}-\eqref{eq:AnsatzPh_p} into Eqs.~\eqref{eq:4coupledDiffEq} and selecting
only the terms with time dependence $\exp(-i\omega_\mathrm{ext}t)$ we obtain a set of coupled algebraic equations
\begin{subequations}
\label{eq:AlgSystemXYMtilde}
\begin{align}
    \omega_\mathrm{ext} X_{ai}
    & =
    (A_{ai,bj} + \Delta_{ai,bj}) X_{bj} + (B_{ai,bj} + \Delta'_{ai,bj}) Y_{bj} - L_{ai,\alpha} \tilde{M}_\alpha + P_\mathrm{ai} , \\
    - \omega_\mathrm{ext} Y_{ai}
    & =
    (A^*_{ai,bj} + \Delta^*_{ai,bj}) Y_{bj} + (B^*_{ai,bj} + \Delta^{\prime *}_{ai,bj}) X_{bj} - L^*_{ai,\alpha} \tilde{M}_\alpha + P^*_\mathrm{ai} , \\
    \label{eq:AlgEqMw2}
    ( -\omega_\mathrm{ext}^2 + \omega_\alpha^2 ) \tilde{M}_\alpha
    & =
    Q'_{jb,\alpha} X_{bj} + Q^{\prime *}_{jb,\alpha} Y_{bj} .
\end{align}
\end{subequations}
Unfortunately, the equation for $\tilde{M}_\alpha$ contains $\omega_\mathrm{ext}^2$, a recurrence of the second order
differential equation~\eqref{eq:q2ndOrderDE} as we would obtain the system of equations~\eqref{eq:AlgSystemXYMtilde}
also by considering Eq.~\eqref{eq:q2ndOrderDE} instead of Eqs.~\eqref{eq:p1stOrderDef} and \eqref{eq:p1stOrderEOM}
for the photonic degrees of freedom.
Since we want to get an equation linear in $\omega_\mathrm{ext}$,
we remove the second power by employing the identity
\begin{equation}
    \frac{1}{\omega_\mathrm{ext}^2 - \omega_\alpha^2}
    =
    \frac{1}{2\omega_\alpha}
    \left(
    \frac{1}{\omega_\mathrm{ext} + \omega_\alpha} - \frac{1}{\omega_\mathrm{ext} - \omega_\alpha}
    \right)
    ,
\end{equation}
which transforms Eq.~\eqref{eq:AlgEqMw2} into
\begin{equation}
    \label{eq:w2trick}
    \tilde{M}_\alpha
    =
    \frac{1}{2\omega_\alpha}
    \left(
    \frac{Q'_{jb,\alpha} X_{bj} + Q^{\prime *}_{jb,\alpha} Y_{bj}}{\omega_\mathrm{ext} + \omega_\alpha} - \frac{Q'_{jb,\alpha} X_{bj} + Q^{\prime *}_{jb,\alpha} Y_{bj}}{\omega_\mathrm{ext} - \omega_\alpha}
    \right)
    .
\end{equation}
We can now divide $\tilde{M}_\alpha$ into two components $\tilde{M}_\alpha = M_\alpha + N_\alpha$
that satisfy Eq.~\eqref{eq:w2trick} term by term
\begin{subequations}
\begin{align}
    M_\alpha
    & =
    -\frac{1}{2\omega_\alpha} \frac{Q'_{jb,\alpha} X_{bj} + Q^{\prime *}_{jb,\alpha} Y_{bj}}{\omega_\mathrm{ext} - \omega_\alpha} , \\
    N_\alpha
    & =
    \frac{1}{2\omega_\alpha} \frac{Q'_{jb,\alpha} X_{bj} + Q^{\prime *}_{jb,\alpha} Y_{bj}}{\omega_\mathrm{ext} + \omega_\alpha}
    ,
\end{align}
\end{subequations}
and both contain $\omega_\mathrm{ext}$ only in the first power.
This way, we can write four coupled algebraic equations linear in $\omega_\mathrm{ext}$ for the variables
$X_{ai}$, $Y_{ai}$, $M_\alpha$, and $N_\alpha$ that collected in a matrix give
\begin{equation}
    \label{eq:CavitySternheimer}
    \begin{bmatrix}
    \begin{pmatrix}
    \mathbf{A} + \mathbf{\Delta}     & \mathbf{B} + \mathbf{\Delta}'             & -\mathbf{L}     & -\mathbf{L}   \\
    \mathbf{A}^* + \mathbf{\Delta}^* & \mathbf{B}^* + \mathbf{\Delta}^{\prime *} & -\mathbf{L}^*   & -\mathbf{L}^* \\
    -\mathbf{Q}                      & -\mathbf{Q}^*                             & \bs{\omega}     & \mathbf{0}    \\
    -\mathbf{Q}                      & -\mathbf{Q}^*                             & \mathbf{0}      & \bs{\omega}   \\
    \end{pmatrix}
    -
    \omega_\mathrm{ext}
    \begin{pmatrix}
    \mathbf{1} &  \mathbf{0} & \mathbf{0} &  \mathbf{0} \\
    \mathbf{0} & -\mathbf{1} & \mathbf{0} &  \mathbf{0} \\
    \mathbf{0} &  \mathbf{0} & \mathbf{1} &  \mathbf{0} \\
    \mathbf{0} &  \mathbf{0} & \mathbf{0} & -\mathbf{1} \\
    \end{pmatrix}
    \end{bmatrix}
    \begin{pmatrix}
    \mathbf{X} \\
    \mathbf{Y} \\
    \mathbf{M} \\
    \mathbf{N} \\
    \end{pmatrix}
    =
    \begin{pmatrix}
    \mathbf{P}   \\
    \mathbf{P}^* \\
    \mathbf{0}   \\
    \mathbf{0}
    \end{pmatrix}
    ,
\end{equation}
where we defined
\begin{equation}
Q_{jb,\alpha}
=
\frac{1}{2\omega_\alpha}Q'_{jb,\alpha}
=
\frac{g_\alpha e \omega_\alpha}{\sqrt{\hbar}} \frac{1}{2\omega_\alpha} (\mathbf{r}_{jb} \cdot \boldsymbol{\epsilon}_\alpha)
=
\frac{1}{2} \frac{g_\alpha e}{\sqrt{\hbar}}(\mathbf{r}_{jb} \cdot \boldsymbol{\epsilon}_\alpha)
,
\end{equation}
and $\bs{\omega}$ is a diagonal matrix with elements $\omega_\alpha$ on the diagonal.
An analogous equation can be derived for the case of external current with the right-hand side
having the form $(\mathbf{0}\ \mathbf{0}\ \mathbf{J}\ \mathbf{J}^*)^\mathrm{T}$.
This is the Sternheimer equation describing the response of a molecular system in a cavity to an external
field of frequency $\omega_\mathrm{ext}$ coupled to the electronic subsystem via operator $\mathbf{P}$.
The algebraic form of the homogeneous equation to the system \eqref{eq:4coupledDiffEq} can now be obtained
from Eq.~\eqref{eq:CavitySternheimer} by discarding the right-hand side with the frequency of the external field
gaining the meaning of a resonance frequency of the system, i.e. $n$-th excitation energy $\Omega_n$, leading to Eq.~\eqref{eq:CavityCasida}.

\section{Solution of the matrix differential equation and the interpretation of photon terms}
\label{sec:AppendixMatrixDiffEq}

It is interesting to note that an alternative approach to Eq.~\eqref{eq:CavityCasida} exists based
on a direct solution of the coupled system of differential equations~\eqref{eq:4coupledDiffEq}.
This system corresponds to a matrix differential equation in the form
$i\dot{\mathbf{x}}(t) = \bm{\mathcal{M}}\mathbf{x}(t) + \mathbf{b}(t)$,
where
\begin{subequations}
    \begin{align}
        \mathbf{x}(t)
        & =
        ( \mathbf{d}^{(1)}(t),\ \mathbf{d}^{(1)*}(t),\ \mathbf{p}^{(1)}(t),\ \mathbf{q}^{(1)}(t) )^\mathrm{T}, \\
        \label{eq:MinMatDiffEq}
        \bm{\mathcal{M}}
        & =
        \begin{pmatrix}
        \mathbf{A} + \mathbf{\Delta}                 & \mathbf{B} + \mathbf{\Delta}'      & \mathbf{0}  & - \mathbf{L}   \\
        - \mathbf{B}^* -  \mathbf{\Delta}^{\prime *} & - \mathbf{A}^* - \mathbf{\Delta}^* & \mathbf{0}  &   \mathbf{L}^* \\
        \mathbf{Q}^{\prime *}                        & \mathbf{Q}'                        & \mathbf{0}  & - \bm{\omega}^2 \\
        \mathbf{0}                                   & \mathbf{0}                         & -\mathbf{1} & \mathbf{0}
        \end{pmatrix}, \\
        \mathbf{b}(t)
        & =
        (\mathbf{P} e^{-i \omega_\mathrm{ext} t}  + \mathbf{P}^* e^{i \omega_\mathrm{ext} t},\
        -\mathbf{P}^* e^{i \omega_\mathrm{ext} t} - \mathbf{P} e^{- i \omega_\mathrm{ext} t},\
         \mathbf{J} e^{-i \omega'_\mathrm{ext} t} - \mathbf{J}^* e^{i \omega'_\mathrm{ext} t},\
         \mathbf{0})^\mathrm{T}
         .
    \end{align}
\end{subequations}
The homogeneous equation takes the form $i\dot{\mathbf{x}}(t) = \bm{\mathcal{M}}\mathbf{x}(t)$ and its general solution
is $\mathbf{x}(t) = \sum_n c_n e^{-i\Omega_n t}\mathbf{u}_n$, where $\mathbf{u}_n$ are the eigenvectors of matrix $\bm{\mathcal{M}}$,
$\Omega_n$ their corresponding eigenvalues and $c_n$ constant coefficients.
The solution of the homogeneous equation thus boils down to finding the eigenvectors and eigenvalues of the matrix
$\bm{\mathcal{M}}$ defined in Eq.~\eqref{eq:MinMatDiffEq}.
Since two matrices related by a similarity transformation have the same set of eigenvalues,
one can alternatively solve the eigenproblem for its similar matrix
$\bm{\mathcal{M}}' = \mathbf{S}\bm{\mathcal{M}}\mathbf{S}^{-1}$
with eigenvectors $\mathbf{u}' = \mathbf{S}\mathbf{u}$,
if the new matrix has some more favorable properties.
For example, matrix $\bm{\mathcal{M}}$ defined in Eq.~\eqref{eq:MinMatDiffEq} can be transformed
into a diagonally dominant matrix by
\begin{equation}
\label{eq:SimTrans1}
    \begin{pmatrix}
    \mathbf{A}    + \mathbf{\Delta}            & \mathbf{B} + \mathbf{\Delta}'      & - \mathbf{L}  & - \mathbf{L}   \\
    -\mathbf{B}^* - \mathbf{\Delta}^{\prime *} & - \mathbf{A}^* - \mathbf{\Delta}^* & \mathbf{L}^*  &   \mathbf{L}^* \\
    -\mathbf{Q}^*             & -\mathbf{Q}       & \bm{\omega}  & \mathbf{0} \\
    \mathbf{Q}^*              & \mathbf{Q}        & \mathbf{0} & -\bm{\omega} \\
    \end{pmatrix}
    =
    \begin{pmatrix}
    \mathbf{1} & \mathbf{0} & \mathbf{0} & \mathbf{0}  \\
    \mathbf{0} & \mathbf{1} & \mathbf{0} & \mathbf{0}  \\
    \mathbf{0} & \mathbf{0} & -\bm{\frac{1}{2\omega}} & \mathbf{\frac{1}{2}}  \\
    \mathbf{0} & \mathbf{0} &  \bm{\frac{1}{2\omega}} & \mathbf{\frac{1}{2}}  \\
    \end{pmatrix}
    \begin{pmatrix}
    \mathbf{A}    + \mathbf{\Delta}            & \mathbf{B} + \mathbf{\Delta}'      & \mathbf{0}  & - \mathbf{L}   \\
    -\mathbf{B}^* - \mathbf{\Delta}^{\prime *} & - \mathbf{A}^* - \mathbf{\Delta}^* & \mathbf{0}  &   \mathbf{L}^* \\
    \mathbf{Q}^{\prime *}                      & \mathbf{Q}'                        & \mathbf{0}  & - \bm{\omega}^2 \\
    \mathbf{0}                                 & \mathbf{0}                         & \mathbf{-1} & \mathbf{0} \\
    \end{pmatrix}
    \begin{pmatrix}
    \mathbf{1} & \mathbf{0} & \mathbf{0} & \mathbf{0}  \\
    \mathbf{0} & \mathbf{1} & \mathbf{0} & \mathbf{0}  \\
    \mathbf{0} & \mathbf{0} & -\bm{\omega} & \bm{\omega}  \\
    \mathbf{0} & \mathbf{0} & \mathbf{1} & \mathbf{1}  \\
    \end{pmatrix}
    ,
\end{equation}
with the same definition of $\mathbf{Q}$ as above and matrices $\pm\bm{\frac{1}{2\omega}}$, $\mathbf{\frac{1}{2}}$,
and $\pm\bm{\omega}$ corresponding to diagonal matrices with $\pm\frac{1}{2\omega_\alpha}$, $\frac{1}{2}$, and $\pm\omega_\alpha$,
respectively, on the diagonal.
This transformation recovers the matrix in Eq.~\eqref{eq:CavityCasida} (after multiplication with the
matrix $\mathrm{diag}(\mathbf{1},-\mathbf{1},\mathbf{1},-\mathbf{1})$ on the right-hand side).
Moreover, the similarity transformation also shows a relationship between
the new photonic variables $M_\alpha$ and $N_\alpha$, and the old
$\mathbf{p}^{(1)}(t)$ and $\mathbf{q}^{(1)}(t)$ in the form
\begin{subequations}
\begin{align}
    M_\alpha
    & =
    -\frac{1}{2\omega_\alpha} p^{(1)}_\alpha + \frac{1}{2} q^{(1)}_\alpha \\
    N_\alpha
    & =
    \frac{1}{2\omega_\alpha} p^{(1)}_\alpha + \frac{1}{2} q^{(1)}_\alpha
    ,
\end{align}
\end{subequations}
which is analogous but not yet identical to the definition of creation and annihilation operators in Eqs.~\eqref{eq:CreationOperator}
and \eqref{eq:AnnihilationOperator}, suggesting an interpretation of $M_\alpha$ as a creation and $N_\alpha$ as an annihilation amplitude, respectively.

Furthermore, we can perform yet another similarity transformation to symmetrize the matrix $\bm{\mathcal{M}}'$
defined in Eq.~\eqref{eq:SimTrans1},
\begin{equation}
\label{eq:SimTrans2}
    \begin{pmatrix}
    \mathbf{A}    + \mathbf{\Delta}            & \mathbf{B} + \mathbf{\Delta}'      & - \mathbf{g}  & - \mathbf{g}   \\
    -\mathbf{B}^* - \mathbf{\Delta}^{\prime *} & - \mathbf{A}^* - \mathbf{\Delta}^* & \mathbf{g}^*  &  \mathbf{g}^* \\
    - \mathbf{g}^*             & -\mathbf{g}       & \bm{\omega}  & \mathbf{0} \\
    \mathbf{g}^*              & \mathbf{g}        & \mathbf{0} & -\bm{\omega} \\
    \end{pmatrix}
     =
    \begin{pmatrix}
    \mathbf{1} & \mathbf{0} & \mathbf{0} & \mathbf{0}  \\
    \mathbf{0} & \mathbf{1} & \mathbf{0} & \mathbf{0}  \\
    \mathbf{0} & \mathbf{0} & \bm{\sqrt{2\omega}} & \mathbf{0}  \\
    \mathbf{0} & \mathbf{0} & \mathbf{0} & \bm{\sqrt{2\omega}}  \\
    \end{pmatrix}
    \begin{pmatrix}
    \mathbf{A}    + \mathbf{\Delta}            & \mathbf{B} + \mathbf{\Delta}'      & - \mathbf{L}  & - \mathbf{L}   \\
    -\mathbf{B}^* - \mathbf{\Delta}^{\prime *} & - \mathbf{A}^* - \mathbf{\Delta}^* & \mathbf{L}^*  &   \mathbf{L}^* \\
    -\mathbf{Q}^*             & -\mathbf{Q}       & \bm{\omega}  & \mathbf{0} \\
    \mathbf{Q}^*              & \mathbf{Q}        & \mathbf{0} & -\bm{\omega} \\
    \end{pmatrix}
    \begin{pmatrix}
    \mathbf{1} & \mathbf{0} & \mathbf{0} & \mathbf{0}  \\
    \mathbf{0} & \mathbf{1} & \mathbf{0} & \mathbf{0}  \\
    \mathbf{0} & \mathbf{0} & \bm{\frac{1}{\sqrt{2\omega}}} & \mathbf{0}  \\
    \mathbf{0} & \mathbf{0} & \mathbf{0} & \bm{\frac{1}{\sqrt{2\omega}}} \\
    \end{pmatrix}
    ,
\end{equation}
where the photon--electron and the electron--photon coupling has the form
\begin{equation}
    g_{ai,\alpha} = \sqrt{\frac{\hbar\omega_\alpha}{2}}\, g_\alpha \, \vec{\mu}_{ai} \cdot \vec{\epsilon}_\alpha
    ,
\end{equation}
and matrices $\bm{\sqrt{2\omega}}$ and $\bm{\frac{1}{\sqrt{2\omega}}}$ are diagonal matrices with
$\sqrt{2\omega_\alpha}$ and $\frac{1}{\sqrt{2\omega_\alpha}}$, respectively, on the diagonal.
Symmetrization in Eq.~\eqref{eq:SimTrans2} recovered the matrix appearing in the Casida equation solved in some other works
on linear response QEDFT.~\cite{Yang2021, Liebenthal2023}
Furthermore, the symmetrizing similarity transform gives
\begin{subequations}
\begin{align}
    M_\alpha
    & =
    \sqrt{\frac{\omega_\alpha}{2}} \left( - \frac{1}{\omega_\alpha} p^{(1)}_\alpha + q^{(1)}_\alpha \right) \\
    N_\alpha
    & =
    \sqrt{\frac{\omega_\alpha}{2}} \left( \frac{1}{\omega_\alpha} p^{(1)}_\alpha + q^{(1)}_\alpha \right)
    ,
\end{align}
\end{subequations}
which exactly corresponds to the definition of the creation and annihilation operators in
Eqs.~\eqref{eq:CreationOperator} and \eqref{eq:AnnihilationOperator}
giving $M_\alpha$ and $N_\alpha$ the meaning of creation and annihilation amplitudes, respectively.

\section{Derivation of the expression for polarizability using left and right eigenvectors}
\label{sec:SpectrumDetails}

To compactify the following discussion, let us adopt the notation common in quantum chemistry where the Sternheimer equation
\eqref{eq:CavitySternheimer} is written term-by-term as $[\mathbf{E}^{[2]} - \omega_\mathrm{ext} \mathbf{S}^{[2]}] \mathbf{Z} = \mathbf{G}$.
and Eq.~\eqref{eq:alphaFromSternheimer} takes the form
$\bm{\alpha}(\omega_\mathrm{ext}) = -\mathbf{G}^\dagger \left[ \mathbf{E}^{[2]}-\omega_\mathrm{ext} \mathbf{S}^{[2]} \right]^{-1} \mathbf{G}$.
The right and left eigenvectors are defined as
\begin{subequations}
\label{eq:LREV}
\begin{align}
    \mathbf{S}^{[2]} \mathbf{E}^{[2]} \mathbf{Z}^\mathrm{R} & = \mathbf{Z}^\mathrm{R} \mathbf{\Omega}, \\
    \mathbf{Z}^\mathrm{L\dagger} \mathbf{S}^{[2]} \mathbf{E}^{[2]} & = \mathbf{\Omega} \mathbf{Z}^\mathrm{L\dagger},
\end{align}
\end{subequations}
where matrix $\mathbf{\Omega}$ is a diagonal matrix containing all the eigenvalues and the columns of matrices $\mathbf{Z}^\mathrm{R}$
and $\mathbf{Z}^\mathrm{L}$ contain all the right and left eigenvectors, respectively.
Moreover, we made use of the fact that $\mathbf{S}^{[2],-1} = \mathbf{S}^{[2]}$.
Left and right eigenvectors corresponding to different eigenvalues are orthogonal,
and can be transformed to be so for degenerate subspaces,
meaning that their matrices satisfy $\mathbf{Z}^\mathrm{L\dagger} \mathbf{Z}^\mathrm{R} = \mathbf{1}$.
Let us note that the definition of $\mathbf{Z}^\mathrm{L\dagger}$ as a left eigenvector of matrix $\mathbf{S}^{[2]} \mathbf{E}^{[2]}$
is due to the computer implementation that solves the eigenproblem for matrix $\mathbf{S}^{[2]} \mathbf{E}^{[2]}$ instead of the
generalized eigenproblem. The alternative formulation of left eigenvectors stemming from the generalized eigenproblem
$\tilde{\mathbf{Z}}^\mathrm{L\dagger} \mathbf{E}^{[2]} = \mathbf{\Omega} \tilde{\mathbf{Z}}^\mathrm{L\dagger} \mathbf{S}^{[2]}$
is related to ours via $\tilde{\mathbf{Z}}^\mathrm{L\dagger} \mathbf{S}^{[2]} = \mathbf{Z}^\mathrm{L\dagger}$
and the eigenvectors satisfy the condition $\tilde{\mathbf{Z}}^\mathrm{L\dagger} \mathbf{S}^{[2]} \mathbf{Z}^\mathrm{R} = \mathbf{1}$.
From Eqs.~\eqref{eq:LREV} it also follows that
$\mathbf{Z}^\mathrm{L\dagger} \mathbf{E}^{[2]} \mathbf{Z}^\mathrm{R} = \mathbf{Z}^\mathrm{L\dagger} \mathbf{S}^{[2]} \mathbf{Z}^\mathrm{R} \mathbf{\Omega}$.

The expression for the polarizability can be worked out as
\begin{align}
\begin{split}
   \bm{\alpha}(\omega) & = -\mathbf{G}^\dagger \left[ \mathbf{E}^{[2]}-\omega \mathbf{S}^{[2]} \right]^{-1} \mathbf{G} \\
          & = -\mathbf{G}^\dagger \mathbf{Z}^\mathrm{R} \mathbf{Z}^\mathrm{L\dagger} \left[ \mathbf{E}^{[2]}-\omega \mathbf{S}^{[2]} \right]^{-1} \mathbf{Z}^\mathrm{R} \mathbf{Z}^\mathrm{L\dagger} \mathbf{G} \\
          & = -\left( \mathbf{G}^\dagger \mathbf{Z}^\mathrm{R} \right)
              \left( \mathbf{Z}^\mathrm{L\dagger} \left[ \mathbf{E}^{[2]}-\omega \mathbf{S}^{[2]} \right]^{-1} \mathbf{Z}^\mathrm{R} \right)
              \left( \mathbf{Z}^\mathrm{L\dagger} \mathbf{G} \right) \\
          & = -\left( \mathbf{G}^\dagger \mathbf{Z}^\mathrm{R} \right)
              \left( \mathbf{Z}^\mathrm{R,-1} \left[ \mathbf{E}^{[2]}-\omega \mathbf{S}^{[2]} \right] \mathbf{Z}^\mathrm{L\dagger,-1} \right)^{-1}
              \left( \mathbf{Z}^\mathrm{L\dagger} \mathbf{G} \right) \\
          & = -\left( \mathbf{G}^\dagger \mathbf{Z}^\mathrm{R} \right)
              \left( \mathbf{Z}^\mathrm{L\dagger} \left[ \mathbf{E}^{[2]}-\omega \mathbf{S}^{[2]} \right] \mathbf{Z}^\mathrm{R} \right)^{-1}
              \left( \mathbf{Z}^\mathrm{L\dagger} \mathbf{G} \right) \\
          & = -\left( \mathbf{G}^\dagger \mathbf{Z}^\mathrm{R} \right)
              \left( \mathbf{Z}^\mathrm{L\dagger} \mathbf{E}^{[2]} \mathbf{Z}^\mathrm{R} - \omega \mathbf{Z}^\mathrm{L\dagger} \mathbf{S}^{[2]} \mathbf{Z}^\mathrm{R} \right)^{-1}
              \left( \mathbf{Z}^\mathrm{L\dagger} \mathbf{G} \right) \\
          & = -\left( \mathbf{G}^\dagger \mathbf{Z}^\mathrm{R} \right)
              \left(  \mathbf{Z}^\mathrm{L\dagger} \mathbf{S}^{[2]} \mathbf{Z}^\mathrm{R} \mathbf{\Omega} - \mathbf{Z}^\mathrm{L\dagger} \mathbf{S}^{[2]} \mathbf{Z}^\mathrm{R} \omega \right)^{-1}
              \left( \mathbf{Z}^\mathrm{L\dagger} \mathbf{G} \right) \\
          & = -\left( \mathbf{G}^\dagger \mathbf{Z}^\mathrm{R} \right)
              \left(  \mathbf{Z}^\mathrm{L\dagger} \mathbf{S}^{[2]} \mathbf{Z}^\mathrm{R} \left[ \mathbf{\Omega} - \omega \mathbf{1} \right] \right)^{-1}
              \left( \mathbf{Z}^\mathrm{L\dagger} \mathbf{G} \right) \\
          & = -\left( \mathbf{G}^\dagger \mathbf{Z}^\mathrm{R} \right)
              \left( \left[ \mathbf{\Omega} - \omega \mathbf{1} \right]^{-1} \mathbf{Z}^\mathrm{L\dagger} \mathbf{S}^{[2]} \mathbf{Z}^\mathrm{R} \right)
              \left( \mathbf{Z}^\mathrm{L\dagger} \mathbf{G} \right) \\
          & = -\left( \mathbf{G}^\dagger \mathbf{Z}^\mathrm{R} \right)
              \left[ \mathbf{\Omega} - \omega \mathbf{1} \right]^{-1}
              \left( \mathbf{Z}^\mathrm{L\dagger} \mathbf{S}^{[2]} \mathbf{G} \right), \\
\end{split}
\end{align}
where we renamed $\omega_\mathrm{ext}$ to $\omega$ both for clarity as well as to indicate its meaning as the frequency
at which the frequency-dependent polarizability is evaluated.
The matrix $\mathbf{\Omega} - \omega \mathbf{1}$ is diagonal and can be easily inverted.
However, the formula can be further simplified by taking into account the fact that the eigenvalues
come in positive--negative pairs, i.e. the matrix $\mathbf{\Omega}$ has the structure
\begin{equation}
    \mathbf{\Omega}
    =
    \begin{pmatrix}
    \mathbf{\Omega}^{+} & \mathbf{0} \\
    \mathbf{0}      & -\mathbf{\Omega}^{+}
    \end{pmatrix}
    ,
\end{equation}
where $\mathbf{\Omega}^{+}$ is a diagonal matrix of the positive eigenvalues labeled from 1 to $n$.
Therefore, the expression for the polarizability tensor becomes
\begin{equation}
\label{eq:alphaAlmostThere}
  \bm{\alpha}(\omega)
  =
  -
  \sum_n
  \left[
  \frac{t^\mathrm{R}_n t^\mathrm{L*}_n}{\Omega_n-\omega}
  -
  \frac{t^\mathrm{R}_{-n} t^\mathrm{L*}_{-n}}{-\Omega_n-\omega}
  \right]
  ,
\end{equation}
where variables labeled by $n$ and $-n$ belong to excitation energies $\Omega_n$ and $-\Omega_n$, respectively, and the right and left transition dipole moments are defined as
\begin{subequations}
\label{eq:RLtransDipMomentsApp}
\begin{alignat}{2}
    t^\mathrm{R}_n
    & =
    \mathbf{G}^\dagger \mathbf{Z}^\mathrm{R}
    && =
    \mathbf{P}^\dagger \mathbf{X}^\mathrm{R}_n + \mathbf{P} \mathbf{Y}^\mathrm{R}_n , \\
    t^\mathrm{L}_n
    & =
    \mathbf{Z}^\mathrm{L\dagger} \mathbf{S}^{[2]} \mathbf{G}
    && =
    \mathbf{P}^\dagger \mathbf{X}^\mathrm{R}_n - \mathbf{P} \mathbf{Y}^\mathrm{R}_n .
\end{alignat}
\end{subequations}
The fact that only the electronic parts $\mathbf{X}^\mathrm{R/L}_n$ and $\mathbf{Y}^\mathrm{R/L}_n$ of the transition vector
$\mathbf{Z}^\mathrm{R/L}_n = (\mathbf{X}^\mathrm{R/L}_n,\ \mathbf{Y}^\mathrm{R/L}_n,\ \mathbf{M}^\mathrm{R/L}_n,\ \mathbf{N}^\mathrm{R/L}_n)^\mathrm{T}$
are used to calculate the transition dipole moment in Eqs.~\eqref{eq:RLtransDipMoments} is the consequence of the fact that the electric dipole moment
is an electron-only property described by $\mathbf{G} = (\mathbf{P},\ \mathbf{P}^*,\, \mathbf{0},\ \mathbf{0})^\mathrm{T}$.
The eigenvector corresponding to an eigenvalue with the opposite sign has
$\mathbf{X}^\mathrm{R/L}_{-n} = \mathbf{Y}^\mathrm{R/L *}_n$
and
$\mathbf{Y}^\mathrm{R/L}_{-n} = \mathbf{X}^\mathrm{R/L *}_n$,
therefore
$t^\mathrm{R}_{-n} = t^\mathrm{R *}_n$
and
$t^\mathrm{L}_{-n} = t^\mathrm{L *}_n$.
After inserting these expressions into Eq.~\eqref{eq:alphaAlmostThere}, we obtain the final expression for the
frequency-dependent polarizability calculated from the distinct right and left eigenvectors
\begin{equation}
  \bm{\alpha}(\omega)
  =
  \sum_n
  \left[
  \frac{t^\mathrm{R*}_{n} t^\mathrm{L}_{n}}{\Omega_n+\omega}
  -
  \frac{t^\mathrm{R}_n t^\mathrm{L*}_n}{\omega-\Omega_n}
  \right]
  .
\end{equation}

\end{appendices}

\section*{Supplemental Material}

Additional spectra, JC model definition, geometry of the Hg@porphyrin.

\section*{ACKNOWLEDGMENTS}

LK acknowledges support by the Research Council of Norway through its Centres of Excellence scheme,
project no.~262695, and its Mobility Grant scheme, project no.~314814.
LK would like to thank Stanislav Komorovsky, Johannes Flick, and Davis Welakuh for helpful discussions.
VK would like to thank Simone Latini for the help in the preparation figure of the cavity.
The calculations were performed on resources provided by Sigma2—the National Infrastructure for
High Performance Computing and Data Storage in Norway, Grant Nos. NN4654K and NN14654K. 



%
\bibliographystyle{unsrt}
\bibliography{references}

\end{document}


\title{Relativistic Linear Response in Quantum-Electrodynamical Density Functional Theory\\
\normalsize{Supplemental Material}}
\author{Lukas Konecny}
\email{lukas.konecny@uit.no} 
\affiliation{Hylleraas Centre for Quantum Molecular Sciences, Department of Chemistry, UiT The Arctic University of Norway, N-9037 Troms{\o}, Norway}
\affiliation{Max Planck Institute for the Structure and Dynamics of Matter, Center for Free Electron Laser Science, Luruper Chaussee 149, 22761 Hamburg, Germany}
\author{Valeriia P. Kosheleva}
\email{valeriia.kosheleva@mpsd.mpg.de}
\affiliation{Max Planck Institute for the Structure and Dynamics of Matter, Center for Free Electron Laser Science, Luruper Chaussee 149, 22761 Hamburg, Germany}
\author{Heiko Appel}
\email{heiko.appel@mpsd.mpg.de}
\affiliation{Max Planck Institute for the Structure and Dynamics of Matter, Center for Free Electron Laser Science, Luruper Chaussee 149, 22761 Hamburg, Germany}
\author{Michael Ruggenthaler}
\email{michael.ruggenthaler@mpsd.mpg.de}
\affiliation{Max Planck Institute for the Structure and Dynamics of Matter, Center for Free Electron Laser Science, Luruper Chaussee 149, 22761 Hamburg, Germany}
\author{Angel Rubio}
\email{angel.rubio@mpsd.mpg.de}
\affiliation{Max Planck Institute for the Structure and Dynamics of Matter, Center for Free Electron Laser Science, Luruper Chaussee 149, 22761 Hamburg, Germany}
\affiliation{Center for Computational Quantum Physics (CCQ), The Flatiron Institute, 162 Fifth Avenue, New York, New York 10010, USA}


\maketitle

\section{Excitation energies without cavity}

\begin{table}[h]
    \centering
    \caption{Excitation energies ($E$ in au and eV) and oscillator strengths ($f$, dimensionless)
    of Group 12 atoms (Dyall's DZ basis, B3LYP functional)
    for the allowed \ce{^1S0} $\rightarrow$ \ce{^3P1} (S--T)
    and \ce{^1S0} $\rightarrow$ \ce{^1P1} (S--S) transitions
    compared to experimental data.}
    \begin{tabular}{cccccccccc}
      Atom & From      & To        & &  $E\ [\unit{au}]$ &  $E\ [\unit{eV}]$ & $f$          & & $E_\mathrm{exp}\ [\unit{eV}]$~\textsuperscript{\emph{a}} & $f_\mathrm{exp}$~\textsuperscript{\emph{a}} \\
      \hline
      Zn   & \ce{^1S0} & \ce{^3P0} & &  0.1463           &  3.9815      &     0.0           & &   4.01     & 0.0           \\
           & \ce{^1S0} & \ce{^3P1} & &  0.1472           &  4.0060      &     0.0000688     & &   4.03     & 0.000053      \\
           & \ce{^1S0} & \ce{^3P2} & &  0.1491           &  4.0562      &     0.0           & &   4.08     & 0.0           \\[0.2cm]
           & \ce{^1S0} & \ce{^1P1} & &  0.2101           &  5.7163      &     0.493         & &   5.80     & 0.4895        \\[0.3cm]
      Cd   & \ce{^1S0} & \ce{^3P0} & &  0.1345           &  3.6594      &     0.0           & &   3.73     & 0.0           \\
           & \ce{^1S0} & \ce{^3P1} & &  0.1370           &  3.7282      &     0.000792      & &   3.80     &  ---          \\
           & \ce{^1S0} & \ce{^3P2} & &  0.1425           &  3.8772      &     0.0           & &   3.95     & 0.0           \\[0.2cm]
           & \ce{^1S0} & \ce{^1P1} & &  0.1937           &  5.2713      &     0.493         & &   5.42     & 0.4159        \\[0.3cm]
      Hg   & \ce{^1S0} & \ce{^3P0} & &  0.1652           &  4.4944      &     0.0           & &   4.67     & 0.0           \\
           & \ce{^1S0} & \ce{^3P1} & &  0.1731           &  4.7089      &     0.00974       & &   4.89     & 0.0077        \\
           & \ce{^1S0} & \ce{^3P2} & &  0.1943           &  5.2877      &     0.0           & &   5.46     & 0.0           \\[0.2cm]
           & \ce{^1S0} & \ce{^1P1} & &  0.2332           &  6.3457      &     0.376         & &   6.70     & 0.3825        \\
    \end{tabular}
    
    \vspace{5pt}
    \textsuperscript{\emph{a}}
      J. E. Sansonetti, W. C. Martin. Handbook of Basic Atomic Spectroscopic Data.
      \textit{J. Phys. Chem. Ref. Data}, 34, 1559–2259, 2005.
    \label{tab:G12AtomsFreeSupp}
\end{table}

\newpage

\section{Additional polaritonic spectra}

\begin{figure}[h]
\centering
\caption{Absorption spectra of Zn and Cd atoms in a cavity strongly coupled ($g_\alpha = \unit[0.01]{au}$) to photonic modes, non-relativistic one-component (1c) vs relativistic four-component (4c) calculations.}
  \begin{subfigure}{0.32\textwidth}
    \includegraphics[width=\textwidth]{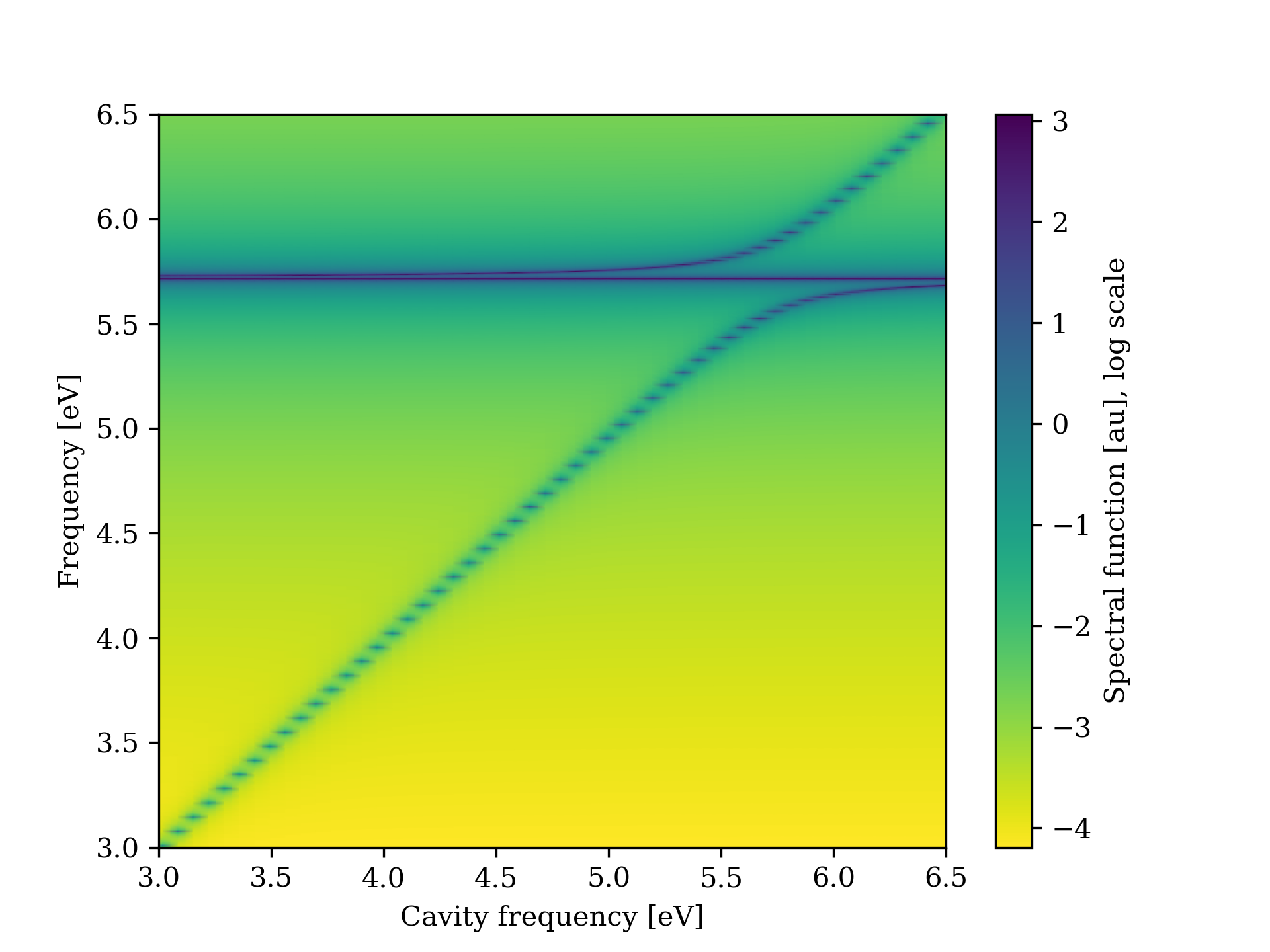}
    \caption{Zn 1c}
    \label{fig:Group12_1cVs4c:Zn1c}
  \end{subfigure}
  \begin{subfigure}{0.32\textwidth}
    \includegraphics[width=\textwidth]{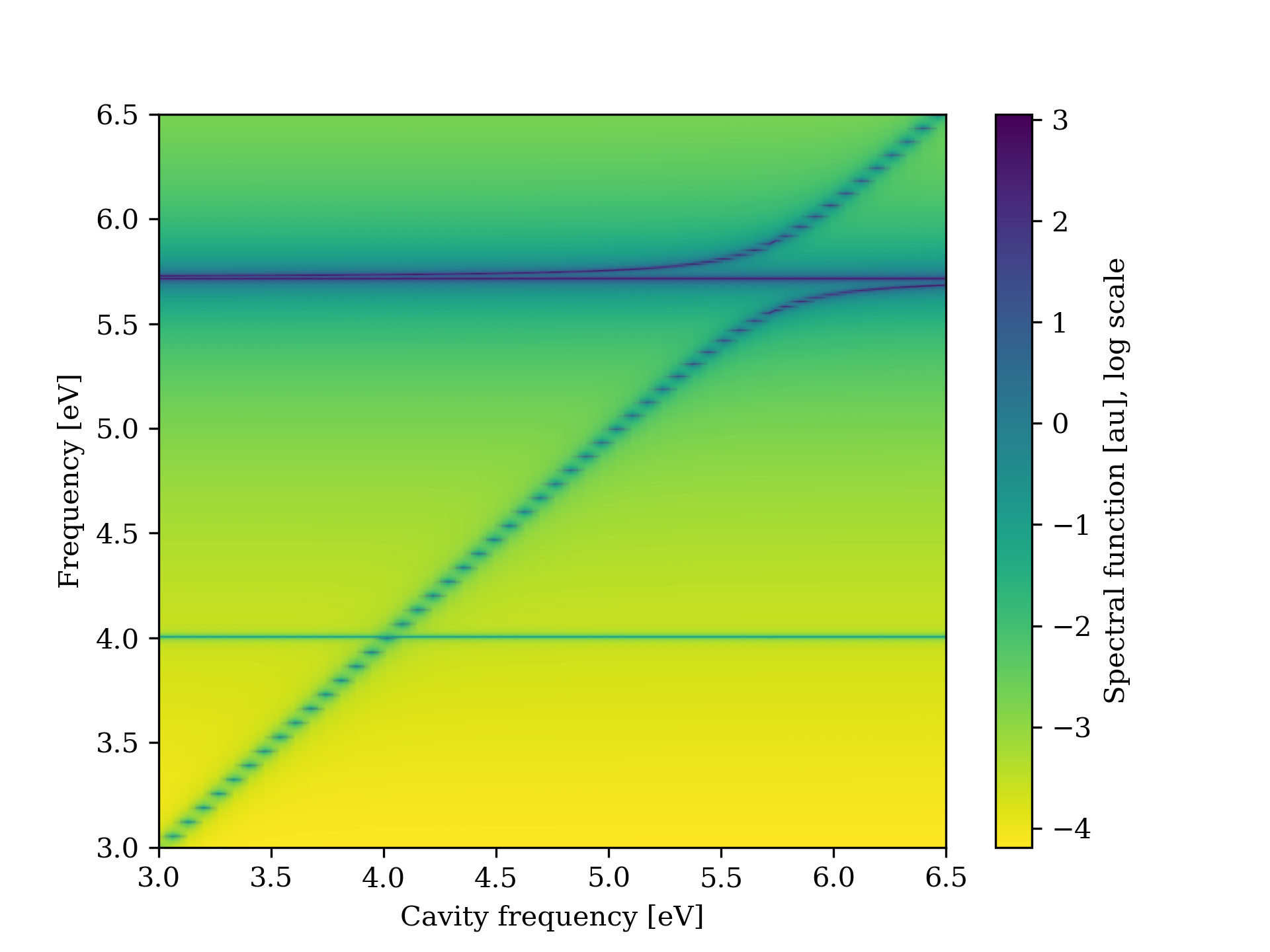}
    \caption{Zn 4c}
    \label{fig:Group12_1cVs4c:Zn4c}
  \end{subfigure}
  \begin{subfigure}{0.32\textwidth}
    \includegraphics[width=\textwidth]{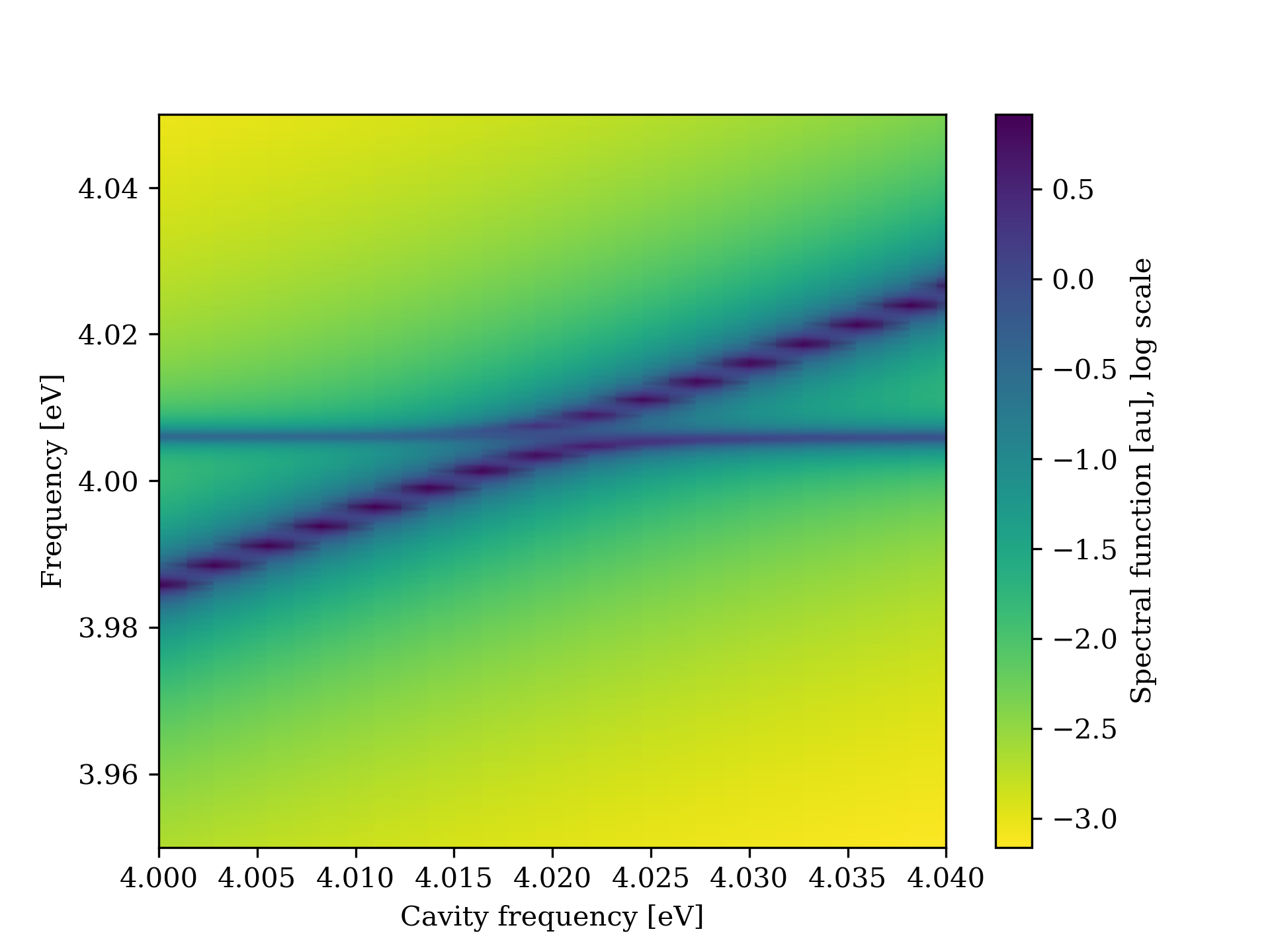}
    \caption{Zn 4c zoom}
    \label{fig:Group12_1cVs4c:Znzoom}
  \end{subfigure}
  \begin{subfigure}{0.32\textwidth}
    \includegraphics[width=\textwidth]{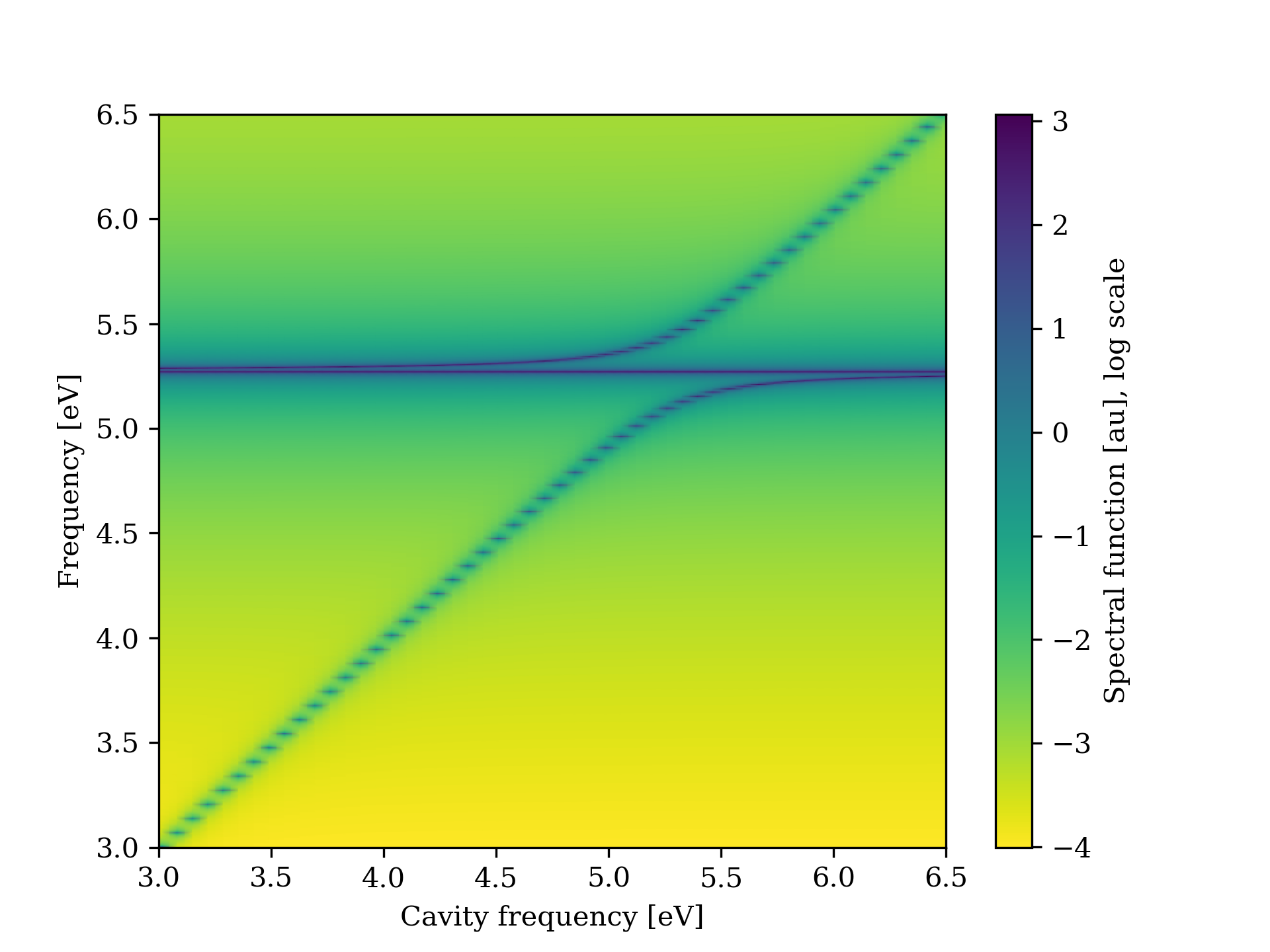}
    \caption{Cd 1c}
    \label{fig:Group12_1cVs4c:Cd1c}
  \end{subfigure}
  \begin{subfigure}{0.32\textwidth}
    \includegraphics[width=\textwidth]{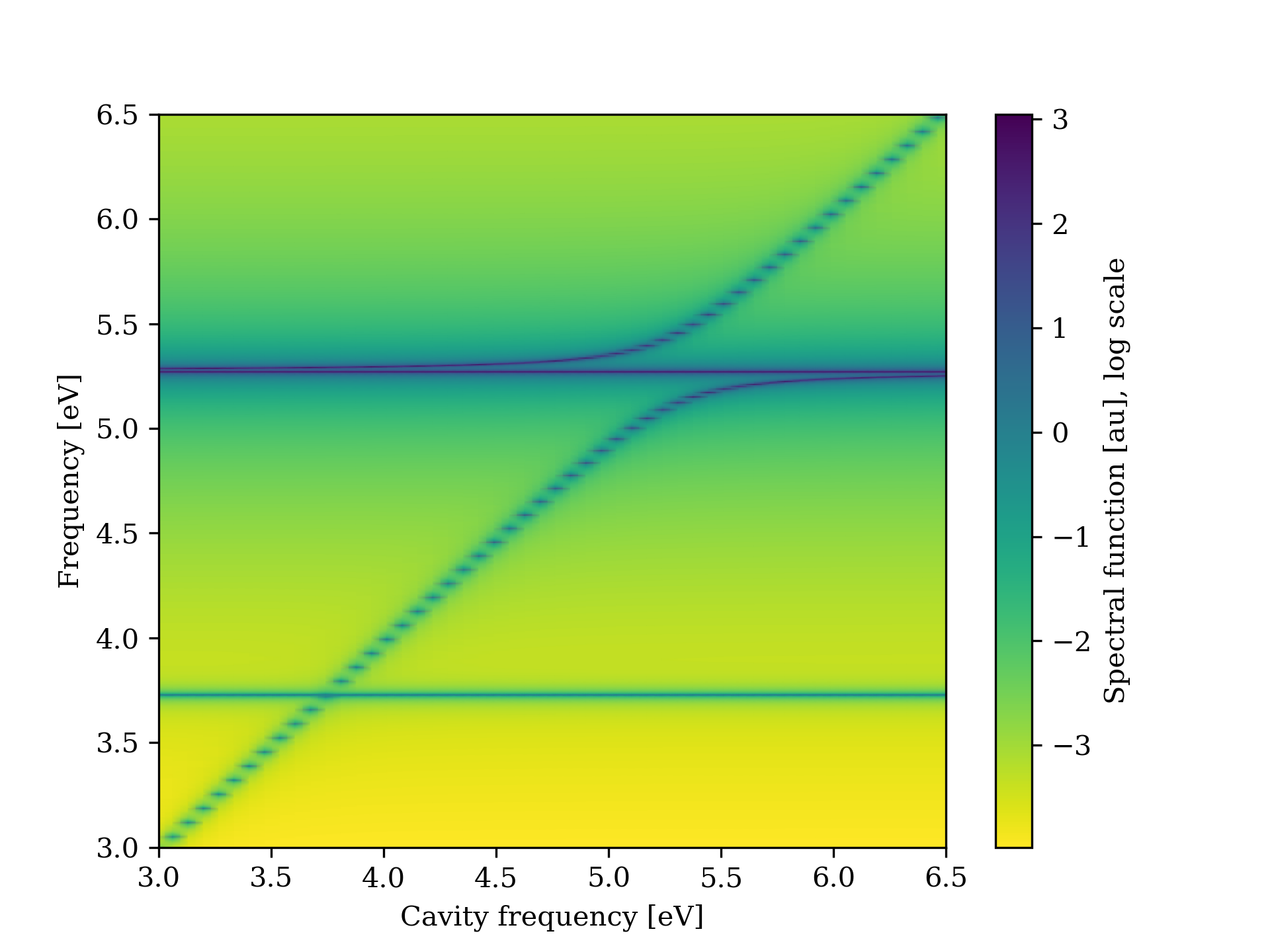}
    \caption{Cd 4c}
    \label{fig:Group12_1cVs4c:Cd4c}
  \end{subfigure}
  \begin{subfigure}{0.32\textwidth}
    \includegraphics[width=\textwidth]{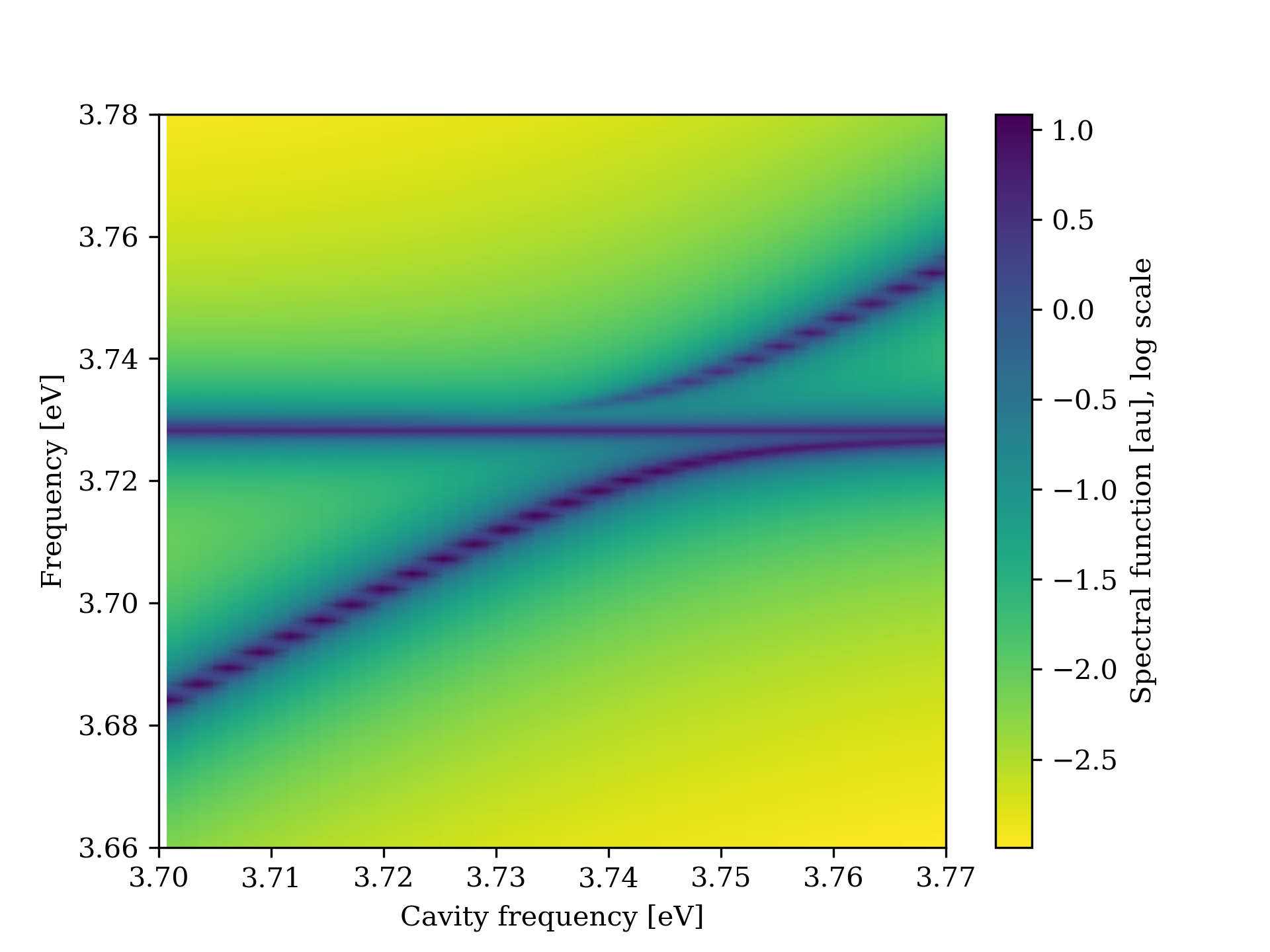}
    \caption{Cd 4c zoom}
    \label{fig:Group12_1cVs4c:Cdzoom}
  \end{subfigure}
  \label{fig:Group12_ZnCd}
\end{figure}

\begin{figure}[h]
 \centering
 \caption{Absorption spectra of Cd atom in a cavity strongly coupled ($g_\alpha = \unit[0.01]{au}$) to a cavity set to effective resonance with the singlet--triplet (S--T) transition defined by 50:50 light--matter mixing of the lower polariton rather than by
 the numerical value compared to reference spectra of free atoms without cavities. The region around the low-intensity S--T transition is magnified to ease reading by a factor specified in each figure.}
    \includegraphics[width=0.5\textwidth]{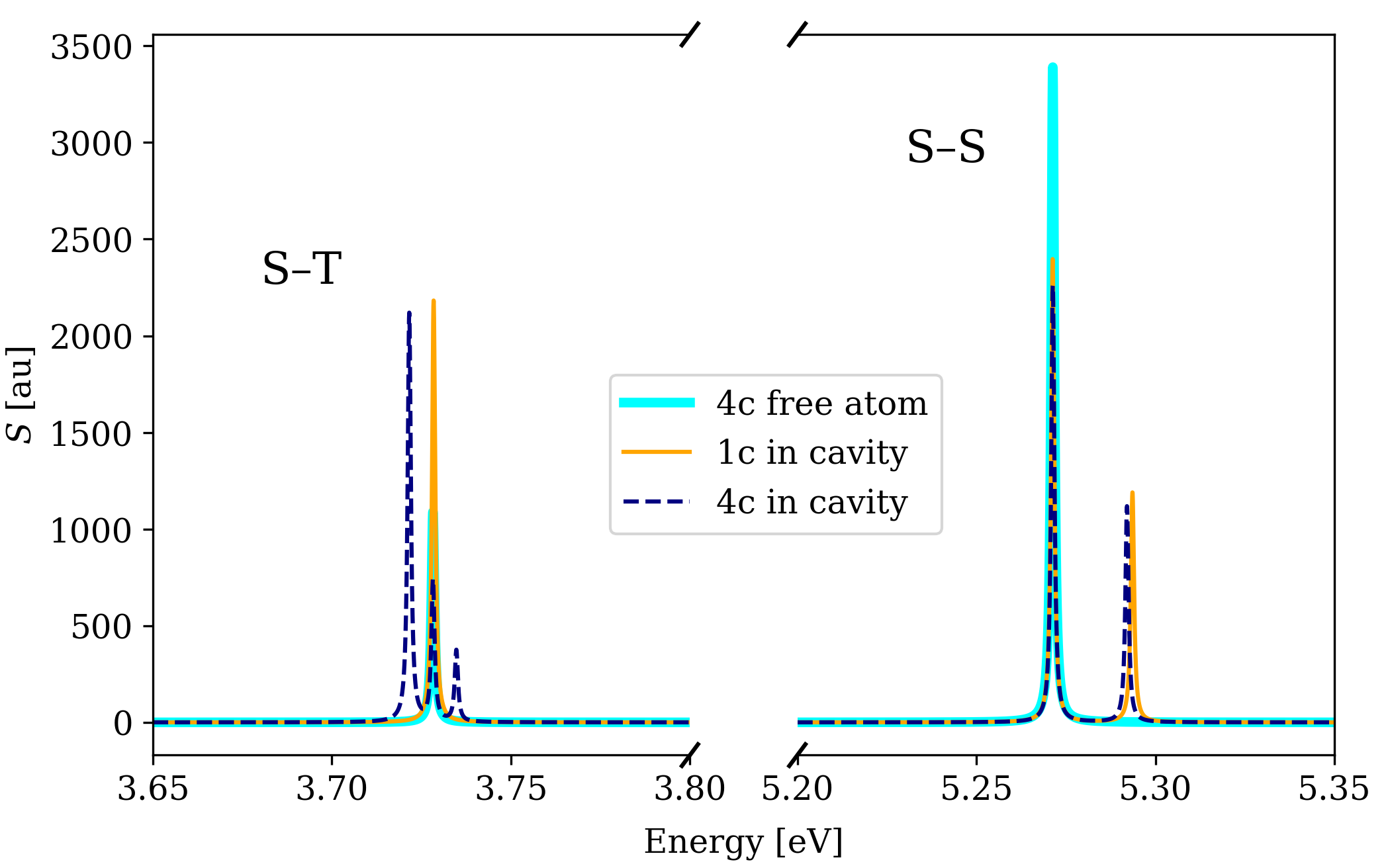}
  \label{fig:Group12_CdAllIn1}
\end{figure}

\newpage

\section{Additional results from the Jaynes--Cummings model}

The Jaynes--Cummings (JC) Hamiltonian describing a two-level system coupled to a single cavity mode is
\begin{equation}
H_\mathrm{JC}
=
\begin{bmatrix}
\hbar \omega_\mathrm{c} - \hbar \Delta & \hbar \Omega / 2        \\
\hbar \Omega / 2                       & \hbar \omega_\mathrm{c} \\
\end{bmatrix}
,
\end{equation}
where
\begin{align*}
\Delta & = \omega_\mathrm{c} - \omega_\mathrm{eg}, \\
\Omega & = 2 g_\mathrm{c}, \\
g_\mathrm{c} & = g \sqrt{\frac{\omega_\mathrm{c}}{2\hbar}} \left\langle e | \boldsymbol{\mu} | g \right\rangle \boldsymbol{\epsilon}_c,
\end{align*}
with $\omega_\mathrm{c}$ and $\boldsymbol{\epsilon}_c$ being the frequency and the polarization of the cavity mode, $g$ the coupling strength,
$\boldsymbol{\mu}$ the electric dipole moment operator, and $|\mathrm{g},1\rangle$ and $|\mathrm{e},1\rangle$ the ground and excited states of the two-level system describing the atom inside the cavity and $\omega_\mathrm{eg}$ the excitation frequency from the ground to the excited state.
Its eigenvalues, i.e. energies for the upper ($+$) and lower ($-$) polariton, are
\begin{equation}
E_{\pm} = (\hbar \omega_\mathrm{c} - \frac{1}{2}\hbar\Delta) \pm \frac{1}{2} \hbar \sqrt{\Delta^2 + \Omega^2} 
\end{equation}
The corresponding polaritonic wave functions, i.e. the eigenfuctions of the JC Hamiltonian are
\begin{align}
|\mathrm{UP}\rangle & = \cos\left(-\frac{\Omega}{2\Delta}\right) |\mathrm{e},0\rangle + \sin\left(-\frac{\Omega}{2\Delta}\right) |\mathrm{g},1\rangle, \\
|\mathrm{LP}\rangle & = \cos\left(-\frac{\Omega}{2\Delta}\right) |\mathrm{g},1\rangle - \sin\left(-\frac{\Omega}{2\Delta}\right) |\mathrm{e},0\rangle,
\end{align}
where UP and LP stand for the upper and lower polariton, respectively, and $|\mathrm{g/e},0/1\rangle$ are the tensor product basis states with atom
in the ground/excited state and the cavity mode populated with $0/1$ photons.
The transition dipole moments from the uncoupled ground state $|\mathrm{g},0\rangle$ to the polaritonic states are
\begin{alignat}{2}
\left\langle UP | \boldsymbol{\mu} | g,0 \right\rangle
& =
\cos\left(-\frac{\Omega}{2\Delta}\right) \left\langle e,0 | \boldsymbol{\mu} | g,0 \right\rangle
+ \sin\left(-\frac{\Omega}{2\Delta}\right)  \left\langle g,1 | \boldsymbol{\mu} | g,0 \right\rangle
&& =
\cos\left(-\frac{\Omega}{2\Delta}\right) \left\langle e | \boldsymbol{\mu} | g \right\rangle, \\
\left\langle LP | \boldsymbol{\mu} | g,0 \right\rangle
& =
 - \sin\left(-\frac{\Omega}{2\Delta}\right) \left\langle e,0 | \boldsymbol{\mu} | g,0 \right\rangle
+  \cos\left(-\frac{\Omega}{2\Delta}\right) \left\langle g,1 | \boldsymbol{\mu} | g,0 \right\rangle
&& =
- \sin\left(-\frac{\Omega}{2\Delta}\right) \left\langle e | \boldsymbol{\mu} | g \right\rangle.
\end{alignat}

The absorption spectrum (i.e. spectral function $S_{\pm}(\omega)$) from the two polaritonic states is then calculated as
\begin{equation}
S_{\pm}(\omega)
=
\frac{4\pi\omega}{3 c} \Im
  \left[
  \frac{|\left\langle UP | \boldsymbol{\mu} | g,0 \right\rangle|^2}{E_{+}+\omega+i\gamma}
  -
  \frac{|\left\langle UP | \boldsymbol{\mu} | g,0 \right\rangle|^2}{\omega-E_{+}+i\gamma}
  +
  \frac{|\left\langle LP | \boldsymbol{\mu} | g,0 \right\rangle|^2}{E_{-}+\omega+i\gamma}
  -
  \frac{|\left\langle LP | \boldsymbol{\mu} | g,0 \right\rangle|^2}{\omega-E_{-}+i\gamma}
  \right]
.
\end{equation}
In addition, we may add the signal resulting from the two triplet states not coupled to the cavity
(due to their transition dipole moment being in perpendicular directions to the cavity mode polarization)
in order to obtain a final spectrum that is similar to the QEDFT resutlt:
\begin{equation}
S(\omega)
=
S_{\pm}(\omega)
+
2 \frac{4\pi\omega}{3 c} \Im
  \left[
  \frac{|\left\langle e | \boldsymbol{\mu} | g \right\rangle|^2}{\omega_\mathrm{eg}+\omega+i\gamma}
  -
  \frac{|\left\langle e | \boldsymbol{\mu} | g \right\rangle|^2}{\omega-\omega_\mathrm{eg}+i\gamma}
  \right]
,
\end{equation}
where the additional factor 2 is for the double degeneracy of the uncoupled states.

When building the JC model for group 12 atoms, we use parametrization based on the results of TDDFT calculations
(see section S1 and Table~\ref{tab:G12AtomsFreeSupp}) with settings mirroring those used in QEDFT calculations.
The particular values of JC parameters are summarized in Table~\ref{tab:JCparam}.
\begin{table}[h]
    \centering
    \caption{Values of JC model parameters obtained from TDDFT calculations.}
    \begin{tabular}{ccc}
      Atom & $\omega_\mathrm{eg}$ & $\left\langle e | \boldsymbol{\mu} | g \right\rangle$  \\
      \hline
      Zn   & 0.14721961           & 0.29059     \\
      Hg   & 0.17305060           & 0.026481     \\
    \end{tabular}
    \label{tab:JCparam}
\end{table}

\begin{figure}[h]
\centering
\caption{Absorption spectra of Hg atom strongly coupled ($g = \unit[0.01]{au}$) to a cavity vs the
predictions of the Jaynes--Cummings model (JC) parametrized by TDDFT results.}
  \begin{subfigure}{0.48\textwidth}
    \includegraphics[width=\textwidth]{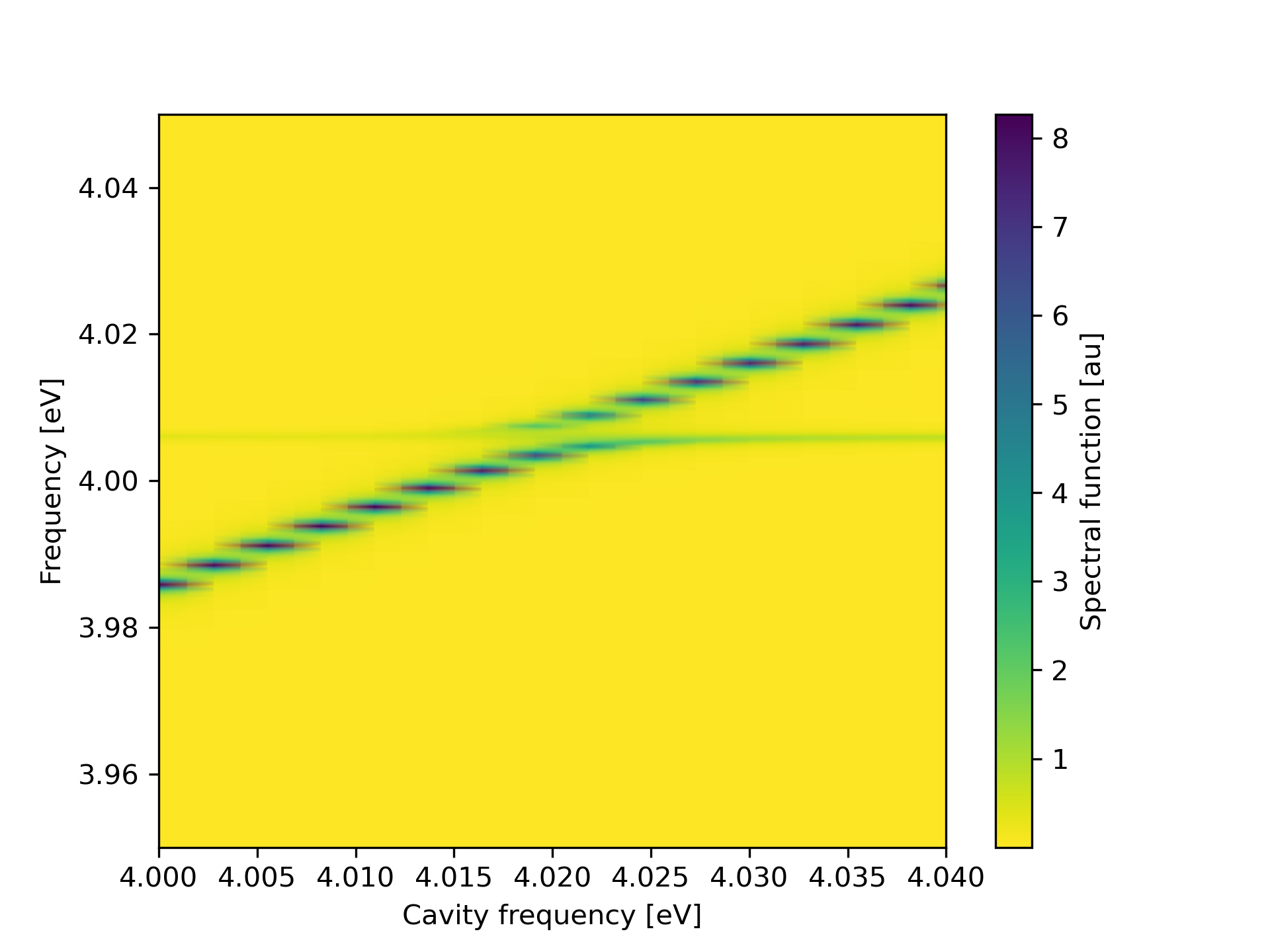}
    \caption{Zn QEDFT}
    \label{fig:Zn_2D_STzoom_abs}
  \end{subfigure}
  \begin{subfigure}{0.48\textwidth}
    \includegraphics[width=\textwidth]{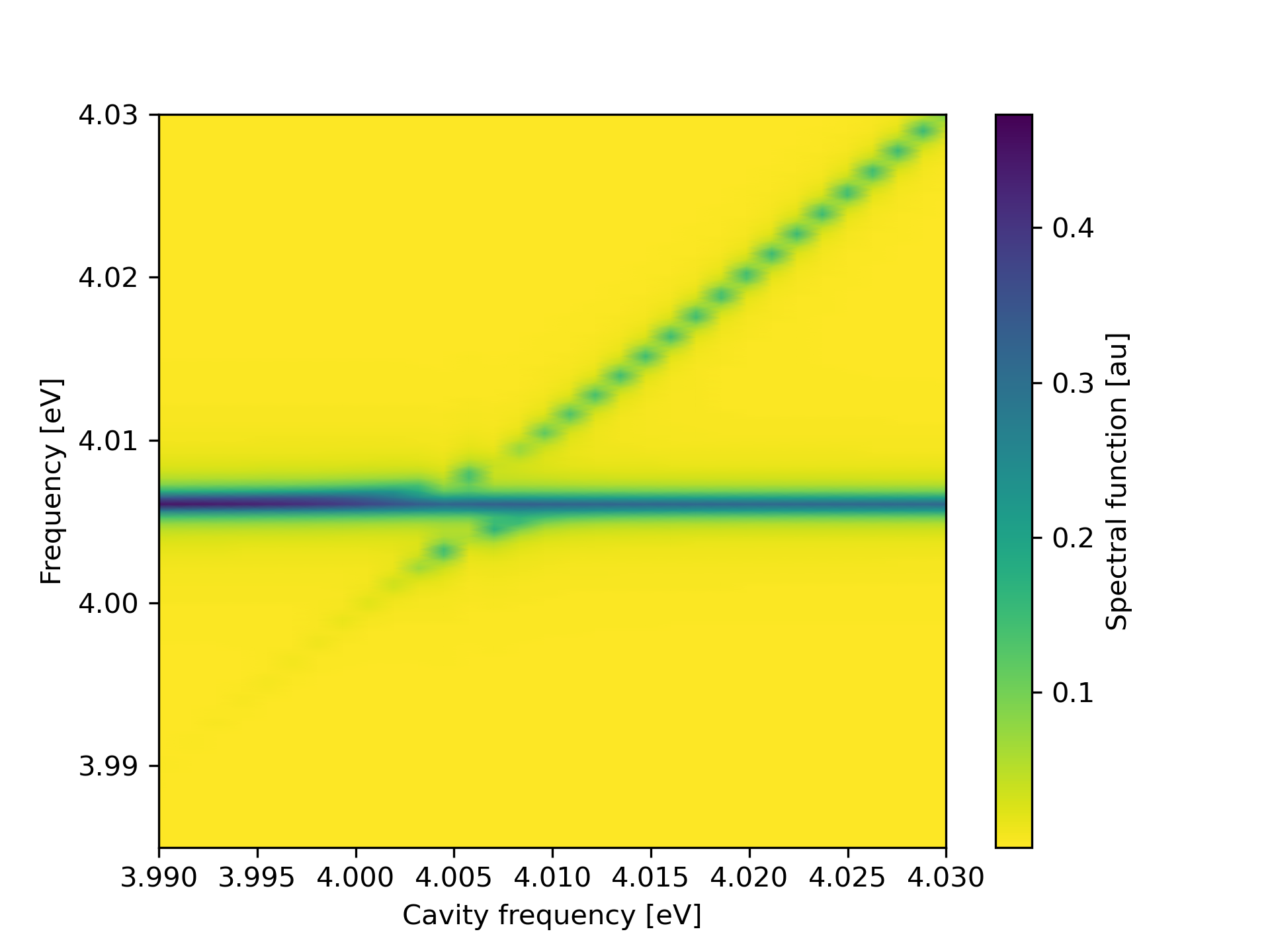}
    \caption{Zn JC}
    \label{fig:JC_Zn_param_2Dspect}
  \end{subfigure}
  \begin{subfigure}{0.48\textwidth}
    \includegraphics[width=\textwidth]{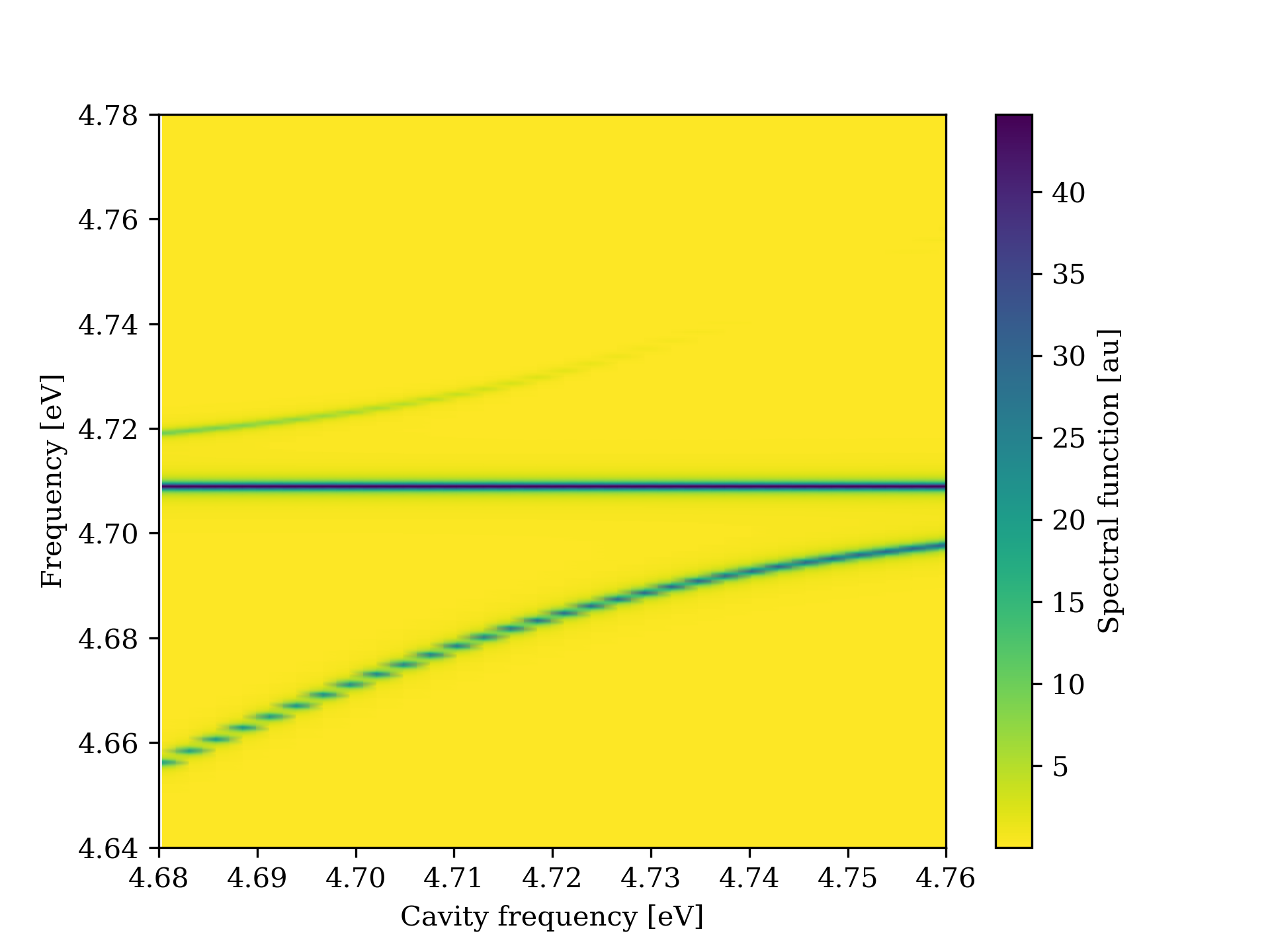}
    \caption{Hg QEDFT}
    \label{fig:Hg_2D_STzoom_abs}
  \end{subfigure}
  \begin{subfigure}{0.48\textwidth}
    \includegraphics[width=\textwidth]{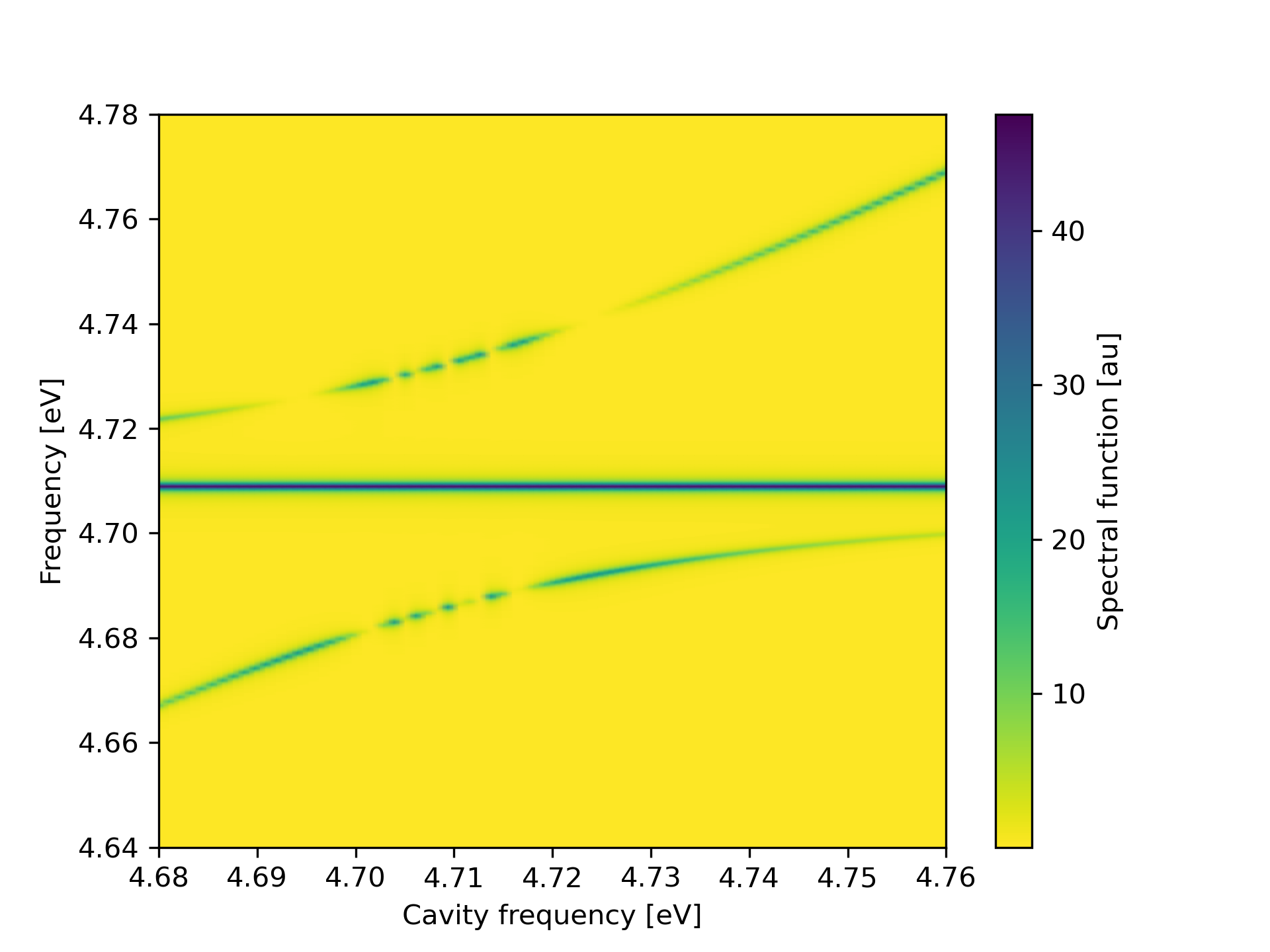}
    \caption{Hg JC}
    \label{fig:JC_Hg_param_2Dspect}
  \end{subfigure}
  \label{fig:JCvsQEDFT}
\end{figure}

\section{Geometry of mercury porphyrin}

\begin{table}[H]
\centering
\caption{ Molecular geometry of Hg@porphyrin. }
\begin{tabular}{lrrrclrrr}
\toprule
Atom & $x$         & $y$         & $z$        & &  Atom & $x$         & $y$         & $z$            \\
\midrule
N   &   -0.011132 &   0.000000  &    2.169275 & &   H   &   -0.445739 &  -5.174990  &    1.344482    \\ 
N   &   -0.011132 &   2.169275  &    0.000000 & &   N   &   -0.011132 &   0.000000  &   -2.169275    \\ 
C   &   -0.164467 &   2.451269  &    2.451269 & &   C   &   -0.164467 &   2.451269  &   -2.451269    \\ 
C   &   -0.313340 &   0.686057  &    4.317622 & &   C   &   -0.313340 &   0.686057  &   -4.317622    \\ 
C   &   -0.135961 &   1.122057  &    2.938936 & &   C   &   -0.135961 &   1.122057  &   -2.938936    \\ 
C   &   -0.135961 &   2.938936  &    1.122057 & &   C   &   -0.135961 &   2.938936  &   -1.122057    \\ 
C   &   -0.313340 &   4.317622  &    0.686057 & &   C   &   -0.313340 &   4.317622  &   -0.686057    \\ 
H   &   -0.280223 &   3.219998  &    3.219998 & &   H   &   -0.280223 &   3.219998  &   -3.219998    \\ 
H   &   -0.445739 &   1.344482  &    5.174990 & &   H   &   -0.445739 &   1.344482  &   -5.174990    \\ 
H   &   -0.445739 &   5.174990  &    1.344482 & &   H   &   -0.445739 &   5.174990  &   -1.344482    \\ 
Hg  &    0.381402 &   0.000000  &    0.000000 & &   C   &   -0.164467 &  -2.451269  &   -2.451269    \\ 
N   &   -0.011132 &  -2.169275  &    0.000000 & &   C   &   -0.313340 &  -0.686057  &   -4.317622    \\ 
C   &   -0.164467 &  -2.451269  &    2.451269 & &   C   &   -0.135961 &  -1.122057  &   -2.938936    \\ 
C   &   -0.313340 &  -0.686057  &    4.317622 & &   C   &   -0.135961 &  -2.938936  &   -1.122057    \\ 
C   &   -0.135961 &  -1.122057  &    2.938936 & &   C   &   -0.313340 &  -4.317622  &   -0.686057    \\ 
C   &   -0.135961 &  -2.938936  &    1.122057 & &   H   &   -0.280223 &  -3.219998  &   -3.219998    \\ 
C   &   -0.313340 &  -4.317622  &    0.686057 & &   H   &   -0.445739 &  -1.344482  &   -5.174990    \\ 
H   &   -0.280223 &  -3.219998  &    3.219998 & &   H   &   -0.445739 &  -5.174990  &   -1.344482    \\ 
H   &   -0.445739 &  -1.344482  &    5.174990 & &       &             &             &                \\ 
\bottomrule
\end{tabular}
\end{table}
